\documentstyle[emulateapj, epsfig,psfig]{article}
\def\ltsima{$\; \buildrel < \over \sim \;$}
\def\simlt{\lower.5ex\hbox{\ltsima}}   
\def\gtsima{$\; \buildrel > \over \sim \;$}
\def\simgt{\lower.5ex\hbox{\gtsima}}

\begin{document}

\title{The metamorphosis of tidally stirred dwarf galaxies}

\author{Lucio Mayer $^1$, Fabio Governato $^2$, 
Monica Colpi $^3$, Ben Moore $^4$,
Thomas Quinn$^1$, James Wadsley$^5$, Joachim Stadel$^1$ \& 
George Lake$^1$}

\affil{$^1$ Department of Astronomy, University of Washington,
Seattle, USA, mayer@astro.washington.edu,trq@astro.washington.edu,
stadel@astro.washington.edu,lake@astro.washington.edu\\
$^2$ Osservatorio Astronomico di Brera,
via Bianchi 46, I--23807 Merate (LC) - Italy, fabio@merate.mi.astro.it\\
$^3$Dipartimento di Fisica, Universit\`a Degli Studi di
Milano Bicocca, via Celoria 16, I-20133 Milano, Italy, Monica.Colpi@mib.infn.it\\
$^4$Department of Astronomy, University of Durham, Durham U.K, 
Ben.Moore@durham.ac.uk\\
$^5$Department of Physics and Astronomy, McMaster University, Hamilton,
Ontario,L8S 4M1, Canada,wadsley@physics.mcmaster.ca}

\begin{abstract}

We present results from high-resolution N-Body/SPH simulations of
rotationally supported dwarf irregular galaxies moving on 
bound orbits in the massive dark matter halo of the Milky Way. 
The dwarf models span a range in disk surface density 
and the masses and sizes of their dark halos are consistent 
with the predictions of cold dark matter cosmogonies. 
 We show that
the strong tidal field of the Milky Way determines severe mass
loss in their halos and disks and induces bar and bending instabilities
that transform low surface brightness dwarfs (LSBs) 
into dwarf spheroidals (dSphs) and high surface brightness dwarfs (HSBs)
into  dwarf ellipticals (dEs) in less than 10 Gyr. The final
central velocity dispersions of the remnants are in the range $8-30$ km/s 
and their final  $v/\sigma$ falls to values $< 0.5$, matching well 
the kinematics of early-type dwarfs. The transformation requires 
the orbital time of the dwarf to be $\simlt 3-4$ Gyr, which implies a halo 
as massive and extended as predicted by hierarchical models
of galaxy formation  to explain the origin of even the farthest dSph 
satellites of the Milky Way,  Leo I and Leo II.
We show that only dwarfs with central dark matter densities as high as those
of Draco and Ursa Minor can survive for 10 Gyr in the proximity
of the Milky Way: this is naturally achieved within hierarchical 
models, where the densest objects should have 
small orbital times due to their early formation epochs. 
Part of the gas is stripped and part
is funneled to the center due to the bar, generating one
strong burst of star formation in HSBs
and smaller, multiple bursts in LSBs. 
Therefore, the large variety of star formation
histories observed in LG dSphs naturally arises 
because different types of dIrr progenitors respond differently to
the external perturbation of the Milky Way.
Our evolutionary model automatically explains the morphology-density relation
observed in the LG and in other nearby loose groups.
Extended low-surface brightness stellar and gaseous
streams originate from LSBs and follow the orbit of the dwarfs
for several Gyr. Due to their high velocities, unbound
stars projected along the line of sight can lead to overestimate 
the mass-to-light ratio of the bound remnant by a factor $\simlt 2$,
but this does not eliminate the need of 
extremely high dark matter contents in some of the dSphs.


\end{abstract}

\section{Introduction}

Since the fundamental review by Paul Hodge in 1971 the known dwarf population
of the Local Group has more than doubled. The current census of dwarf
galaxies, faint objects with luminosities $-18 < M_B < -9$,
is around 40, including all objects that lie within or very close to the
zero-velocity surface of the Local Group
(Mateo 1998; Van den Bergh 1999).
These tiny galaxies are either dwarf irregulars (dIrrs), 
dwarf ellipticals (dEs) or dwarf spheroidals (dSphs) 
(Grebel 1998; Van den Bergh 1999).
dIrrs ($M_B \le -11$) are irregularly shaped galaxies
with recent or ongoing star formation and generally quite low
surface brightness ($\mu_B \sim 23$ mag arcsec$^{-2}$).
The dEs have ellipsoidal shapes, are fairly luminous ($-17 < M_B < -15$) and 
typically have a central surface brightness ($\mu_B \le 21$ mag arcsec$^{-2}$)
higher than that of bright dIrrs like the LMC
or IC10 (Mateo 1998). Similarly
to their counterparts found in galaxy clusters (Ferguson \& Binggeli 1994;
James 1991) some of them have blue nuclei 
(Bica et al. 1990; Jones et al. 1990).
They contain mainly old ($> 10$ Gyr) and intermediate-age 
(1-10 Gyr) stellar populations and show in part recent star formation.
The dSphs number some of the least luminous ($M_B \sim - 9$) 
and lowest surface brightness galaxies known to date 
($\mu_B \ge 24$ mag arcsec$^{-2}$).
They do not have nuclei nor a significant
central concentration and are dominated by old or intermediate-age 
stellar populations. Star formation in some of them has ceased a few Gyr ago, 
while in others it has been more extended in time and episodic
(Hurley-Keller et al. 1998;Grebel 1999).
A few galaxies have properties intermediate between dIrrs and dSphs 
and may be evolving from dIrrs to dSphs (Grebel 1999).
A clear progression from dIrrs to dSphs is evident
in their HI-to-total baryonic mass (stars + gas), ranging between 7\% and
50 \% for dIrrs, between 1\% and 10\% in transition systems, while being
lower than 0.1 \% in dSphs and dEs.

The Local Group dwarf galaxies shows a striking morphology-density relation
(Grebel 1998), which seems to be featured even in other nearby loose groups 
(Karachentsev et al. 2000): dIrrs are found far
away from either the Milky Way or M31, while the early-type
dwarfs, dSphs and dEs, lie  
within 300-400 kpc from the center of the primary galaxies.
The distances of dSphs and dEs from the center of the giant spirals
are sufficiently small to postulate that
they are satellites of the dominant spirals, moving on
bound orbits in their large dark matter halos (Peebles et al. 1989).

Kinematical analysis of dwarf galaxies in the Local Group highlights 
an important distinction between
early-type dwarfs and dIrrs, 
namely {\it their different angular momentum content}.
The stars in dSphs and dEs are supported by velocity 
dispersion (typically $v/\sigma < 0.5$, see Mateo 1998 and Bender et al.
1992), while the neutral hydrogen is known to rotate in nearly all
dIrrs ($v/\sigma > 1$). The kinematics of the stellar
component is still not available for dIrrs but equilibrium arguments suggest 
that the stars should be arranged in a rotating disk component like the gas.
Only the faintest LG dIrrs, like SagDIG and GR8 ($M_B > -12$), are
dominated by rotation inside their half-right radius but
are pressure supported at large
radii (Carignan et al. 1990; Hoffman et al. 1996; Lo et al. 1993).

Despite these differences, the surface brightness profiles of dIrrs and
early-type dwarfs
are quite similar. Traditionally, fits with exponential profiles are preferred for dIrrs while King profiles are used for dSphs and dEs
(Ferguson \& Binggeli 1994). However, in most cases an exponential law also reproduces
the surface brightness profiles of early-type dwarfs well
(Mateo 1998; Ferguson \& Binggeli 1994; Faber \& Lin 1983; Hodge
et al. 1991a,b; Irwin \& Hatzidimitriou 1995). 
In addition, the luminosity grows with increasing surface brightness
in dSphs and dEs, similarly
to  dIrrs and spirals (Ferguson \& Binggeli 1994; 
Binggeli \& Cameron 1991; Bender et al. 1992).

Moreover, both dSphs and dIrrs are dark matter dominated and show an
anti-correlation between their mass-to-light ratio and their
luminosity (Mateo 1998): Draco and Ursa Minor, the faintest dSphs
($M_B > -10$),
have $M/L > 25$ (Lake
1990)) and the dimmest dIrrs, GR8 and SagDIG ($M_B > -12$) 
have $M/L \sim 30$.

In summary, the similarities existing between dSphs and dIrrs
are enough to postulate an evolutionary link between them.
For this to happen one needs
a mechanism which not only removes the cold gas from dIrrs but also 
drains internal angular momentum and heats the stellar disks.

Supernovae feedback is by far the most widely quoted mechanism to
expel gas from dwarf galaxies. In the model by Dekel \& Silk (1986)
supernovae can induce substantial gas removal in galaxies with a low
circular velocity, $V_c \le$ 100 km/s. For a dwarf
dominated by its dark matter halo, any gas loss would have a 
marginal dynamical effect on the stellar system that is left 
behind (Navarro et al. 1996; Burkert \& Silk 1997; Gelato \& 
Sommer-Larsen 1999): as a consequence, after star formation has stopped
and the galaxy has faded, the ``newly born'' dSph will maintain the 
profile and the correlation between luminosity and surface brightness
present in its former dIrr state.
However, Mac Low and Ferrara (1999), who perform a detailed
calculation of the energy balance,
have shown that complete blow out of gas occurs only in very small halos, with $V_c <15$ km/s, while, for example, all bright dEs likely reside 
in substantially more massive halos (Mateo 1998).

An alternative scenario has been proposed in which the gas is kept hot and
star formation is inhibited at high-redshift by the photoionizing cosmic 
UV background in dwarfs with halos corresponding to $V_c \le 30$ km/s
(Babul \& Rees 1992; Efstathiou 1992; Quinn et al. 1996; Bullock et al. 2000): 
the stellar component observed in dSphs would have formed prior
to the reionization epoch, while dIrrs would have appeared
at low redshift, when the intensity of the background has sufficiently
dropped.

However, neither of these mechanisms acts preferentially close 
to the primary galaxies and thus 
no explanation is provided for the morphology-density relation.
Moreover, they offer no way to explain the different angular
momentum content of the two types of dwarfs.

Among mechanisms related to the local environment, the
ram pressure stripping scenario has been
invoked frequently for the dwarfs in the Local Group (Einasto et
al. 1974, Faber \& Lin 1983, Van den Bergh 1999, Blitz \& Robishaw
2000): a hot gaseous halo surrounding the disk of the Milky Way 
or M31 would strip the gas from the shallow potential
wells of dwarfs moving through 
(Gunn \& Gott 1972). Sophisticated numerical simulations have 
shown the effectiveness of this mechanism
in the inner regions of galaxy clusters like Coma (Abadi et al. 1998;
Quilis et al. 2000).
However, observations indicate that the hot gaseous
corona of the Milky Way is three
order of magnitude less dense than the core of rich clusters.
A recent work by Murali (2000) has shown that a coronal gas density
$n_H< 10^{-5}$ cm$^{-3}$, at $50$ kpc, is necessary
for the Magellanic Stream not to be dissolved by ram-pressure
driven evaporation in less than 500 Myr.
Halo models derived from ROSAT X-ray observations
also yield comparable upper limits (Kalberla \& Kerp 1998), while
models in which LG dwarfs lose most of their gas due to ram pressure
(Gallart et al. 2000; Einasto et al. 1974) 
require gas densities more than one order of magnitude higher.
In addition, even if the coronal gas density were substantially larger
than currently estimated (e.g. Blitz \& Robishaw 2000),                        ram pressure stripping would
be unable to affect the structure and kinematics of the
pre-existing stellar component. Thus, although we cannot exclude
that ram pressure is at play, it hardly seems to be the key evolutionary
driver for LG dwarfs.

In galaxy clusters, fast fly-by collisions of galaxies with the 
largest cluster members plus a secondary contribution of the tides
raised by the global potential of the cluster,
are capable of transforming disk galaxies in
objects resembling S0 or spheroidal galaxies (Moore et al. 
1996, 1998, 1999) by removing both mass and angular momentum.
This mechanism, known as {\it galaxy harassment} affects all galaxy 
components and is purely gravitational.
For harassment to be effective, 3-4 
collisions per galaxy with a galaxy as massive as the Milky Way,
namely having a circular velocity $V_c$ $\sim 220$ km/s, 
corresponding to roughly $0.15 V_{clus}$ (where $V_{clus}$ is the
circular velocity of a Coma-sized cluster, $\sim 1500$ km/s), are required.
This occurs in clusters, where the collision rate ${\cal {R}}$ is
around 1/Gyr, as follows from ${\cal{R}}= N  \sigma S$, where N 
is the number density of Milky Way-sized galaxies ($\sim 30$ Mpc$^{-3}$), 
in a rich Coma-like cluster (of radius $\sim 1.5$ Mpc), $\sigma$ is 
the velocity dispersion of the cluster ($\sim 1000$ km/s) and $S$
is the cross section of the target galaxy (here we are considering 
the extent of a typical $L_*$ galaxy out to the virial radius of 
the dark matter halo, $\sim 300$ kpc, as it will take a few Gyr for 
a galaxy to approach the cluster core and be significantly
tidally truncated).
Rescaling the calculation to the case of an LG dwarf satellite of either
the Milky Way or M31 
colliding against a fairly large satellite with $V_c \ge 30$
km/s (namely comparable to the Small Magellanic Cloud
and having $V_c \sim 0.15 V_{MW}$, where  $V_{MW}$ is the circular velocity
of the Milky Way)
one obtains only  $\cal{R} <$ 0.1/Gyr: this is because the number of
observed satellites in the halo of the Milky Way or M31 
is much lower than the number of MW-like galaxies in clusters (Moore
et al. 1999; Klypin et al. 1999). Thus, unless we are missing a substantial
population of satellites, either because they are completely dark 
or because their baryons have not yet been turned into stars due to
some cooling-halting mechanism (Blitz et al. 1999), harassment is negligible 
in the Local Group.

If close-by encounters between dwarfs are rare in the Local Group, only
the tidal field of the primary halos can substantially reshape these
small galaxies.
Gnedin et al. (1999) found, using a generalized semi-analytical 
formulation based on the impulsive approximation and including
adiabatic corrections, that the heating of a satellite by the
tides of the primaries can be very strong when the orbits are 
eccentric or have pericenters close to the center of the primary.
These results are also supported by the more sophisticated
time-dependent perturbation theory of gravitational shocks developed 
by Weinberg (1994a,b,c). Typical orbits of subhalos 
in structure formation models
are fairly eccentric, having an average  apocenter to pericenter 
ratio $\sim 5$ 
(Ghigna et al. 1999). 
The orbital time in the Milky Way halo is roughly
the same as in a Coma-like cluster, as $T_{orb} \sim {2 \pi R}/V_c$ and
$R \sim  V_c$ for virialized structures (White \& Frenk 1991), and is
$\sim 3$ Gyr for an object on a circular orbit halfway from the
center of the primary halo. However, the Milky Way satellites
should have performed a higher number of orbits than galaxies in virialized 
clusters because the latter typically appeared 5 Gyr ago (Rosati et al. 2000), 
while the Galaxy was already in place at least 10 Gyr ago (Van den Bergh 1996).

Semi-analytical approaches limit the analysis to the amount of energy
that can be injected by tides into a satellite, being unable to handle
with the nonlinear mechanisms  possibly excited by such energy input, such as
gravitational instabilities in the stellar component, that can lead 
to substantial morphological evolution.
In Mayer et al. (2001) we investigated for the first time the fate of
``tidally stirred'' dwarf irregulars orbiting the halo of the Milky Way
using high-resolution N-Body/SPH simulations. We found that the 
general properties of their remnants match well those of the dSphs and
dEs in the Local Group, additionally providing a natural explanation
for their observed morphological segregation.
In this paper we carefully describe the techniques 
used to construct realistic replicas of dwarf irregulars, and
we study in depth their dynamical evolution and the internal structure 
of their remnants. In addition, by carrying out an unprecedented number 
of high-resolution N-Body simulations (more than 50), 
we explore a much wider parameter space in terms of both
the orbital parameters and the internal structure of the satellites,
obtaining a statistical sample for a detailed  comparison with observations.
Finally, with our present study we aim to identify the basic conditions
for tidal stirring to be effective as well as at testing 
its predictions within the framework of hierarchical structure formation
against observations. 
The simulations were performed with PKDGRAV (Stadel \& Quinn, in preparation)
, a fast,
parallel binary treecode, or, when a gas component is also included, 
with a newer version of the same code, GASOLINE, which 
implements an SPH algorithm 
for solving the hydrodynamical equations and radiative cooling
(Wadsley, Stadel \& Quinn, in preparation).

The outline of the paper is as follows. In section 2 and 3 we illustrate,
respectively, the initial dwarf galaxy models and orbital configurations.
In section 4 we follow the dynamical evolution of the satellites, while
section 5 is devoted to the structural properties of the remnants.
In section 6 we test the tidal stirring scenario, varying the
internal structure of the satellites, the orbits and
the structure of the primary halo.
In section 7 we derive the star formation history of the satellites,
and section 8 concerns the gross observational
properties of the remnants. Finally, in section 9 we discuss the main
results and draw our conclusions.

\section{Models of satellites}

The models of dwarf irregular galaxies are constructed using the technique
of Hernquist (1993), which has been extensively shown to produce stable
configurations for systems with more than one component. In the present
work, galaxies normally comprise an exponential stellar disk embedded in
an extended  truncated isothermal dark matter halo. 
In some cases a cold gas component is also present in the disk.
The profile of the dark matter halo is given by:

\begin{equation}
     \rho _h (r) \, = \, {{M_h}\over{2\pi ^{3/2}}} {{\alpha}\over
      {r_c}} \, {{\exp (-r^2/{r_t^2})}\over{r^2+ {r_c}^2}} 
\end{equation}

where $M_h$ is the halo mass, $r_t$ serves as a cut--off radius,
$r_c$ is the ``core'' radius, and $\alpha$ is a normalization
constant defined by 
     $\alpha \, = \, \left [1 \, - \, \sqrt{\pi} q \exp (q^2)
      \left ( 1 - {\rm erf} (q) \right ) \right ] ^{-1} \ $,
where $q=r_c / r_t$.  
The stellar disks initially follow the density profile (in cylindrical
coordinates):

\begin{equation}
 \rho _d (R,z) =  {{M_d}\over{4\pi {r_h}^2 z_0}} \exp (-R/r_h) {\rm sech}^2 \left ( {{z\over{z_0}}} \right ) 
\end{equation}

where $M_d$ is the disk mass, $r_h$ is the radial scale-length, and
$z_0$ is the vertical scale-thickness.  

The particles used to represent the gaseous disk follow roughly
the same density profile of the disk stars but their
vertical scale--height depends on the temperature of the gas;
as an example, for a gas temperature $T=10^4$ K the gas layer is
thinner than the stars for vertical stellar dispersions similar to 
those  in the Milky Way disk.

Our procedure for ``building'' dwarf galaxies 
involves the joint use of observations and theoretical models of structure
formation. We build two basic galaxy models
, a high surface brightness (HSB) dwarf
and a low  surface brightness (LSB) dwarf (HM1 and LM1, respectively,
described in section 2.1).
We then construct other models by keeping fixed
the mass of the halo and disk while varying the other
structural parameters (section 2.1).
We also construct models with different total masses (section 2.2 and 2.3) 
using the scaling relations for objects forming in cold dark matter 
cosmogonies (e.g. Mo, Mao \& White 1998).

\subsection{Models at fixed mass}

In a scenario in which galaxies form from the cooling of gas inside
dark matter halos (White \& Rees 1978, White \& Frenk 1991), the baryons settle
into a disk rotating in centrifugal equilibrium within the halo, and thus
the rotational velocity equals the circular velocity of the halo.
Therefore, once we
fix the circular velocity $V_c$ of the halo,
we determine the luminosity of the embedded disk
according to the Tully-Fisher relation.
For our basic models, LM1 and HM1,
we choose a halo circular velocity $V_c=75$ km/s.
Zwaan et al. (1995) have shown that HSB
and LSB disk galaxies obey
the same Tully-Fisher (T-F) relation in the B-band 
(although LSBs show more scatter), $L_B \sim {V_c}^4$. 
A rotational velocity of $75$ km/s implies $M_B \sim -18$
according to the T-F relation; these parameters correspond to the brightest
dwarf irregular galaxy in the LG, the LMC (Mateo 1998; Sofue 1999).

\medskip
\epsfxsize=9truecm
\epsfbox{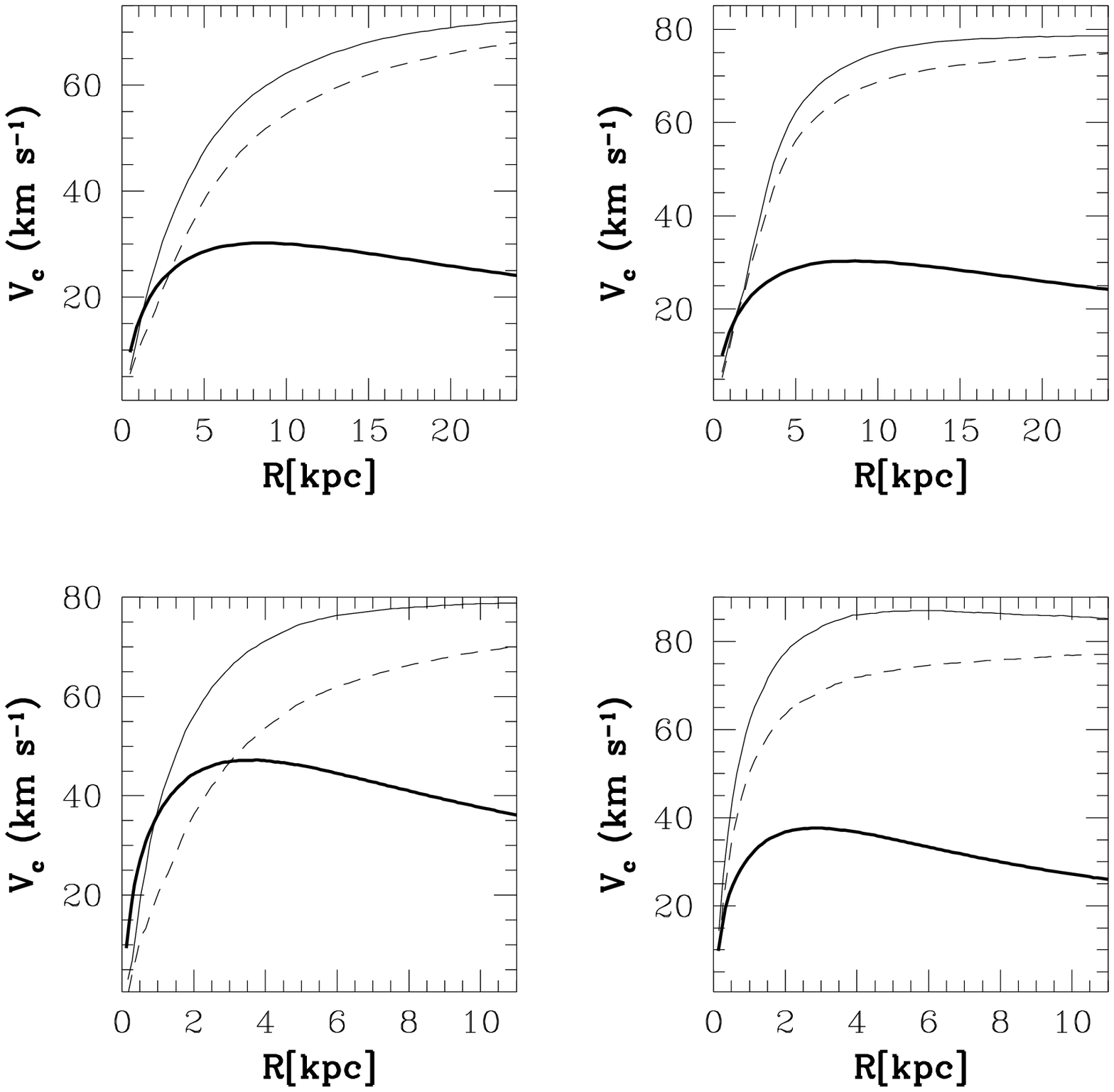}
\figcaption[f1.eps]{\label{fig:asymptotic}
\small{Rotation curves of LSB and HSB models plotted out to 5$r_h$. From
top left to bottom right, we show models LM1, LM1rc03, HM1 and HM1rc03.
Rescaled models (see text) have rotation curves like those
of models LM1 and HM1, albeit with different intrinsic scales.
The thin dashed line and thick solid line represent 
the separate contribution of, respectively,
dark matter and stars, while the thin
solid line represents their sum (see Table 1 for details).
}}
\medskip

In order to  satisfy the same T-F relation 
LSBs must have larger disk scale-lengths compared to equally luminous HSBs,
because $L \propto \Sigma_d {r_h}^2$ (where $\Sigma_d$ is the surface
luminosity).
For the disk scale length $r_h$ of the HSB we chose a value of  $2$ kpc,
while the LSB model has a disk scale
length of $4.8$ kpc. Both values are in agreement with the bimodal $V_c - r_h$
relation in Zwaan et al. (1995). 
To fix the disk mass of both HM1 and LM1 we assume
$(M/L)_{*B} = 2$, following Bottema (1997), which yields a mass of
$3.22 \times 10^{9} M_{\odot}$.

Once the circular velocity is known, the virial mass $M_{200}$ and 
virial radius $R_{200}$ of the halo (which we always equate to the 
tidal radius of the halo, $r_t$, that appears in eq.(1)) follow 
immediately from 
 
\begin{equation}
M_{200}=\frac{V_{200}^3}{10 G H(z) },  \;\;\;\; \mbox{and}\;\;\;\;\;
R_{200}=\frac{V_{200}}{10 H(z)} 
\end{equation}

where, for the basic models, we assume $z=0$ and 
$H(z)= H_0 = 50$ km s$^{-1}$ Mpc$^{-1}$
The normalization in the above equations follows from assuming that the
critical overdensity for the collapse of objects is $200\rho_c$, where
$\rho_c$ is the critical density of the Universe, as appropriate for
a SCDM cosmology. In other more currently favored models,
like in LCDM, the value of the critical overdensity is
marginally different, and thus the final sizes and masses of halos
are only slightly affected (Eke et al. 1996). 
According to (1) we assign to our basic dwarf galaxy models a (halo)
virial mass
$M_{200} = 1.6 \times 10^{11} M_{\odot}$ and a virial radius 
$R_{200}=r_t=150$ kpc

By changing the
{\it concentration} $c=r_t/r_c$ at fixed $r_t$ and halo mass we
reproduce the rotation curves of both LSB and HSB dwarfs, as done
by de Blok \& McGaugh (1997) for a large sample of galaxies of both types
(see also Persic \& Salucci 1997).
Following again the same authors, we assume that
the halo and disk scale lengths are related and set $r_c= r_h$, thus making
HSBs halos more than twice as concentrated as LSB halos given the 
difference in their disk scale lengths. We also build two models in
which $r_h$ and $r_c$ are not equal to further explore the parameter
space. The rotation curves of the basic HSB and LSB models 
(LM1 and HM1) and of the latter two models (LM1rc03 and HM1rc03)
are shown in Figure 1.

\medskip
\epsfxsize=8truecm
\epsfbox{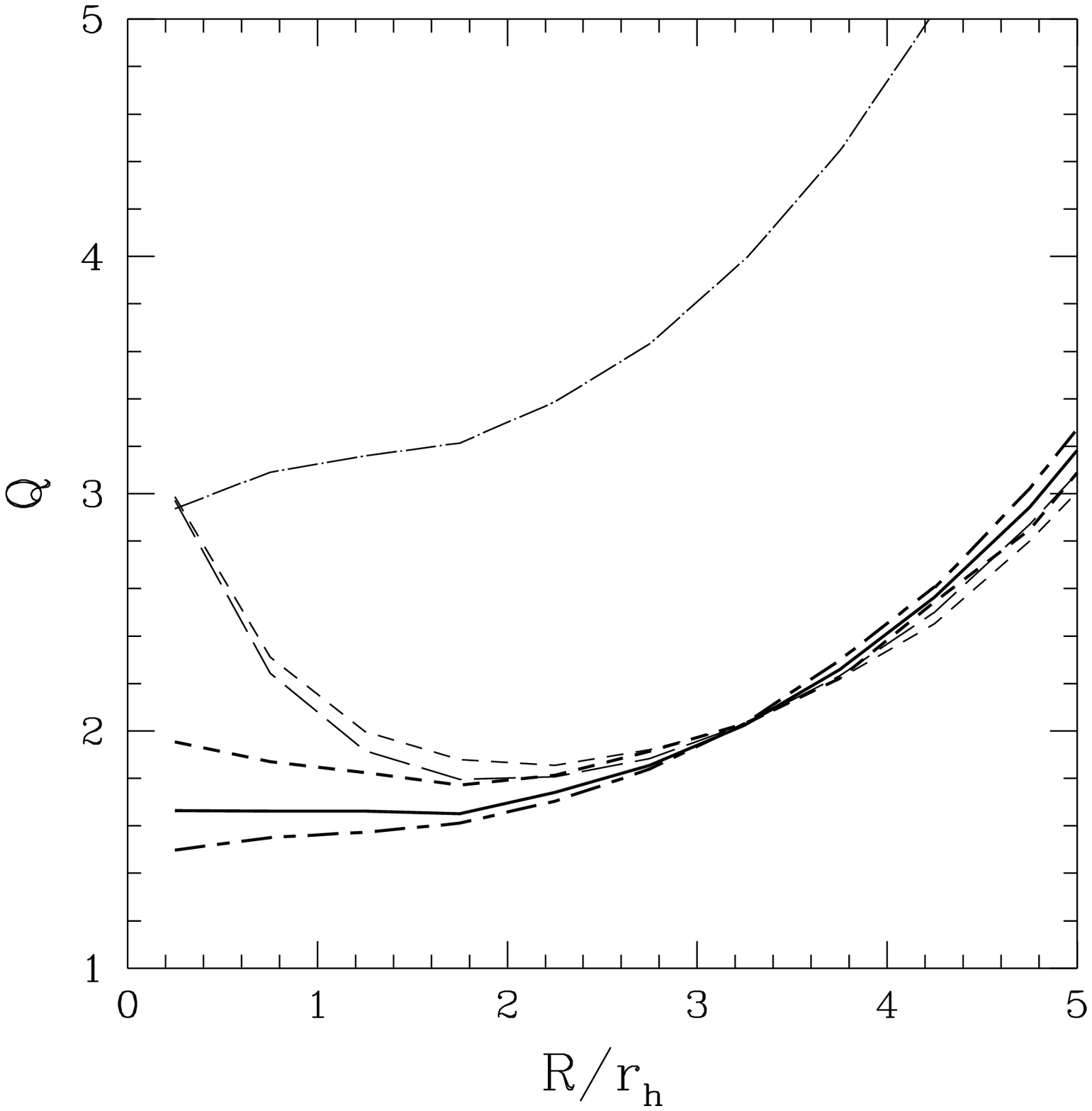}
\medskip
\epsfxsize=8truecm
\epsfbox{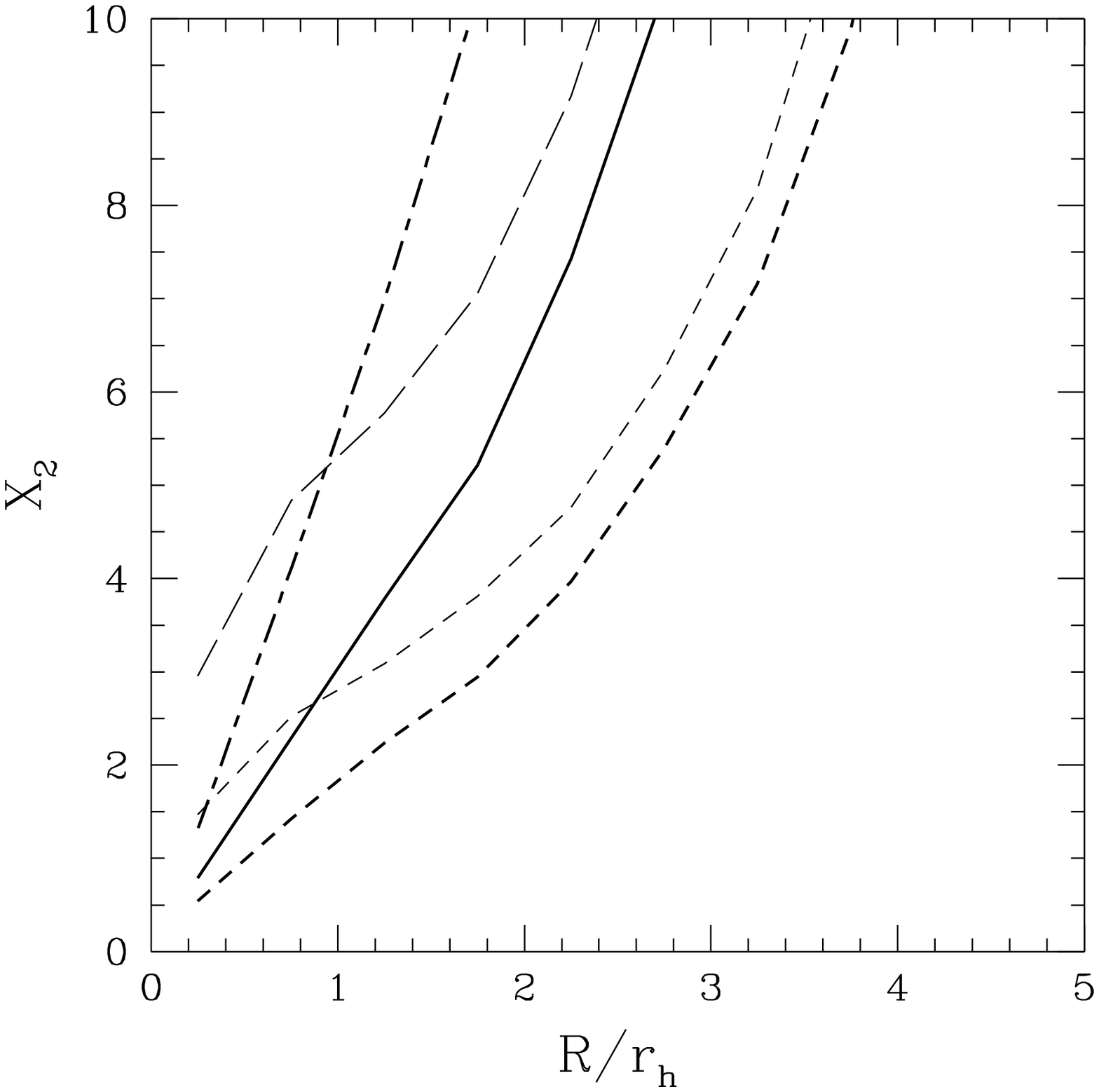}
\medskip
\figcaption[f2b.ps]{\label{fig:asymptotic} 
\small{Q and $X_2$ parameters of the various models as a function of
distance in units of the disk-scale length. The thick solid and short-dashed
lines refer, respectively, to models LM1 and HM1 (as well as to all 
their rescaled versions, see text). 
The thick long-short dashed line is used for the
GR8 model, the thin long-dashed line is for model LM1rc03, the thin short-dashed line is for model HM1rc03 and the dot-dashed line refers to model LM1Q4 
(see Table 1 for details on all the models).
}}
\medskip

In the models described so far the scale height of the disk is
fixed at $0.1 r_h$, as suggested by observations of late-type galaxies
(Van der Kruit \& de Grijs 1999, de Grijs et al. 1997), and we set $Q=2$ at $r=2.5 r_h$; detailed numerical studies have shown that 
a Toomre parameter $Q \geq 2$ is a necessary condition
for avoiding the growth of spontaneous bars 
(Athanassoula \& Sellwood 1986; Friedli 1999).

Different choices  of $Q$ and $z_s$ can determine, respectively, a different
susceptibility to the growth of bars and 
to bending instabilities, the latter being vertical
oscillations responsible, for instance, for the buckling 
of bars (Merritt \& Sellwood 1994; Raha et al. 1991).
Given the chosen parameters, the basic models have structurally normal 
disks. To further explore the issue of instabilities
we construct a small set of models with different values of $Q$ and $z_s$.
The rotation curves are not affected by the latter variation of parameters,
and they look the same as those of models LM1 and HM1 shown in Figure 1.

\medskip
\epsfxsize=8truecm
\epsfbox{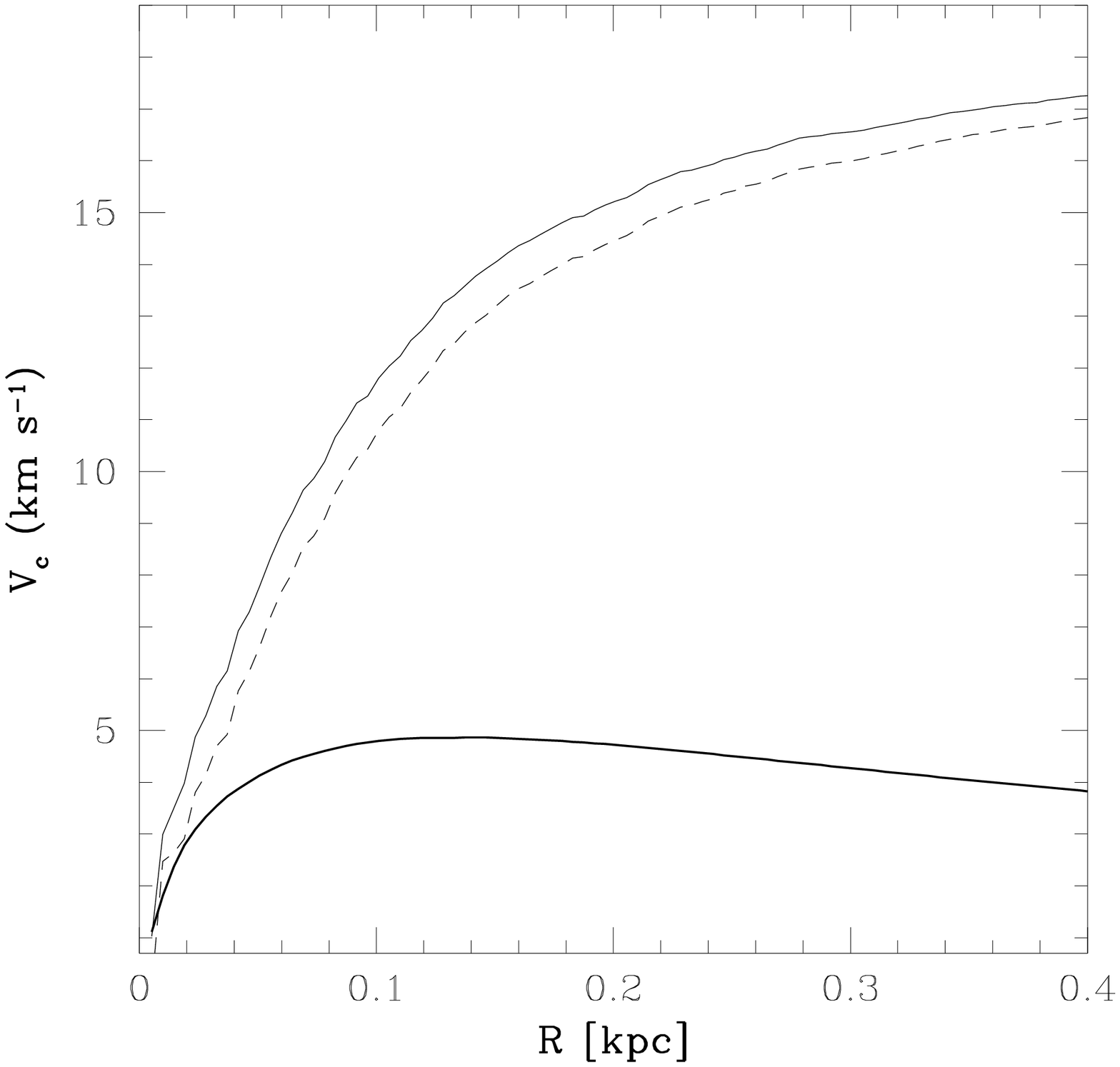}
\figcaption[f3.eps]{\label{fig:asymptotic}
\small{Rotation curve of the GR8 model plotted out to 5$r_h$. The
thin dashed line and thick solid line represent
the separate contribution of, respectively,
dark matter and stars, while the thin
solid line represents their sum (see Table 1 for details).
}}
\medskip

Figure 2 shows the stability properties of the various models employed
(including the model of GR8 described in subsection 2.3)
in terms of $Q$ and $X_2={{\kappa}^2 R \over 4 \pi G \Sigma_d}$ ($\kappa$ 
is the epicyclic frequency, $R$ is the disk radius and $\Sigma_d$ is the disk 
surface density), the latter being related to the swing amplifier
mechanism that can determine a rapid growth of the $m=2$ modes
responsible for the bar instability (Binney \& Tremaine 1987).
In Table 1 the structural parameters of the models
are listed; note that by assuming a smaller halo core radius $r_c$,
(smaller than the disk scale length $r_h$) the $X_2$ parameter increases.
N-Body simulations by Mihos et al. (1997) have shown
that in encounters of equal-mass galaxies,
$X_2 > 1$ in the inner
disk is enough to stabilize systems with slowly rising rotation 
curves like those of LSBs, to be stable against bar formation,
while $X_2 > 3$ is required for disks with 
steep rotation curves like those of HSBs. As in our simulations the perturber's
mass (the Milky Way) is at least 30 times larger than the mass of the target 
dwarf galaxy, we will be able to test the validity of this criterion in
a radically different situation. 
Finally, we will test the susceptibility to vertical bending instabilities
with model LMH (Table 1), identical to model LM1 except for
a larger scale-height ($z_s = 0.3r_h$); this 
should suffer little vertical heating (see Raha et al. 1991).

\subsection{Rescaled models}

To fix the parameters of model galaxies with different (smaller) masses we 
choose a value for the new mass and then we make use of equations (3)
to determine the new circular velocity $V_{200}$ and virial radius $R_{200}$
of the dark matter halo {\it assuming $z=0$ and thus $H(z)=H_0$}.
The disk masses and scale
lengths are assumed to simply follow the halo scaling 
(Mo et al. 1998). The resulting models are HM2 and LM2 (see Table 1).

As shown by eqs.(3), masses and radii
of objects formed in hierarchical cosmogonies depend on redshift {\it at fixed
circular velocity}. The scaling is contained in the Hubble constant:

\begin{equation}
H(z)=H_0{\left[\Lambda _0 +(1-\Omega_0-\Lambda_0){(1+z)}^2+\Omega_0
{(1+z)}^3\right]}^{1/2}
\end{equation}

and implies that radii and masses decrease at the same rate, i.e. $\sim {(1 + z)}^{-3/2}$. As a consequence the density of objects grows as $({1 + z)}^3$. 
Some of the dwarf spheroidal galaxies inhabiting the Local Group  
must have formed at $z=1$ or higher given the very old age of the bulk 
of their stellar populations.
Therefore, we also build models with
masses and radii rescaled as they were 
formed at $z=1$ (models LZ and HZ, see Table 1)
to investigate how the scaling with redshift implied
by the hierarchical scenario can affect their evolution.

The intrinsic stability properties of all these rescaled models are unchanged
with respect to the basic models because equations 3 and 4
define an invariant transformation with respect to both $Q$ and $X_2$.
Moreover, their rotation curves look exactly the same as those of the 
basic models shown in Figure 1, albeit on a different scale.

\subsection{The GR8 model}

So far we have used the Tully-Fisher relation to link circular velocities
of halos to luminosity.
Hoffman et al. (1996) have shown that the
T-F relation exhibits a large scatter at $M_B \ge -12$. 
The cosmological scaling
offers an attractive alternative to set the physical parameters for models
of the faintest dwarf irregular galaxies.
Examples of very faint dIrrs are found in the LG; both GR8 and SagDIG 
have $M_B > -12$.
It turns out that galaxies that faint should correspond
to halos with $M_{200} < 10^{9} M_{\odot}$, and that such halos 
form typically at $z \geq 2$
in cold dark matter cosmogonies (Lacey \& Cole (1993, 1994)).

We build a model resembling the GR8 dwarf using the kinematics of
the gas at large radii to construct its global potential.
At 2-3 disk scale lengths
the neutral hydrogen in GR8 is
pressure supported, with a peak velocity dispersion  $\sigma \sim 12-13$ km/s
(Carignan et al. 1990).  Assuming that the gas is in virial
equilibrium with a surrounding isothermal dark matter halo, the latter
should have a virial circular velocity  
$V_{200} = {\sqrt 2} \sigma \sim 17$ km/s: this fixes
the virial mass and virial radius of the halo through
equations (3) and equation (4), which are evaluated for $z=2$.
The disk scale length ($76$ pc) and stellar disk mass ($1.6 \times 
10^{6} M_{\odot}$) are the observed values (Carignan et al. 1990).
The chosen $r_h$ and $R_{200}$, assuming $r_c = r_h$,
imply a very concentrated halo 
and a very high halo/disk mass ratio (see Table 1).

The resulting central dark matter density is very high, $\sim
0.3 M_{\odot}/pc^2$. This value is slightly higher than inferred for
GR8 ($0.07 M_{\odot}/pc^2$) but similar to that inferred for Draco
and Ursa Minor (Lake 1990; Mateo 1998). The mass-to-light ratio at
the optical radius ($\sim 3 r_h$) is $\sim$ 32, like that of GR8
not counting the contribution of the gaseous disk. The
rotation curve of the ``GR8'' model is shown in Figure 3.

\subsection{Models with a gas component}

We also construct an LSB and an HSB model with a 
cold gaseous disk which should represent the neutral hydrogen
present in observed dwarf irregular galaxies. 
Dwarf irregular galaxies are generally gas-rich, with an average total
HI-to-stellar mass ratio larger than one (Hoffman et al. 1996).
However, the gaseous disks are typically two times more extended
than the stellar disks (Cote et al. 1997; de Blok \& McGaugh, 1997;
Young \& Lo 1997),
whereas within the optical radius ($\sim$ 3 disk scale lengths), 
the neutral hydrogen fraction often drops to less than $50\%$
of the stellar mass (Jobin \& Carignan 1990; 
Cote et al. 1990).

\medskip
\epsfxsize=9truecm
\epsfbox{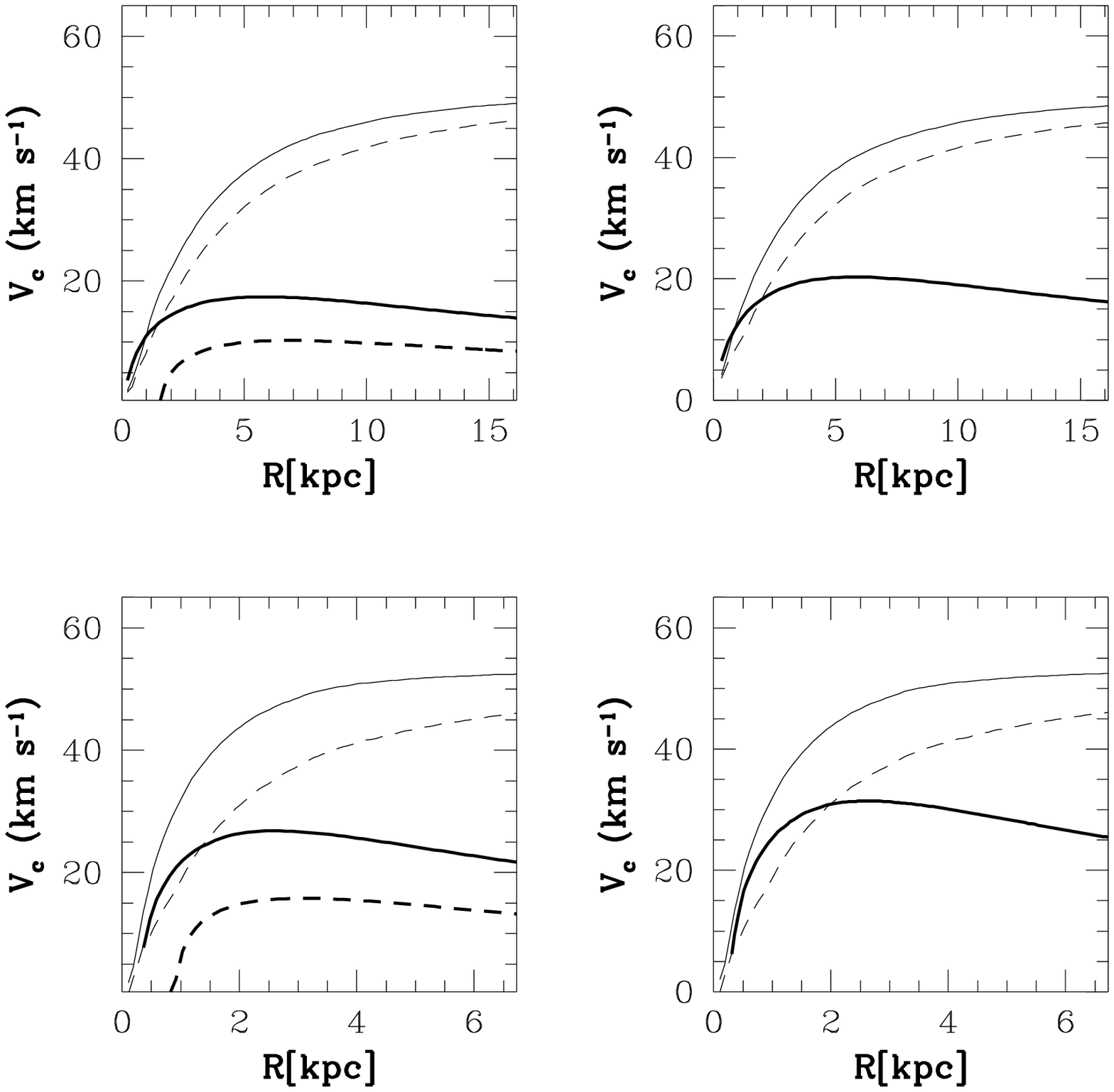}
\figcaption[f4.eps]{\label{fig:asymptotic} 
\small{Rotation curves of  models including a gaseous component along
with those of
their gas-free counterparts plotted out to 5$r_h$. At the top models
LMg2 (left) and LM2 (right) are shown, while models HMg2 (left) and HM2
(right) are shown at the bottom.
The thin dashed line and thick solid line represent
 the separate contribution
of, respectively,
dark matter and stars and the thick dashed line represents the
separate contribution of gas, while the thin solid line represents
their sum (see Table 1 for details).
}}
\medskip

As we will show, material lying outside $3r_{h}$ is almost entirely stripped
when the satellites orbit in the
Milky Way halo and thus would not affect the bound components.
Therefore, the gaseous disk included in our model galaxies
extends only out to the radius of the stellar disk
and its mass is 30\% of the total baryonic mass. The total disk 
mass (stars + gas) as well as all the other structural parameters
are shown in Table 1 (models LMg2 and HMg2).
The density drops to zero at a radius $R < 0.5r_{h}$ to mimic the 
'holes' in neutral gas distribution existing in the center of 
dwarf irregular galaxies (Cote et al. 1997).
The temperature of the gas is fixed at 5000 K after
stability tests in the adiabatic regime.
The rotation curves of both models are shown in Figure 4.

\subsection{Numerical stability of the models}

N-Body simulations are subject to
two-body relaxation effects. These can be particularly severe when
systems of different masses or a system of several components are present.
Massive halo particles may heat up the disk even in absence of
any external perturbation by colliding with lighter stellar or gas particles:
a similar numerical effect is known to be responsible for the premature
evaporation and disruption of small subhalos in cosmological simulations
with insufficient resolution (Moore, Katz, \& Lake 1996).
To minimize numerical disk heating we use lighter halo particles
in the central part of the halo (precisely within 10 disk scale lengths)
where orbital times are shorter and therefore collisions with
disk particles are more frequent. We perform several
tests in isolation in order to
find the optimal balance between reasonable stability and a moderate
total number of particles (see Figure 5).

The final models have a disk of $50.000$ particles while the halo is sampled
by 208350 in the LSB model and 180860 in the HSB model (the hi-res halo
component in the LSB extends out to a larger fraction of the virial
radius because the stellar disk does as well, resulting in a higher 
total number of particles).
For the GR8 model we had to use more than 
$3 \times 10^6$ particles in the halo due to the exceptionally high halo/disk
ratio: this is the {\it highest resolution model ever made for a single
galaxy} with a particle mass of only $\sim 50 M_{\odot}$.
In the simulations we use a spline kernel for the gravitational softening
as well as for the smoothing length in the SPH calculations 
(Hernquist \& Katz 1989). The softening length is a fixed fraction of 
the disk scale length $r_h$,
thus ensuring the same {\it intrinsic} spatial resolution for HSBs and LSBs
and for all their rescaled versions. In code units (for which $r_h = 1$)
the softenings are $s_d = 0.06$ for the stars, $s_{h1} = 0.32$ for the dark
matter particles in the hi-res component of the halo and
$s_{h2} = 0.4$ for those in the low-res component.  We adopt an opening angle 
$\theta=0.7$ for the force calculation and expand multipoles to
hexadecapole order. The timestep $\Delta t$ is assigned
using a multistepping criterion based on the local acceleration of particles,
namely $\Delta t = \eta \sqrt{s/a)}$, where $\eta$ is chosen
to be 0.3.

\medskip
\epsfxsize=5truecm
\figcaption[diskiso.ps]{\label{fig:asymptotic}
\small{Evolution in isolation of the disk model LM1. Shown is the
edge-on colour coded logarithmic density (the darker the colour, 
the lower the density). At T=3.5 Gyr a weak warp is apparent. The boxes are 30 kpc on a side.
}}
\medskip

In the collisionless runs the timestep varies in the
range $0.01 t_{cd} < \Delta t < t_{cd}$, where $t_{cd}$  is the crossing time
of the disks at $r=r_h$.
In the gasdynamical runs we use a Courant parameter = 0.4 and 
the timesteps can be as small as $2 \times 10^{-4} t_{cd}$.
Radiative cooling is implemented for a primordial mixture of hydrogen
and helium.
We use 20.000 particles to sample the gas component and 40.000 particles
for the stellar disk, minimizing any eventual collisional heating of the 
gaseous disk by the star particles since the ratio of the total masses
of the two components is $M_{gas}/M_{stars} \sim 0.3$
(otherwise two-body heating by stellar or dark particles may suppress 
radiative cooling of the gas, see Steinmetz \& White 1997).
Halos, in this case, have 250.000 particles in total.

\section{Initial Conditions}

The satellites are placed on bound orbits in the potential of a Milky-Way
sized primary halo. The latter is modeled as the fixed potential of an
isothermal sphere with a total mass $4 \times 10^{12} M_{\odot}$
inside a virial radius $R_{vp}= 400$ kpc, consistent with both recent measures
based on radial velocities of distant satellites (Wilkinson \& Evans
1999) and with generic models of structure formation (Peebles et
al. 1989).  The core radius is $4$ kpc and the resulting circular
velocity at the solar radius is $220$ km/s. Dynamical friction is
neglected because orbital decay times are expected to be longer than
the Hubble time given the small mass of the satellites (they are at least
$\sim 30$ time less massive than the primary) and the delay
introduced by tidal stripping (Colpi, Mayer, \&
Governato, 1999).

In the majority of the runs, which we will call  the ``standard runs'', 
the orbits have an apocenter close to the virial radius of the
Milky Way halo, in agreement with the results of
N-Body simulations of structure formation (Ghigna et al. 1998).
The same simulations show also that the orbits of satellites
have an average apo/peri $\sim 5$, while sporadic measurements of
the orbits of LG dSphs yield, although with large errors, nearly circular
orbits, with average apo/peri=2 (Schweitzer et al. 1995; Ibata \& Lewis 1998).
We explore mainly orbits with apo/peri between 4 and 10
but we also consider orbits with apo/peri as low as 1.5-2.
The orbital
energy $E_{orb}$ is usually  such that the corresponding circular orbit 
would  have a radius $\sim 0.5 R_{vp}$ and a typical period of $\sim 3$ Gyr.
We will refer to these orbits as to the {\it standard orbits}.
As the orbits considered have typically pericenters larger
than 40 kpc, the contribution of the Milky Way disk to the primary potential
should be negligible; we will test the validity of this hypothesis 
by adding an axisymmetric component to the external potential in a few runs.

\medskip
\epsfxsize=9truecm
\epsfbox{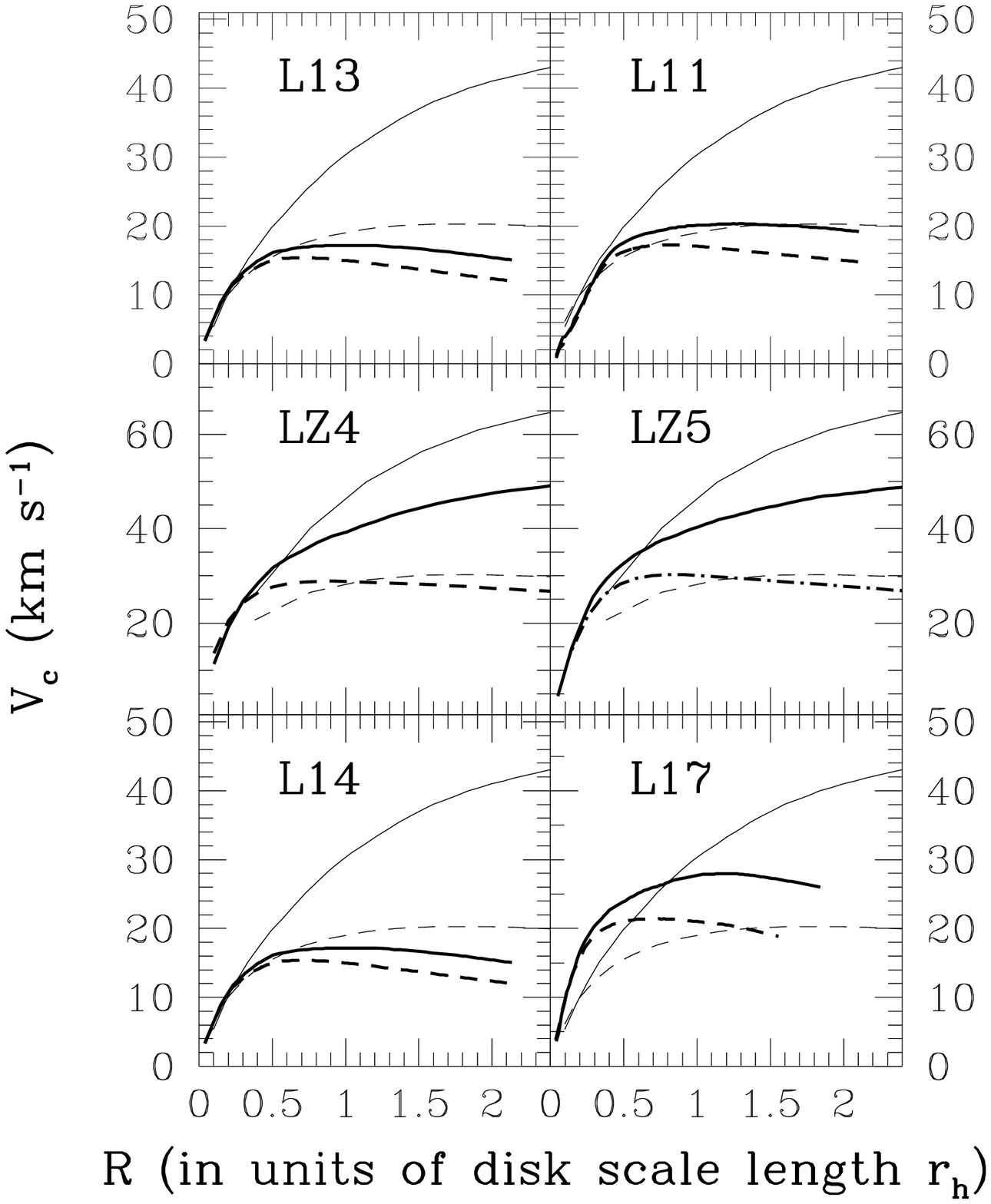}
\figcaption[f6.eps]{\label{fig:asymptotic} 
\small{Evolution of circular velocity profiles of LSB satellites on various
orbital configurations, labelled with the run names (refer to Table 2 and 
Table 3
for details). Thin and thick solid lines are, respectively, the initial
and final overall rotation curve, while thin and thick dashed lines
represent the separate contribution of stars.
It clearly appears that mass loss is less efficient 
in runs with LZ models.}}
\medskip

Satellites that entered the Milky Way halo at high redshift ($z \geq 1$)
became bound on orbits with a significantly smaller apocenter 
compared to satellites infalling later because the primary halo
itself was typically 2-3 times smaller (see equations (3) and (4)).
Therefore, we also consider the regime of {\it tightly bound orbits}, 
with apocenters ranging from $1/2$ to $1/3$ of the present-day 
virial radius of the Milky Way halo, thus having a orbital energy $E_{orb}$ lower
than that of the standard orbits.

The inclination of the disk spin with respect to the orbital angular momentum
is also varied, going from direct to retrograde spin-orbit orientations. 
The combination of different orbital parameters and different models
of satellites has led us to perform more than 50 high-resolution runs.
The orbital parameters and satellite models employed
in the various runs are indicated in Table 2 and Table 3.

\section{Dynamical evolution}

\subsection{Collisionless runs}

The dwarf galaxy models are evolved for several Gyrs in the external 
potential of the Milky Way halo.
The strong tidal disturbance exerted by the primary halo produces a dramatic
morphological evolution of the small satellites after 2-3 pericenter passages,
corresponding to 7 Gyr on the standard orbits. The satellites are first 
tidally truncated down to the radius imposed by their pericenter distance
and then continue to lose mass as each subsequent tidal shock 
pumps energy into them and decreases their potential wells 
(Gnedin \& Ostriker 1999). In this section we will focus on the
evolution of the models with the ``standard'' initial stability
properties (i.e. $Q=2$, $z_s=0.1r_h$ and $r_c=r_h$), while we
will discuss the effects of changing such properties in section 6.1.

In general, LSBs lose more than $90 \%$ of their 
halo mass and $\sim 60 \%$ of their stellar mass, while HSBs, although
still losing most of their halo mass, retain, on average, 
at least $70 \%$ of their initial disk mass.
A high-density dwarf model like GR8 is barely perturbed on a standard orbit
(run GR81), while it loses $\simlt 5 \%$ of its disk mass but most of its
outer halo mass if it falls on a short-period, tightly bound orbit ($T_{orb}
 \sim 1.3$ Gyr) as expected at $z \ge 1$ (run GR82),
performing five pericenter passages in
less then 7 Gyr.

\medskip
\epsfxsize=9truecm
\epsfbox{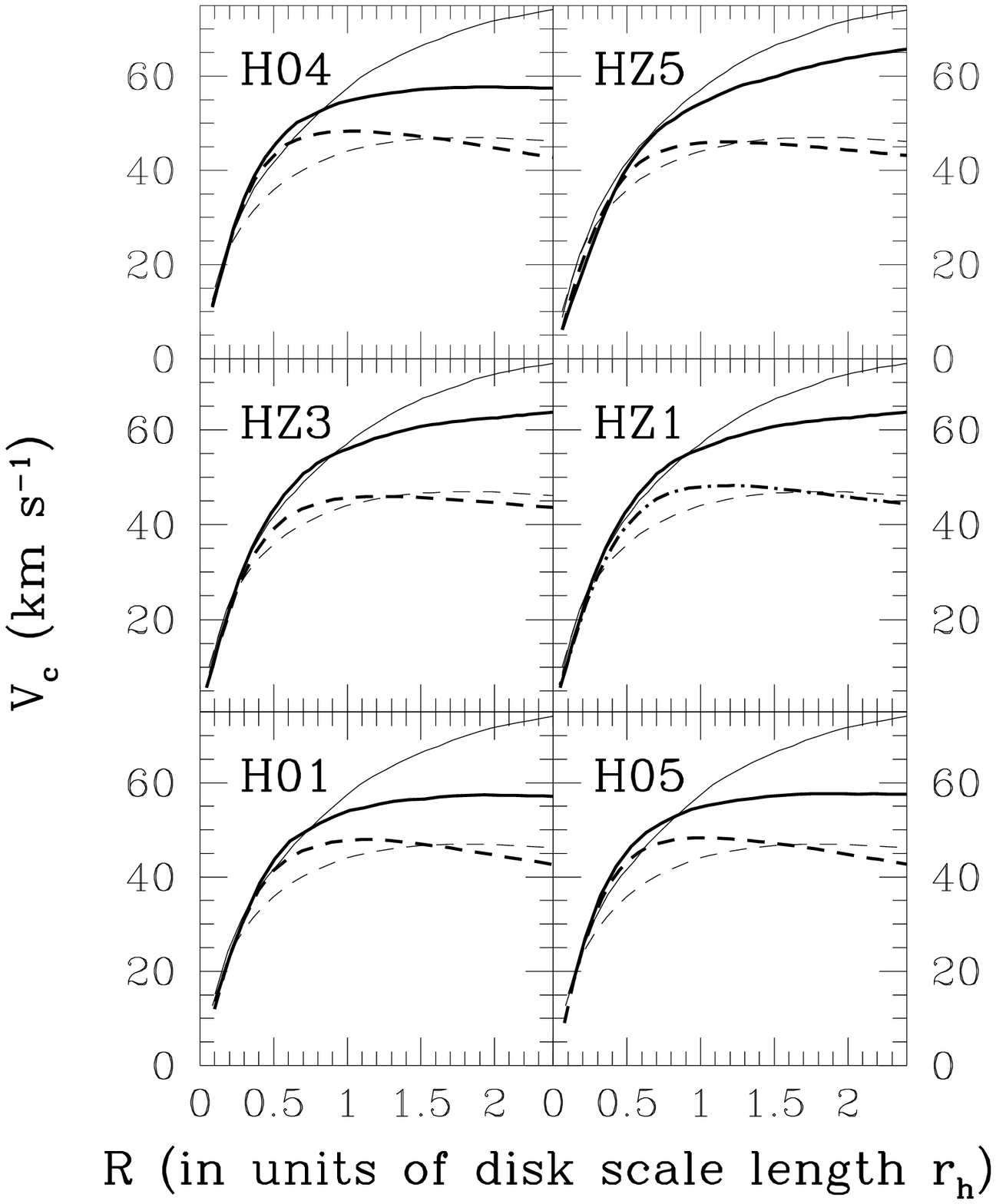}
\figcaption[f7.eps]{\label{fig:asymptotic}
\small{Evolution of circular velocity profiles of HSB satellites on various
orbits, labelled with the run names (refer to Table 3 for details).
Thin and thick solid lines are, respectively, the initial
and final overall rotation curve, while thin and thick dashed lines
represent the separate contribution of stars.
Tidal mass loss is less efficient in HZ runs.}}
\medskip

The evolution of the mass distribution for the various models is
reflected in the variation of the circular velocity profiles (Figures 6,7,8).
After 7 Gyr LSBs have stellar masses  $\sim 10^8 M_{\odot}$ while
HSBs have stellar masses closer to  $10^9 M_{\odot}$. 
LSBs are more fragile than HSBs
both because they have a low-concentration halo
and because they have large disk scale lengths. For the same disk mass,
the difference in scale length is a factor 2.4. Correspondingly, if
$t_c$ is the crossing time at the disk half-mass radius and 
$t_{coll}=R_p/V_p$ is the collision time (i.e. the time spent close
to pericenter $R_p$ while moving at the velocity $V_p$), for
$R_p \sim 40$ kpc as in most of the cases (see Table 2 and Table 3) 
we have $t_c/t_{coll} \sim 1$ for the LSBs, while  $t_c/t_{coll}
< 1$ for the HSBs, and thus the latter will respond more adiabatically
to the perturbation (Gnedin et al. 1999). Orbits with progressively larger
pericenter obviously produce a lower damage even in LSBs. 
Finally, limited stripping occurs in HZ and LZ models 
that have a  disk scale length 2.83 times smaller than their z=0 counterparts 
(Figures 6,7).

Close to the first pericentric passage both LSBs and HSBs develop a central
bar, its scale length being always $\sim r_h$.
The higher intrinsic stability of LSBs to bar formation 
(suggested by the $X_2$ parameter in Figure 2)
is thus not indicative when the perturber, namely
the Milky Way, is several times more massive than the target  satellite,
as was also argued by Miwa \& Noguchi (1998).
The GR8-like model develops 
very extended spiral arms after the first tidal shock. 
The second shock triggers the formation of
a central bar  which is more compact than usual,
being $\sim 0.5 r_h$ in length.

\medskip
\epsfxsize=8truecm
\epsfbox{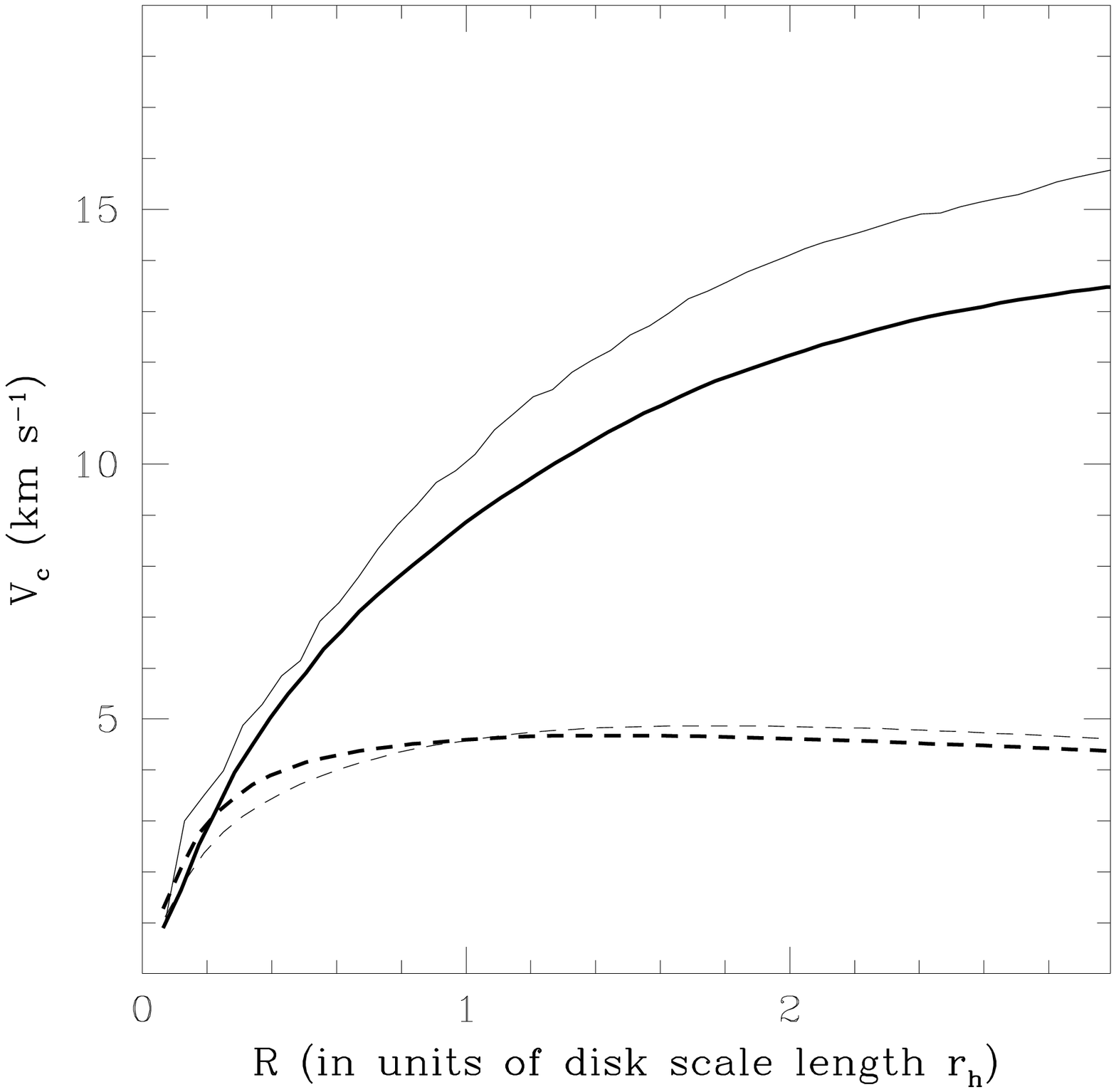}
\figcaption[f8.eps]{\label{fig:asymptotic}
\small{Evolution of circular velocity profile of the GR8 model
in run GR82 (see Table 3).
 Thin and thick solid lines are, respectively, the initial
and final overall rotation curve, while thin and thick dashed lines
represent the separate contribution of stars.}}
\medskip

The appearance of the bar structure corresponds to a drastic change
of the orbits of stars inside the bar radius: nearly circular orbits
turn turn into nearly radial box or tube orbits supporting the
elongation of the bar.
This change is illustrated in Figure 9, where the specific angular 
momentum of stars inside $r_h$ is plotted against time.
The bar leads to an increase in the central concentration of the stellar
density profiles, as explained in the next section. 
However, the degree of central concentration may vary depending on the strength
of the bar. In the case of LSBs the bar is weakened by 
tidal stripping as stars are removed even inside the bar radius when the 
pericenter distance is below 50 kpc (e.g. run L01).
Instead, for larger pericenters or for satellite
models with smaller initial disk scale length and/or higher central densities
(all HSB models and also LZ models), the bar is quite stable to stripping.

The rotating  bar soon slows down
shedding angular momentum outwards to the halo and outer
stellar material  as a result 
of dynamical friction (Hernquist \& Weinberg 1989: Fux et al. 1995).
In the meantime, stripping removes the
high-angular momentum stellar material outside the bar radius.
The bar starts to buckle soon after the first orbit, when the
radial anisotropy has increased in such a way that
$\sigma_z/\sigma_p \le 0.4$, where $\sigma_z$ and $\sigma_p$
are, respectively, the stellar velocity dispersion normal and parallel to the
plane defined by the two longest principal axes of the remnant
(see also Raha et al. 1991 and Merritt \& Sellwood 1994).  
The buckling gradually erases the bar symmetry, ultimately leading to 
a more isotropic, spheroidal configuration: the transition is
highlighted by the evolution of both $\sigma_z$
and the minor/major axis ratio of the stellar mass distribution, $c/a$ 
(Figure 9 and Figure 10).
Once excited, the tidally induced bar and bending modes grow on the dynamical
time scale and thus the morphological transition is
faster in high-redshift satellites (HZ and LZ),
which reach a spheroidal
configuration $\sim 1$ Gyr after the bar appears while, 
other satellites remain in a transitory, bar-like configuration for
a few Gyr.

\medskip
\epsfxsize=8truecm
\epsfbox{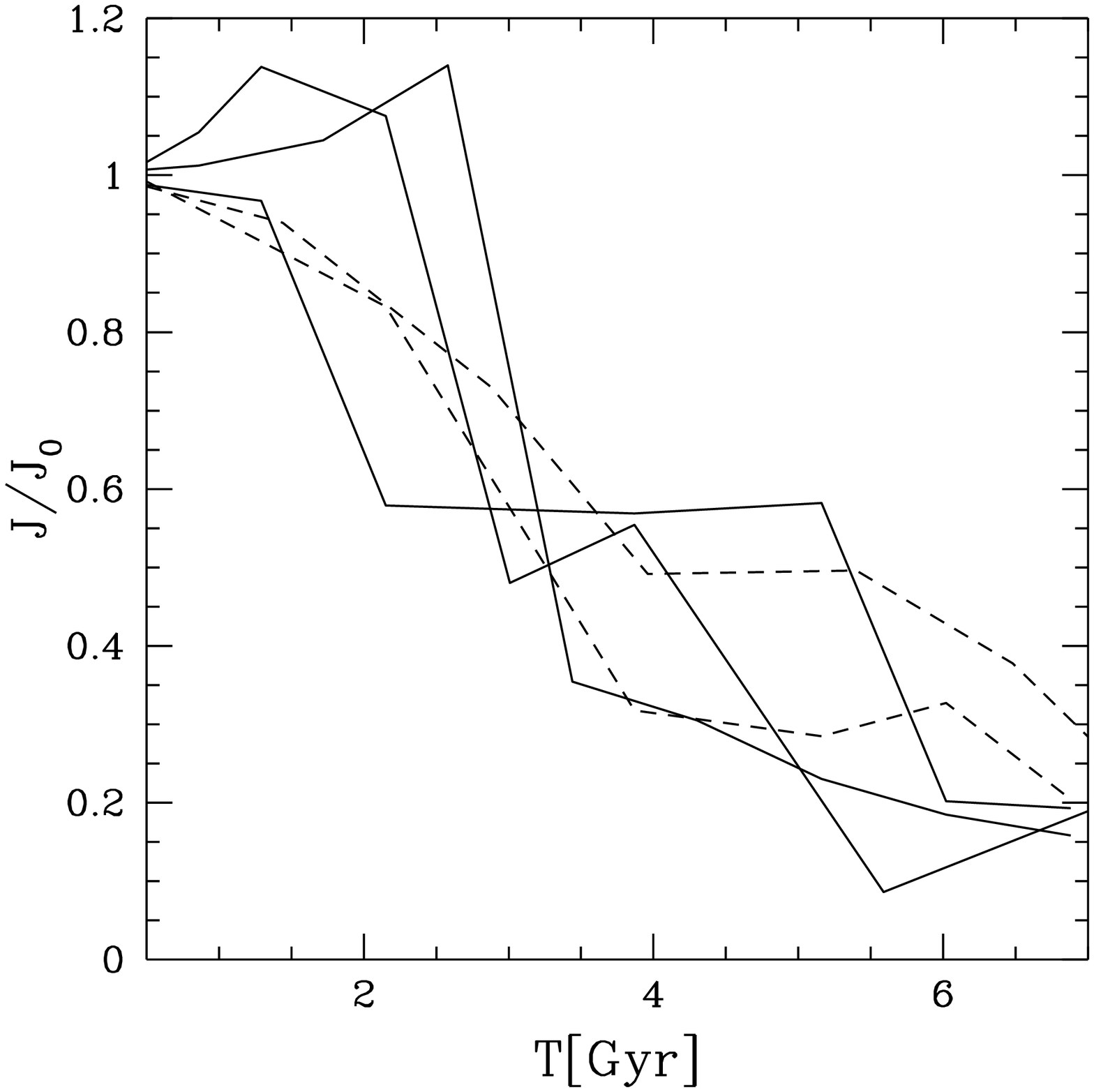}
\medskip
\epsfxsize=8truecm
\epsfbox{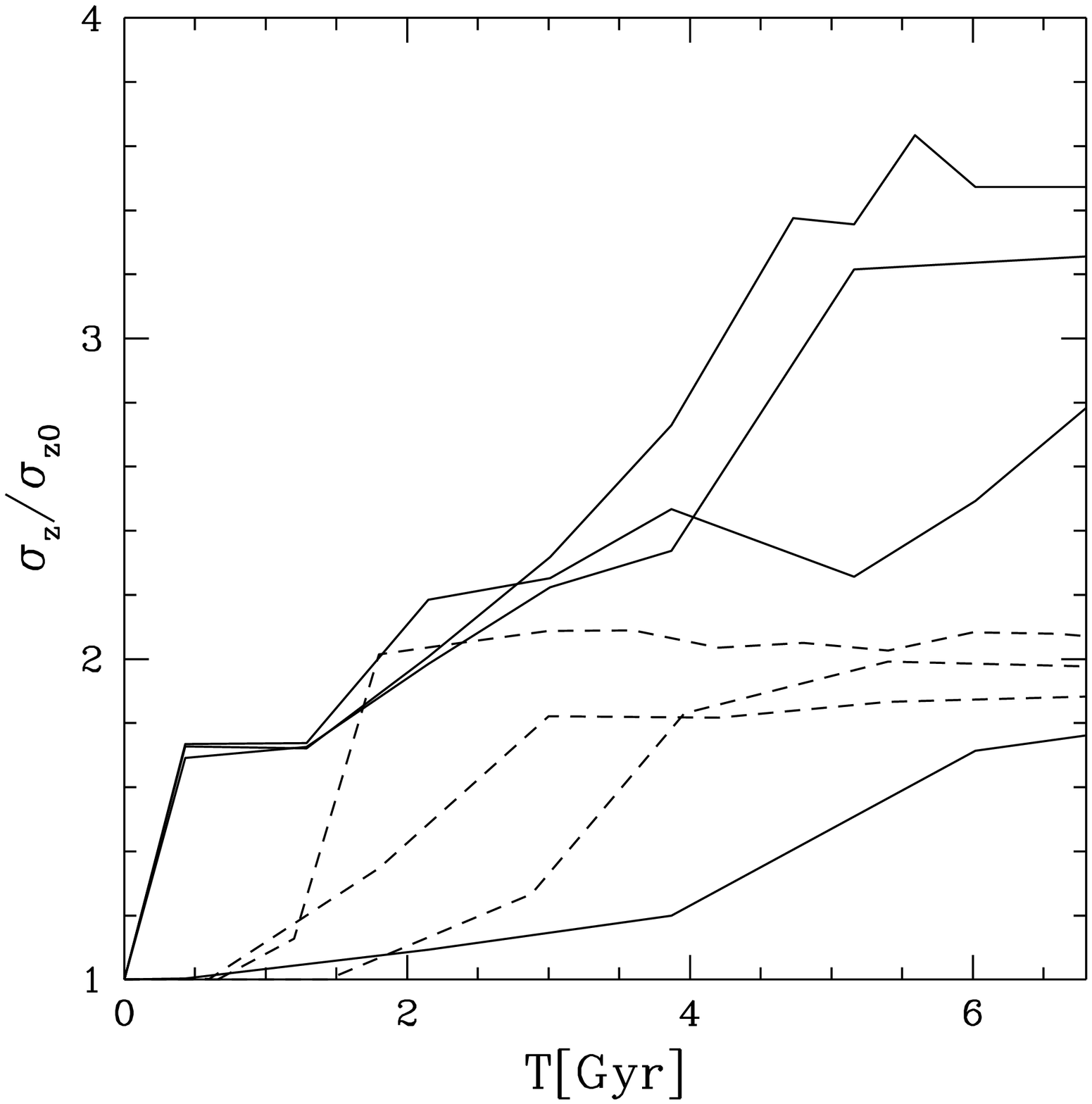}
\figcaption[f9b.ps]{\label{fig:asymptotic}    
\small{The evolution of the specific angular momentum of the bound stellar
component, normalized to the initial value, is shown on top, and that
of the velocity dispersion normal to the disk plane, $\sigma_z$, is shown
in on bottom. Results are from various runs (dashed lines are used for
HSB satellites, solid lines for LSB satellites). Measurements
were done within $\sim 1.5 r_h$, which corresponds 
to the typical size of the remnants. 
Note that the solid line in the lower part of the plot of $\sigma_z$ 
corresponds to run L15 (see Table 1), in which the disk is initially 
inclined by  90 degrees with respect to the orbital plane, 
while in all other cases the inclination is $\le 65$ degrees.}}
\medskip

After $\sim 7$ Gyr of evolution, nearly all the disks have been transformed
into spheroidals (Figure 11).
Heating induced by bar buckling is more important
than direct heating by the tidal field: indeed
satellites inclined by 90 degrees with
respect to their orbital plane (which suffer the strongest direct
tidal heating in the direction normal to the disk plane) 
undergo less vertical heating compared to those on a coplanar orbit 
(where a stronger bar is induced), as shown in Figure 9 (see caption).
It is thus understood why even a very stiff object like the GR8
model, that is barely stripped, is heated into a spheroidal configuration.

The joint action of loss of angular 
momentum and induced vertical heating determine a marked decrease of 
the $v/\sigma$ of the bound stellar component (Figure 10)
After the morphological transformation the satellites remain in
a nearly stable dynamical state as was verified by pushing a few simulations
to over 10 Gyr.

\medskip
\epsfxsize=8truecm
\epsfbox{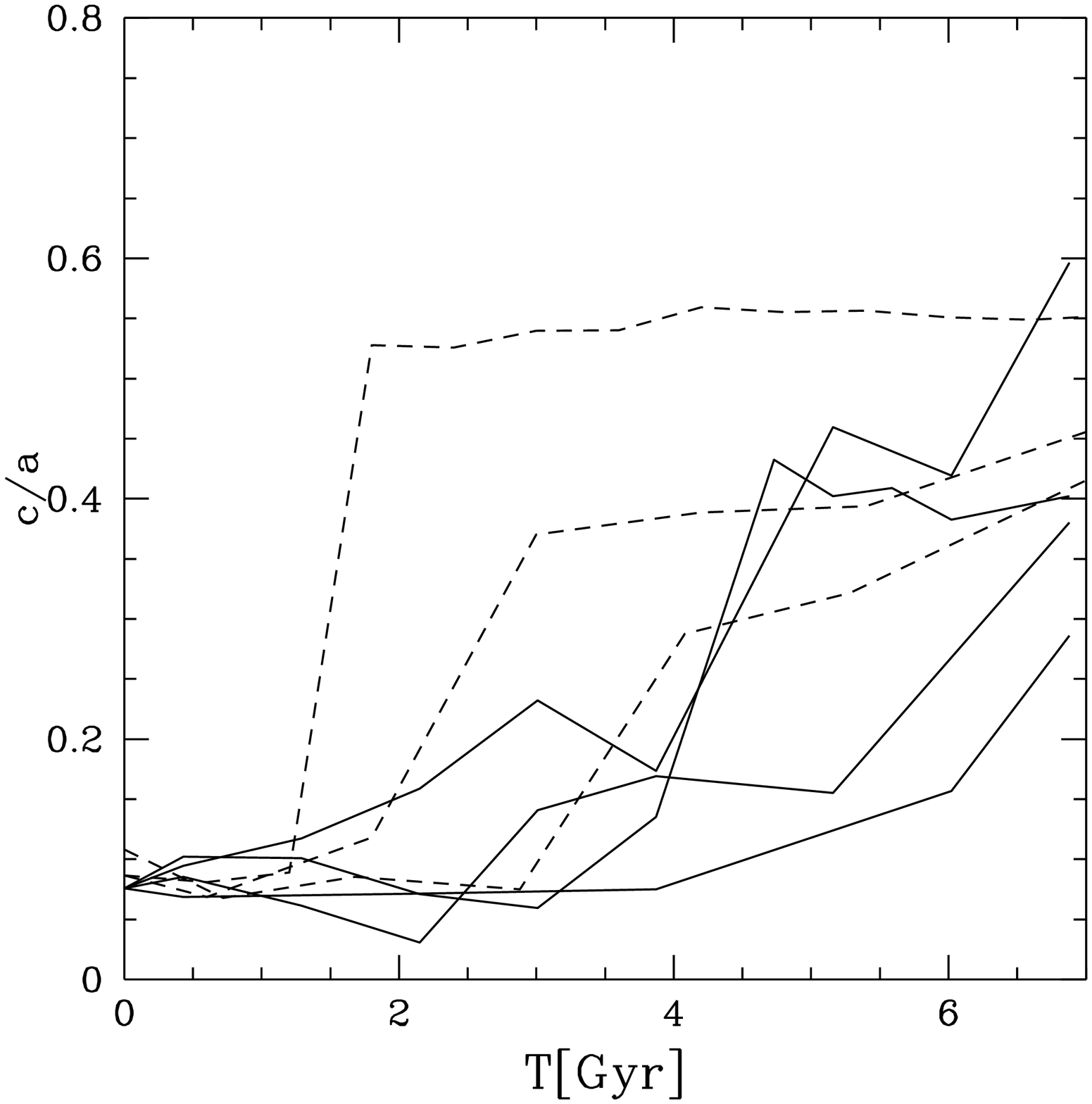}
\medskip
\epsfxsize=8truecm
\epsfbox{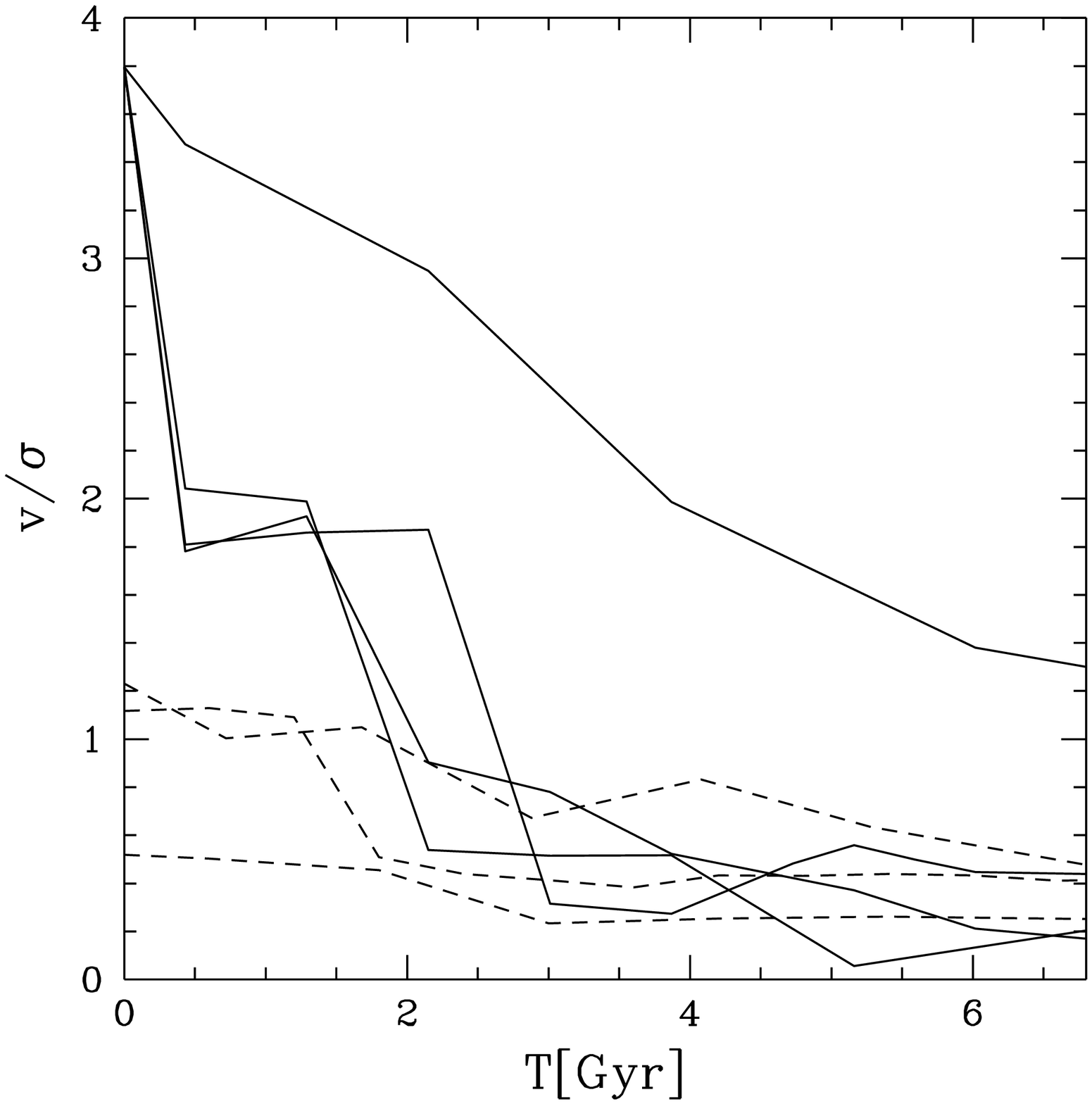}
\figcaption[f10b.ps]{\label{fig:asymptotic}    
\small{The evolution of the minor/major axis ratio of the bound stellar
component is shown on top, and that of $v/\sigma$ is shown
on bottom. Results are from various runs (dashed lines are used for
HSB satellites, solid lines for LSB satellites) and 
were made within $\sim 1.5 r_h$, which corresponds 
to the typical size of the remnants}}
\medskip

When the satellites are placed  on retrograde orbits the initial 
disk structure is partially preserved. Except for the case of LSBs
on orbits with the smallest pericenters, tidal stripping is 
remarkably reduced and weaker tidal tails are observed due to an anti-resonant 
coupling between orbital and internal motions 
(Toomre \& Toomre 1972; Barnes 1988; Springel \& White 1998;
Mihos et  al. 1997). The bar forms  when the satellite performs
the second pericenter passage, after $\sim$ 5 Gyr. 
Vertical heating is not as strong
as usual and the satellite remains quite flattened.
The final $v/\sigma$ within $R_e$ is still close to 1 at the end.

\subsection{Gasdynamical simulations}

We perform two dissipational runs placing 
models LMg2 and HMg2 on orbits with apo/peri=9 (runs L26 and H14 in Table 2
and Table 3, respectively).
Figure 12 shows two important results
: the gas is either stripped or funneled to the
center of the galaxy. The gas loses angular momentum after the first 
pericentric passage, due to the torquing by the stellar bar, which is 
misaligned with respect to the gas 
(Heller \& Shlosman 1994; Mihos et al. 1997).
In the HSB only $\sim 10
\%$ of the initial gas mass is tidally stripped, while all the rest is
driven to the center soon after the first pericentric passage. As a result,
a central spike appears in the gas density profile
(the maximum density increases by more than two orders
of magnitude, as seen in Figure 13)
and then the profile evolves little for the following Gyr.
Instead, in the LSB more than half of the gas is stripped,
while the remaining part is still funneled towards the center but at a slower
pace, the density growing by only one order of magnitude (Figure 13).

\medskip
\epsfxsize=5truecm
\figcaption[HSBall.ps]{\label{fig:asymptotic} 
\small{The evolution of an HSB 
dwarf disk galaxy inside a Milky Way-sized
halo (run H01).
Colour coded logarithmic density plots of the
stellar component in the satellite - the darker the colour the
lower the density - are shown. The boxes are about 30 kpc on each
side.
From top to bottom: the first four panels show a face-on projection of
the
satellite at every 2 Gyr starting from $t=0$ (top left) and moving
clockwise; the two bottom
panels show the initial (left) and final (right) appearance of the disk
seen edge on.}}
\medskip

Stripping dominates in the LSB because the gaseous disk
is more loosely bound (initial gas profiles follow the stellar 
density profile in our models) and the central inflow is moderate
as a result of both the weak bar and the long dynamical time.
In the LSB, as the gas slowly drifts inwards, it
has time to feel the effect of tidal compression at the second 
pericenter passage, which leads to the largest increase in the 
central gas density (Figure 13).
Tidal compression occurs at the turning
points of the orbit because the direction of the tidal force
relative to orbital motion is reversed: the leading edge of the galaxy
is thus pushed towards the trailing edge.
The compression will be
larger at pericenter because there the magnitude of the tidal force is 
greater.

\medskip
\epsfxsize=8truecm
\epsfbox{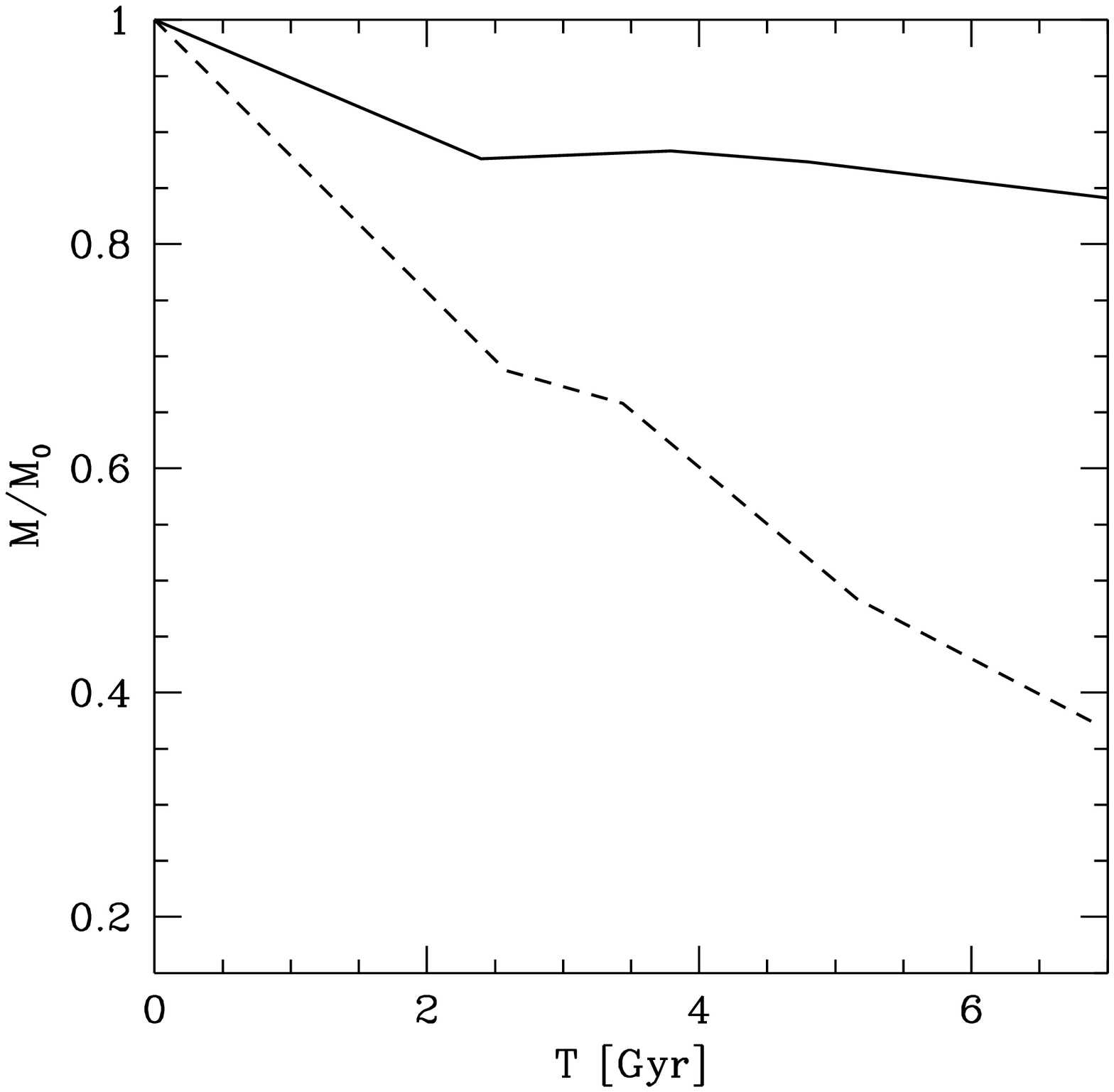}
\medskip
\epsfxsize=8truecm
\epsfbox{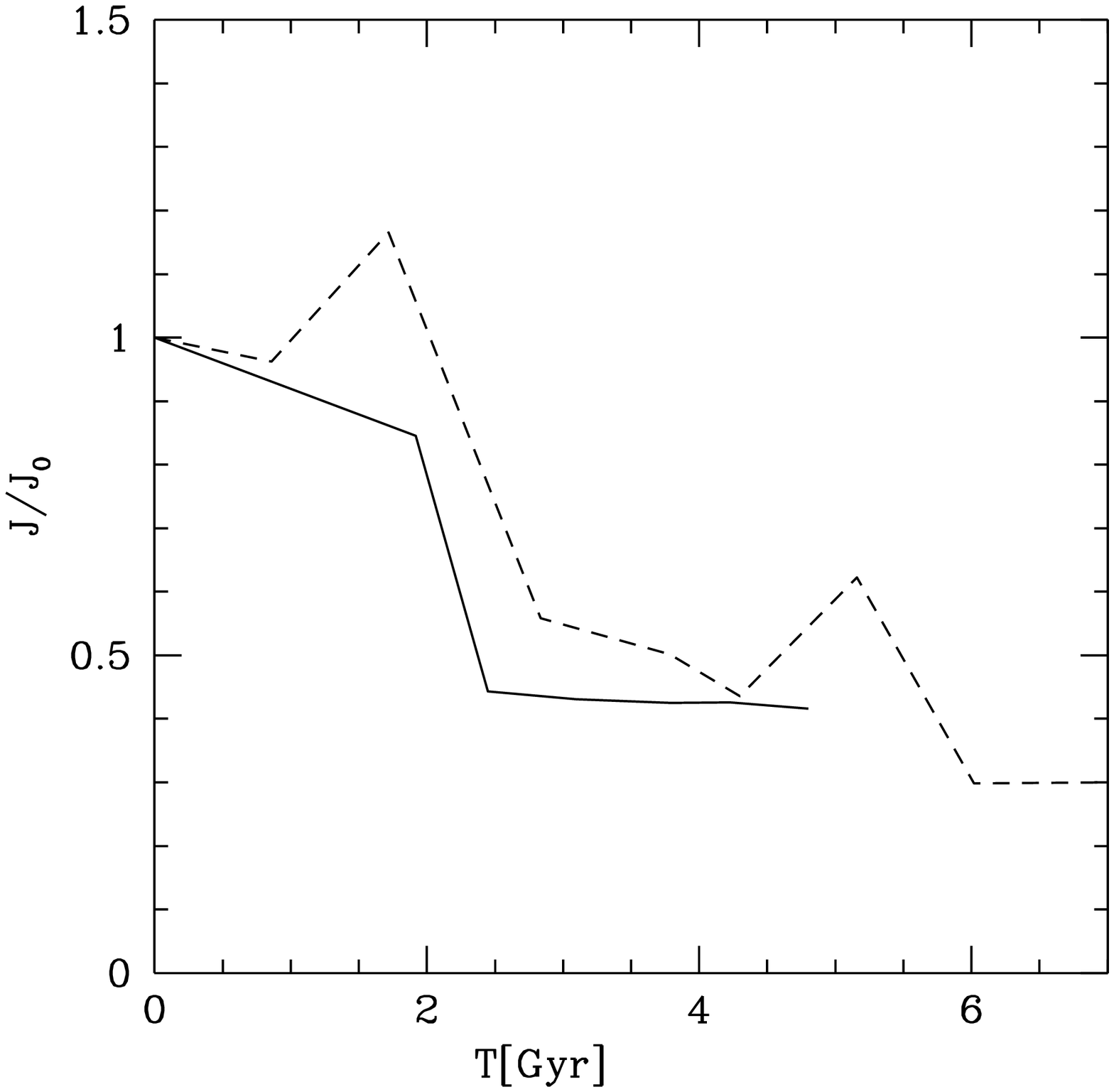}
\figcaption[f12b.ps]{\label{fig:asymptotic}
\small{Evolution of the bound gas mass (top) and of the
specific angular momentum of the gas (bottom) for the HSB (solid line) and 
for the LSB satellite (dashed line).}}
\medskip

\section{Structure of the remnants}

\subsection{Sizes and density profiles of the stellar remnants}

The final stellar surface brightness profiles do not deviate substantially from
the initial exponential form, although they tend to become steeper
(Figures 14 and 15).
In general, when the galaxy is structurally stiff or 
the pericenter distance is fairly large ($R_p > 70-80$ kpc), the profile 
is bi-modal, having a steeper slope inside the half mass radius $R_e$ 
of the galaxy (see Figures 14 and 15): this is because the bar instability 
leads to a more concentrated stellar component, which persists unless the
subsequent tidal shocks are strong enough to strip the system down to the
central part.
As a result, both single exponential and bi-modal profiles are present
among LSB satellites, while the stiffer HSBs or the high-redshift 
dwarf models, in particular GR8, are always fairly stable to tides within
the bar radius and always exhibit bi-modal profiles.

\medskip
\epsfxsize=8truecm
\epsfbox{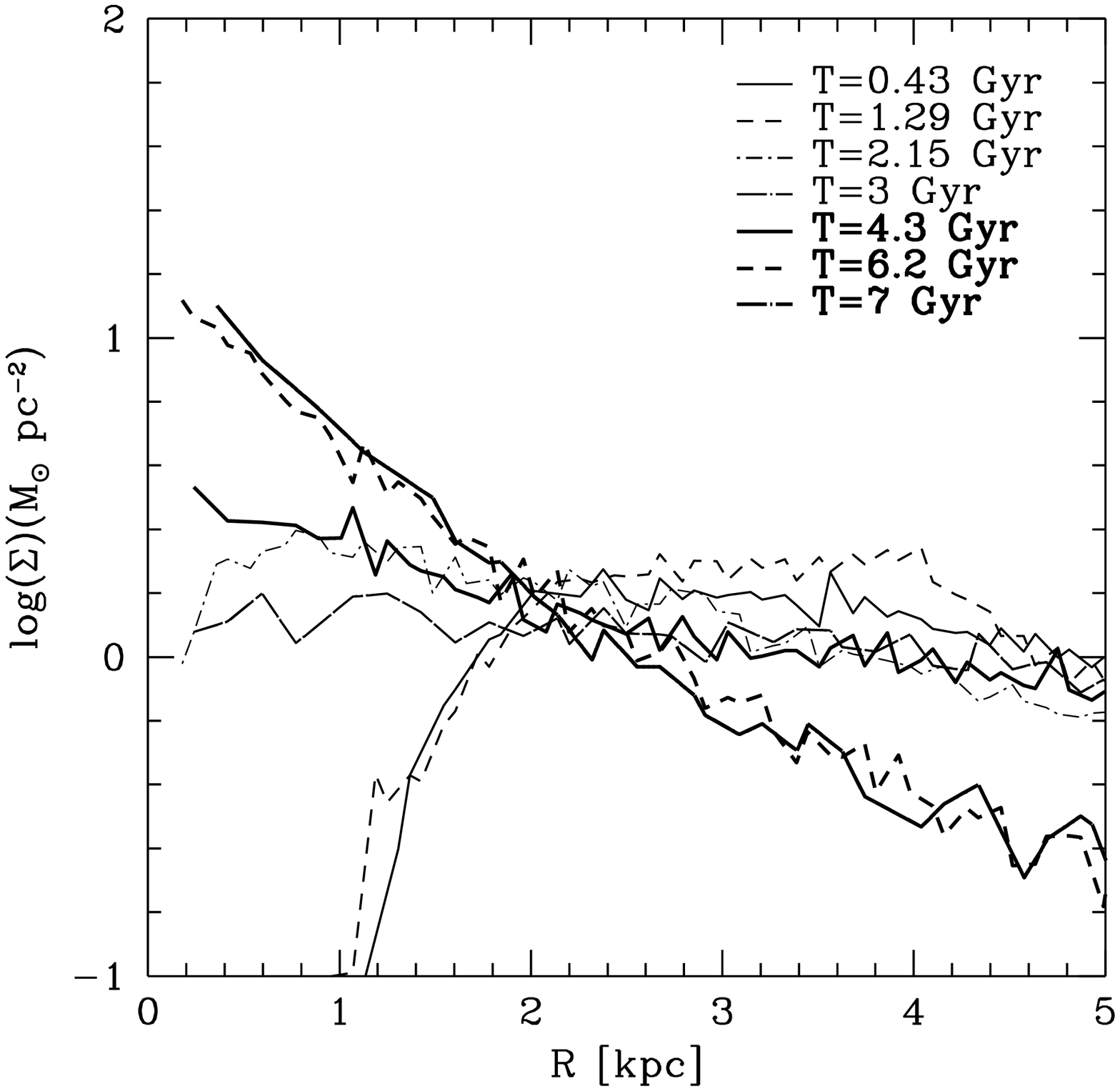}
\epsfxsize=8truecm
\epsfbox{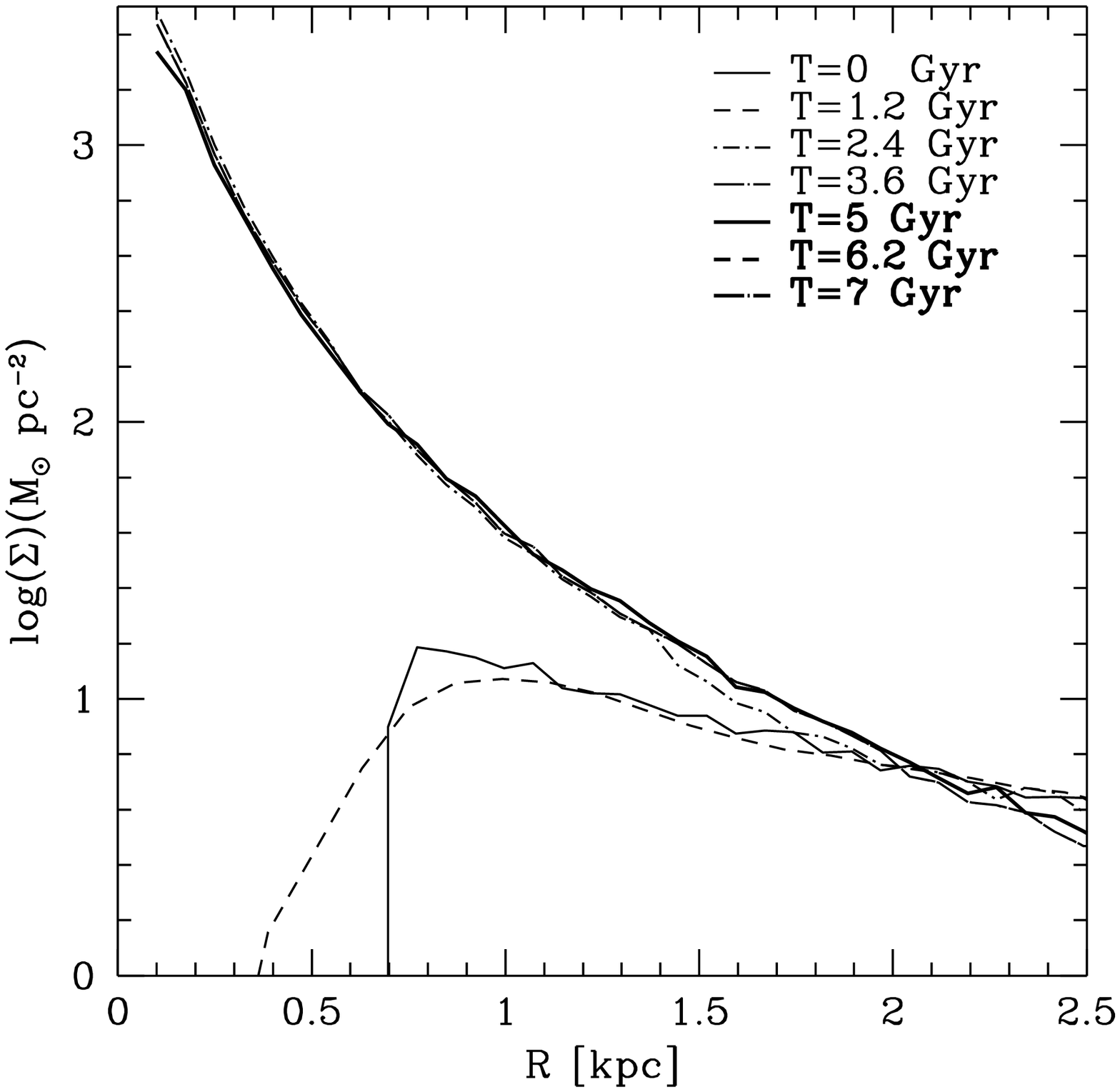}
\figcaption[f13b.ps]{\label{fig:asymptotic}
\small{Evolution of the gas surface density in the LSB satellite (top) 
and in the HSB satellite (bottom).}}
\medskip

The typical sizes
of the remnants can be measured by means of their final half-mass
radius $R_e$ (Figure 16). 
The more pronounced compactness of the remnants of HSBs versus those of LSBs
resembles an analogous difference seen in dEs versus dSphs 
(Ferguson \& Binggeli 1994, Kent 1991, Carter \& Sadler 1990).
At the end HSBs have a central surface brightness only slightly
smaller than the initial one (even including fading according to the
model described in section 9), while
their stellar mass is decreased by a factor of 3 or more: therefore,
they turn out to have a surface brightness higher than any dIrr with
similar luminosity, just as it is typically observed for dEs in the LG 
and in nearby groups and clusters (Mateo 1998, James 1991).

\medskip
\epsfxsize=9truecm
\epsfbox{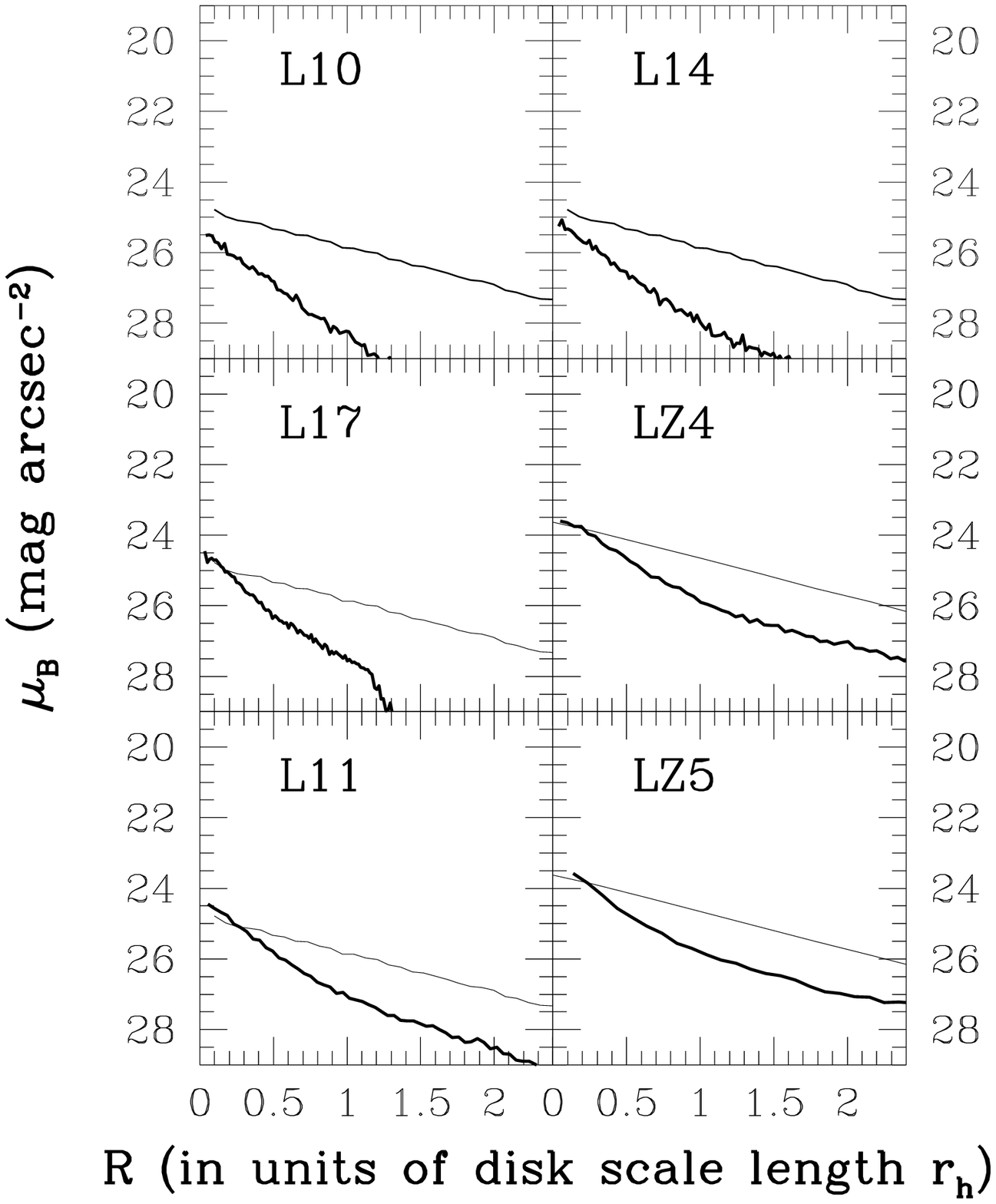}
\figcaption[f14.eps]{\label{fig:asymptotic}
\small{Evolution of the surface brightness profiles of LSB satellites:
the thin line is used the initial profile, the thick line is used the
final profile. The various panels are labelled with the run names
(see Table 2 and Table 3 for details).
The normalization of the final profiles includes the correction
for fading derived in section 9.}}
\medskip

When fitting the profile, for some of the LSBs a single exponential law
works pretty well, while
King profiles with a concentration
$c < 1$ reproduce better the bi-modal profiles
of the remnants of HSBs  and of GR8 (see Figure 17).
Because of the greater compactness of HSB remnants, their core radii 
resulting from the fits with  King profiles are typically 
as small ($\sim 200-300$ pc)
as those of LSB remnants ten times less massive. This matches the 
observations well: bright dEs like NGC147 and NGC185 have core radii 
as small as those of  dSphs like Carina and Leo I (Mateo 1998),
the latter being fainter by $\sim 4$ magnitudes.

\medskip
\epsfxsize=9truecm
\epsfbox{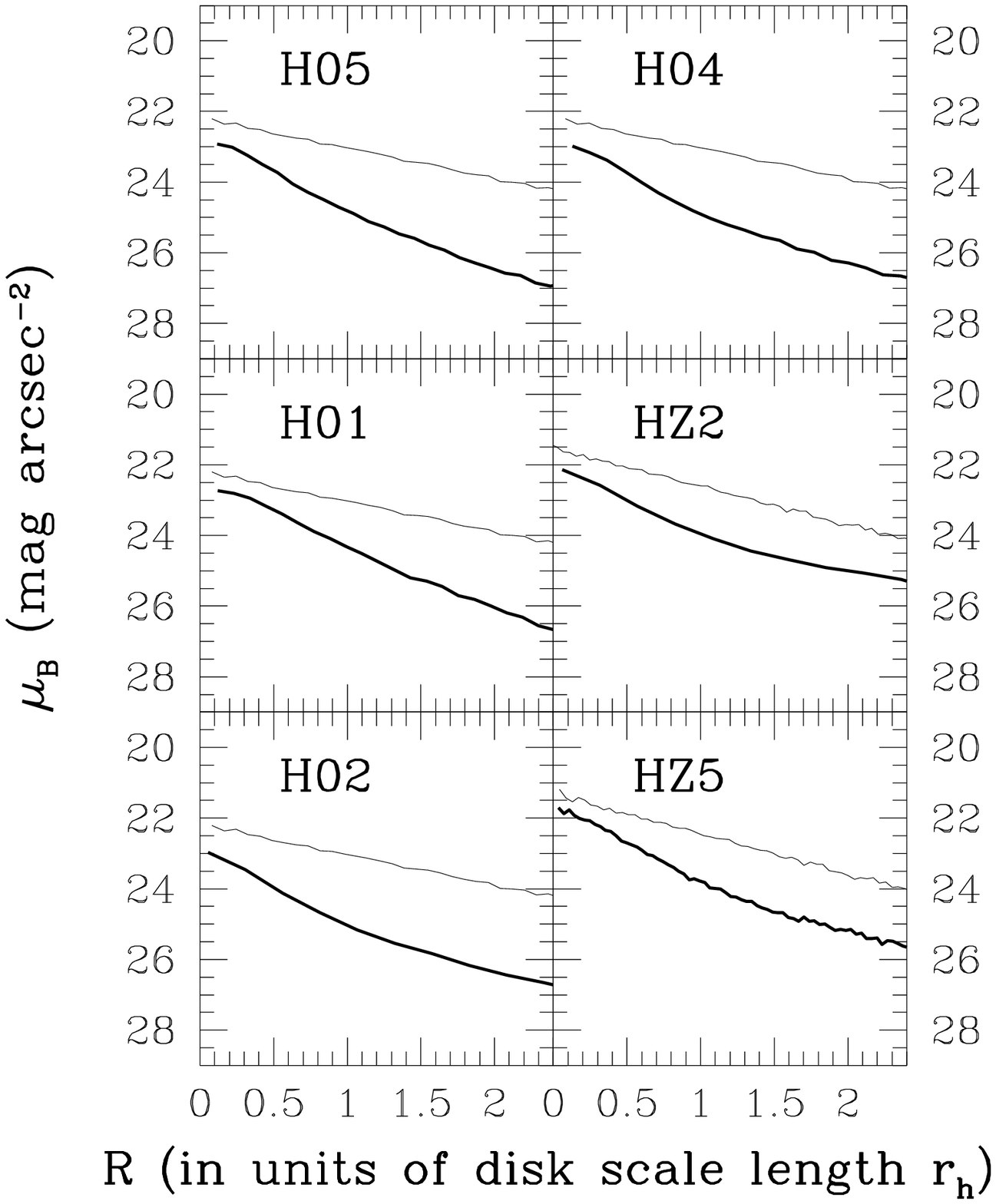}
\figcaption[f15.eps]{\label{fig:asymptotic}
\small{Evolution of the surface brightness profiles of HSB satellites:
the thin line is used for the initial profile, the thick line is used for the 
final profile.  The various panels are labelled with the run names
(see Table 3 for details).
The normalization of the final profiles includes the correction
for fading derived in section 9.}}
\medskip

In the dissipative runs some dynamical feedback of the gas on the stellar
component occurs. The latter is different in HSBs vs. LSBs owing
to the different dynamical evolution of their gas component.
The strong central gas concentration that 
appears in the HSBs soon after the first pericentric passage immediately
produces a response of the  stellar potential and the final surface
brightness profile is 
remarkably steeper in the center compared to the collisionless
case, as shown in Figure 18. 
On the other end, in the LSB
the stellar density profiles evolve  nearly 
in the same way with and without gas
because  the weaker bar does not create a dense central gas knot and 
a substantial fraction of the gas is stripped away; the
gas-to-stellar mass ratio inside $\sim R_e$ is roughly the same
as at the beginning.

\subsection{Kinematics and shapes}

The kinematics of the satellites changes drastically by the end 
of the simulations. The final
$v/\sigma$ measured inside $R_e$ is typically lower
than $0.5$, this being a distinctive feature of dSphs and dEs not only
in the Local Group but even in galaxy clusters
(Mateo 1998; Bender \&
Nieto 1990, Bender et al. 1992; Ferguson \& Binggeli 1994),
A slightly larger (by $20-30 \%$) residual rotation around the minor axis
is present at larger radii, due to high angular momentum, loosely
bound material which is about to be tidally ejected.

The transition in the dwarfs' kinematics is
the major success of our model as it provides an explanation
for the {\it most important dynamical difference} between
dIrrs and dSphs: {\it their angular momentum content}.
The bulk of the result is independent from the viewing projection, 
as shown in Figure 19, which also shows that all the remnants fall
below the theoretical prediction for rotationally flattened spheroidals,
$v/\sigma \sim \sqrt{(\epsilon / 1 - \epsilon)}$, where $\epsilon =
1 -b/a$ is the ellipticity projected onto the plane of the sky 
(Binney \& Tremaine 1987), which means that any flattening is 
due to anisotropy in their velocity dispersion, like in giant ellipticals,
as suggested by observations (Ferguson \& Binggeli 1994).

\medskip
\epsfxsize=8truecm
\epsfbox{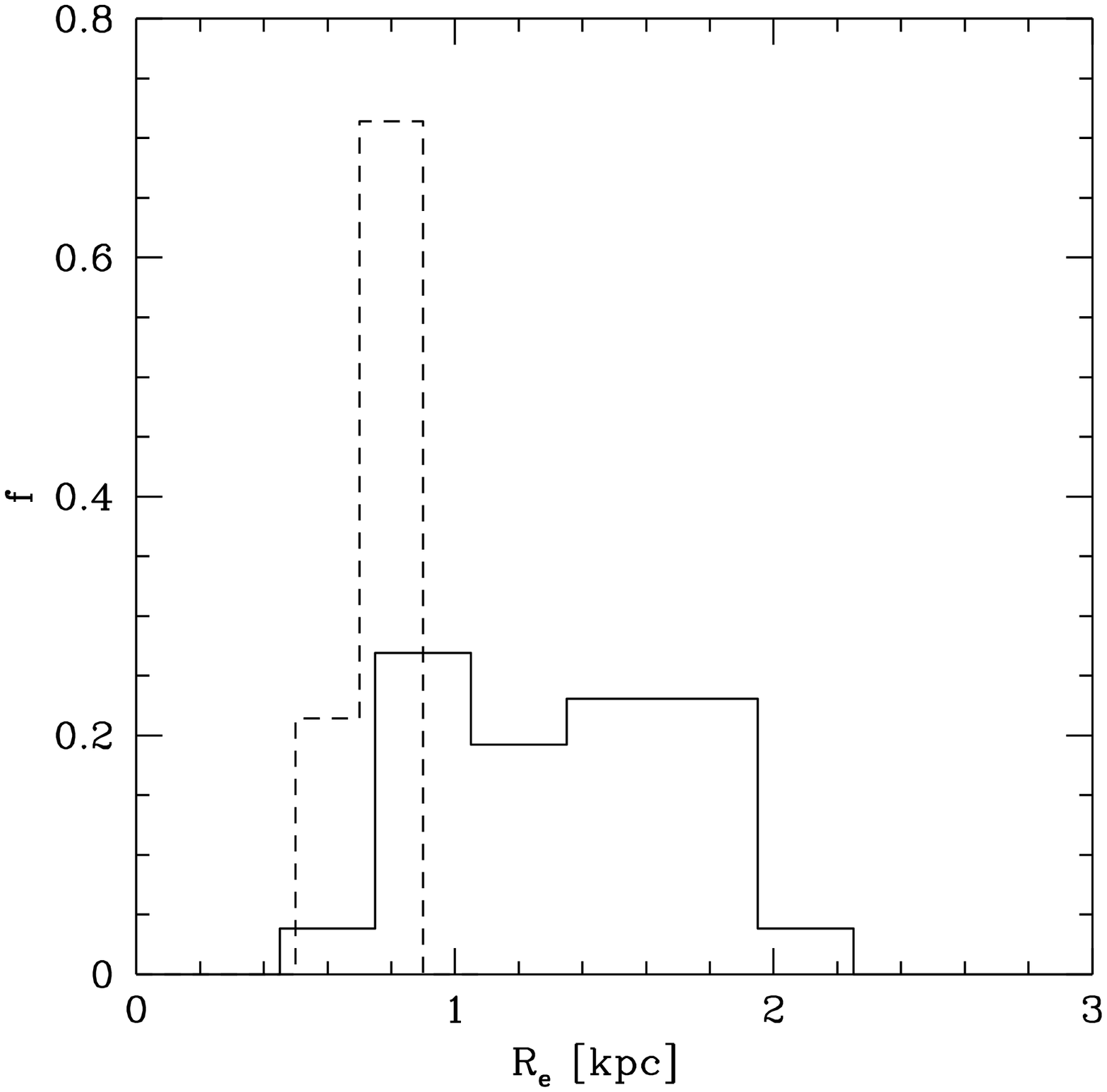}
\figcaption[f16.eps]{\label{fig:asymptotic}
\small{Histogram showing the fraction of the remnants, $f$, whose half-mass
radius falls within the values indicated on the x-axis, for both
LSBs (solid line) and HSBs (dashed line)
.}}
\medskip

The final average velocity dispersions vary from 10 to 20 km/s among
the remnants of LSBs
, while they are around 30-35 km/s for those of HSBs
(Figure 20). The initial models of LSBs and HSBs span exactly
the same range of masses and both have initial velocity dispersions of
a few km/s (only $\sim 30\%$ larger in the HSBs), being rotationally 
supported. It is the larger mass retained by the HSB satellites together
with their typically more concentrated final stellar profile that determines
such a kinematical distinction by the end of the simulations.
The velocity dispersions of the remnants of LSBs are in
good agreement with those inferred for dSphs, while those of the remnants
of HSB match those of dEs (Mateo 1998).
The measured velocities
are very close to the expectation
from virial equilibrium when the residual rotation is negligible.
In
other words, we do not find that tidal forces inflate the
velocity dispersions in the bound remnants, in agreement with Piatek
\& Pryor (1995) and Pryor (1994),
who did a similar analysis on spherical satellites.

\medskip
\epsfxsize=8truecm
\epsfbox{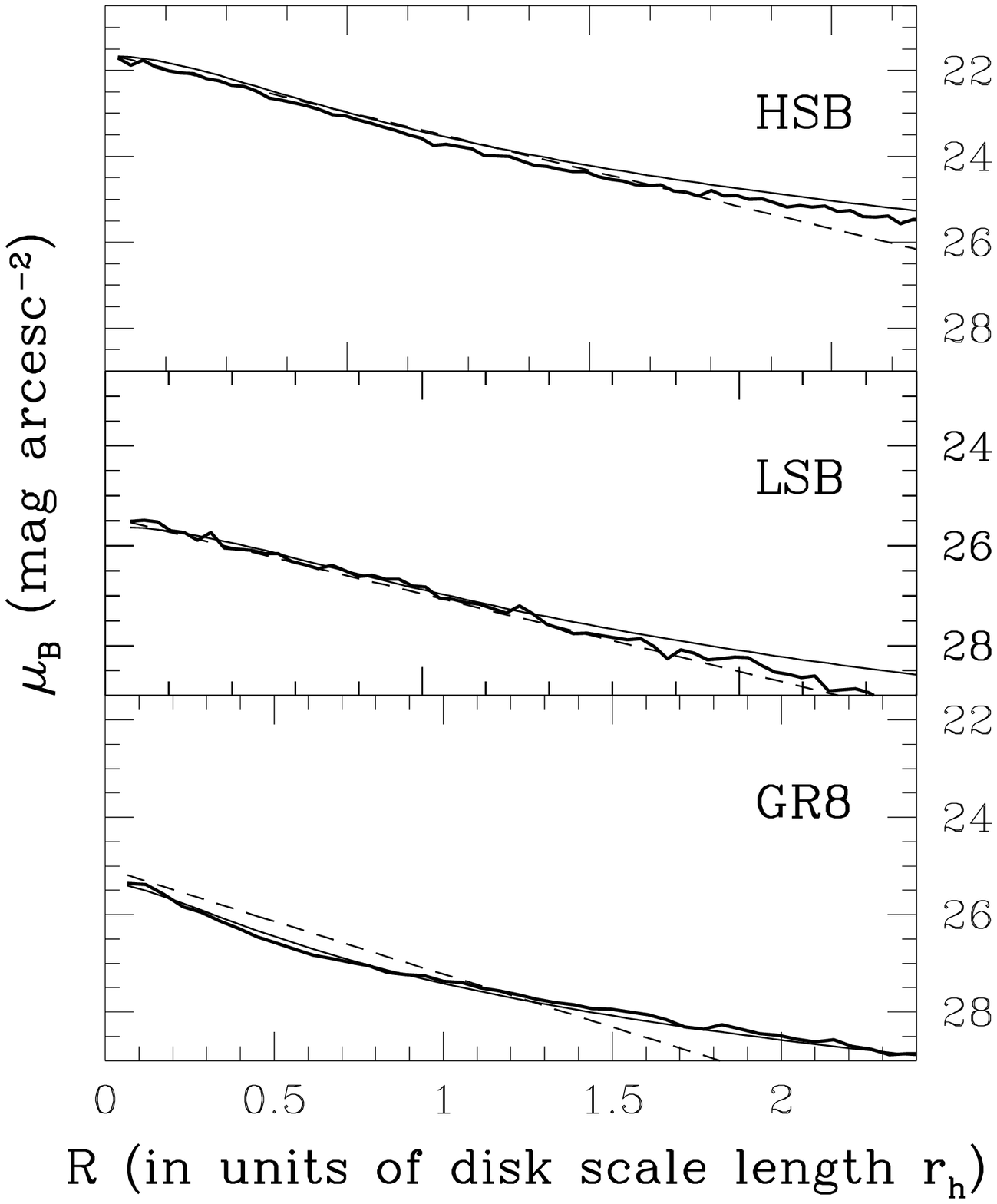}
\figcaption[f17.eps]{\label{fig:asymptotic}
\small{Final surface brightness profiles of ``prototype'' satellites (thick
solid lines) fitted by exponential laws (thin dashed lines) or King profiles
(thin solid lines).
The exponential scale lengths used are 400 pc (HSB, run H08), 1 kpc (LSB, run
L10) and
50 pc (GR8, run GR82), while the core radii in the fits with King profiles
are  325 pc (HSB), 800 pc (LSB)and 50 pc (GR8).}}
\medskip

The velocity dispersion profiles of the remnants are typically slowly 
decreasing (Figure 21), as found in some of the dSphs (Hargreaves
et al. 1994a, b, 1996) and dEs (NGC205, Held et al. 1990, 1992;
Carter \& Sadler 1990).
The final radial velocity anisotropy, measured by the parameter $\beta=
1 - \sigma_z/\sigma_p$,
is  higher in the more flattened LSBs (Figure 22).
The average value of $\beta$ is $\sim 0.15$ for HSBs and $0.35$ for
LSBs: thus remnants are on average only mildly anisotropic.

To measure the shape of the remnants we use the triaxiality parameter
$T_t = (a^2 - b^2/ a^2 - c^2)$ (Franx, Illingworth \& de Zeeuw 1991), 
where $a$, $b$ and $c$ are, respectively, the major, 
intermediate and minor axis of the
remnants. $T_t=1$ represents a purely prolate halo, while we have $T_t=0$ for
a purely oblate halo. Objects can be defined
nearly prolate if  $2/3 < T_t < 1$, nearly oblate 
if $0 < T_t < 1/3$ and triaxial if $1/3 < T_t < 2/3$. In measuring
 the shape of the remnants within $R_e$ we find that the
majority of the remnants are nearly prolate spheroidals and a fair
fraction is triaxial, while oblate shapes are very rare.
The tendency towards prolateness is due to the residual radial
anisotropy originated during the bar stage of the evolution and
can be enhanced by tidal distortion, especially for the loosely bound LSBs.

\medskip
\epsfxsize=8truecm
\epsfbox{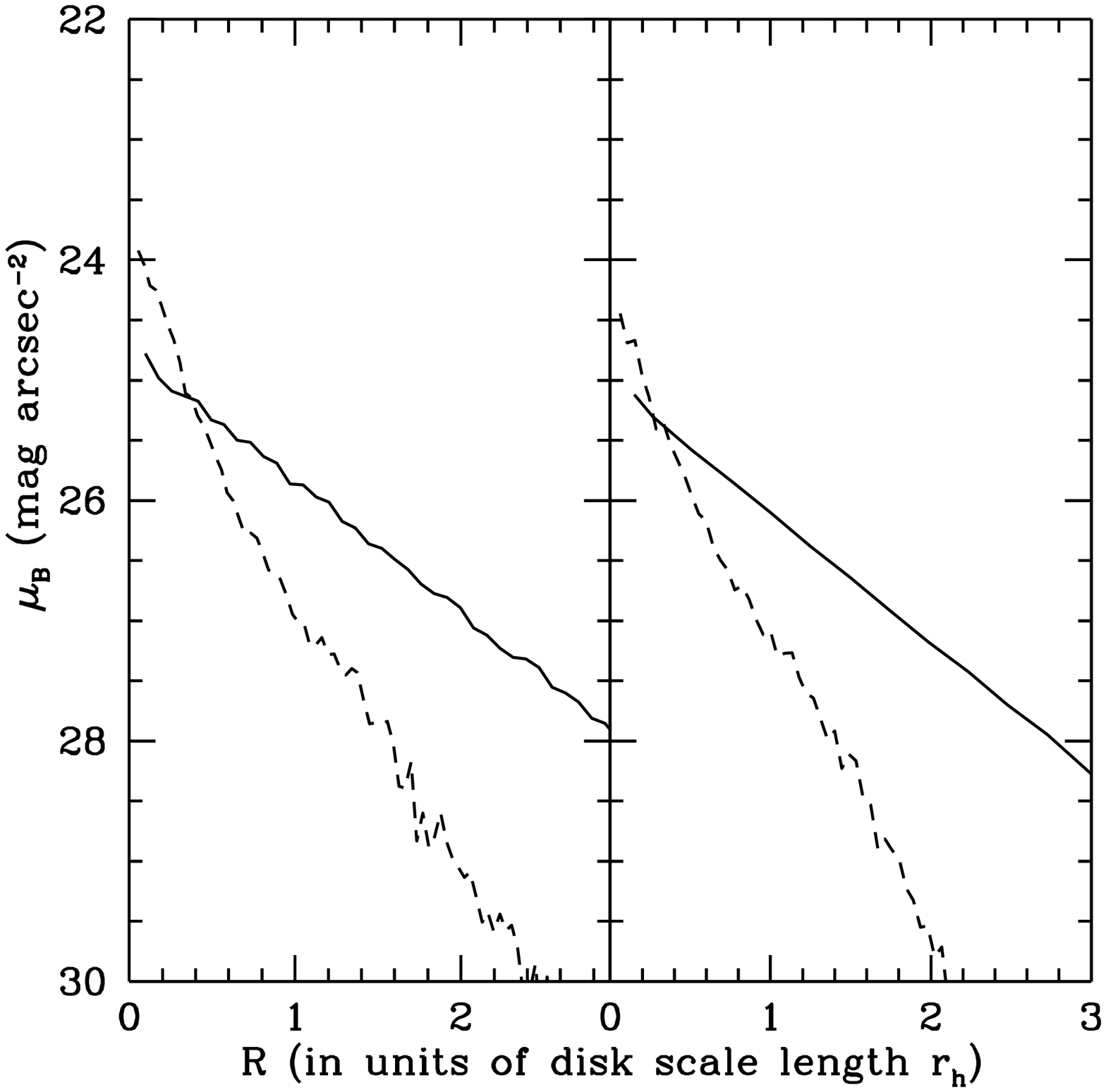}
\epsfxsize=8truecm
\epsfbox{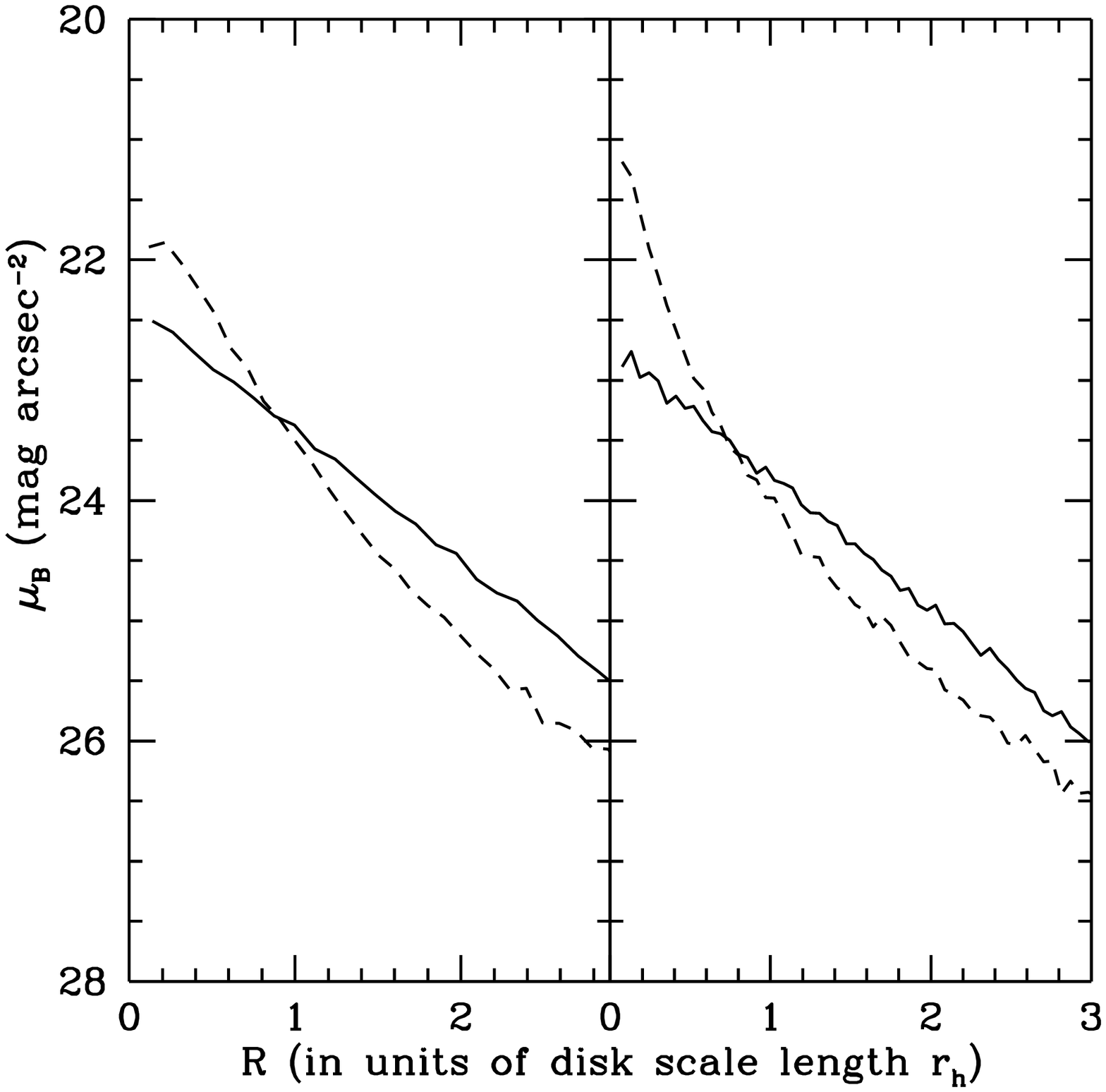}
\figcaption[f18b.eps]{\label{fig:asymptotic}
\small{Comparison of the final surface brightness profiles (with no
fading correction) for the satellites with gas (right panels) and without
gas (left panels). The orbital configurations are exactly the same in all
four runs (the collisionless runs are L13 and H08, see Table 2 and Table 3).
On top the LSBs are shown, on bottom the HSBs.
Solid lines refer to the profiles at T=0, dashed lines to those
after 7 Gyr. Note that even the initial surface brigh
tness profiles are not
the same in collisionless and gasdynamical runs because the effective
stellar surface density is lower when a part of the same total disk mass is
assinged to a gas component.}}
\medskip

The parameter $c/a$ provides a measure of the maximum observable flattening
of the remnants: the higher the flattening the lower will be $c/a$.
Figure 22 clearly indicates that the remnants of LSBs are
more flattened than those of HSBs, with average $c/a$ of $0.35$  and $0.5$,
respectively. This is expected because vertical heating occurs
faster in HSBs due to their smaller dynamical times.
In addition, the more flattened remnants of both HSBs and LSBs 
correspond to cases of nearly retrograde encounters that better preserve
the original disk structure.

An anti-correlation between final surface
brightness and final flattening is present in the remnants.
Interestingly, there are indications that bright dEs,
both in the LG and in galaxy clusters, 
are rounder than low surface brightness dSphs (Van den Bergh 1986, 
Ferguson \& Sandage 1991).
However, the comparison with observations is not trivial because we
have to take into account projection effects.  For example, if viewed
along the major axis all the remnants would appear rounder than
when viewed along either of the two other axes,
as seen in Figures 22. 
The mean ratio between two of the principal axes, 
obtained averaging over the three projections, is thus higher than that 
associated with the measure of $c/a$, 
being around $0.5$ for the remnants of LSBs and around $0.7$
for the remnants of HSBs: we can consider the latter values
as yielding the {\it mean observable flattening}.

\medskip
\epsfxsize=8truecm
\epsfbox{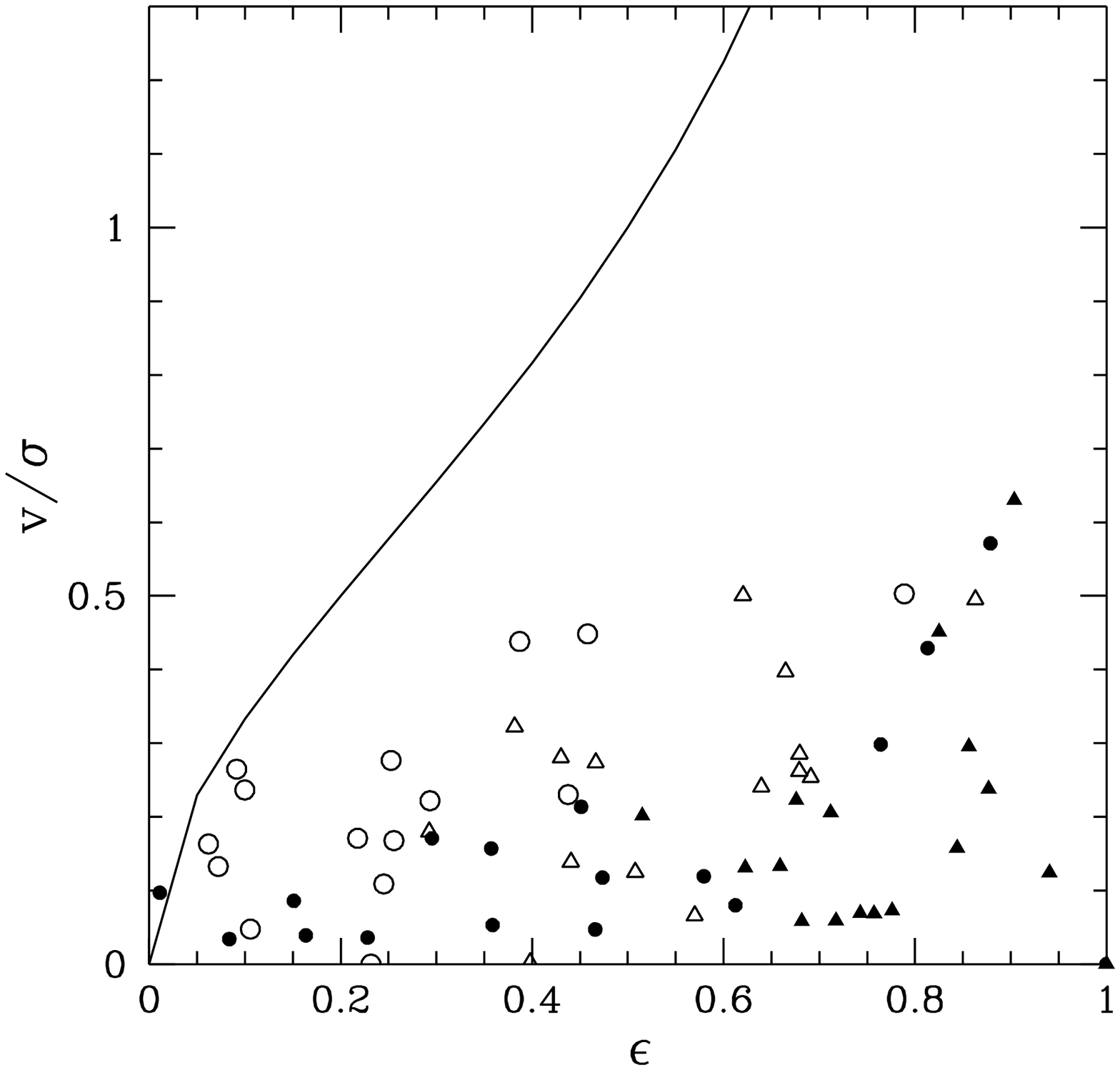}
\figcaption[f19.eps]{\label{fig:asymptotic}
\small{Scatter plot showing the relation between final $v/\sigma$
and projected ellipticity (see text, both measured inside $R_e$,
for LSBs (filled symbols) and HSBs (open symbols) for two
viewing projections, one along the intermediate axis (circles) and the
other along the major axis (triangles) of the remnants 
(rotation perpendicular to the minor axis is always close to zero, thus
results for the viewing projection along the latter axis are not shown).
The solid line shows the relation expected for rotationally flattened
spheroidals (Binney \& Tremaine 1987).}}
\medskip

Ellipticities of LG dSphs 
range from the extreme value of $0.8$
in Sagittarius and $0.65$ for Ursa Minor to
$0.2-0.3$ for Sextans, with  the average being around $0.4$
(Mateo 1998). 
We can safely conclude that the ellipticities of the remnants
fall in the observed range.

When the gas is included in the model satellites we observe an increase
in the final rotation of the stars inside $R_e$, by $\sim 30 \%$  in the LSB
and by about $50 \%$ in the HSB.
The reason is that the gas spins up the
stellar bar while losing angular momentum (Friedli 1999; Friedli
\& Benz 1993).
This mechanism is suggested by Figure 23, where we compare the 
evolution of the specific angular momentum
of the bound stellar component with and without gas. 
Because rotation is higher, the remnant of the LSB is also flatter ($c/a=0.32$
instead of $0.48$) than in the collisionless case. In the case of the
HSB, the dominant effect is instead the stronger central concentration of both
the gas and the stellar components, and thus the remnant is rounder,
notwithstanding the higher rotation ($c/a = 0.67$ instead of $c/a=0.55$).
Overall, the final $v/\sigma$ increases slightly for the LSB while it is almost
unchanged for the HSB because a higher velocity dispersion is achieved
due to the more concentrated potential. A rounder shape of gaseous
remnants was also found by Moore et al. (1998) for harassment in galaxy clusters as in that case the model galaxies were structurally more similar to
our HSB models.

\subsection{Dark matter content}

The circular velocity profiles (Figures 6, 7 and 8) show
the relative contribution of dark matter and stars after 7 Gyr of
evolution inside the Milky Way potential. Remnants of LSBs, being severely
stripped down to their inner parts, have a ``central''
$M_{dark}/M_{stars}$ (measured inside $R_e$) around 1-2, while
their high-redshift counterparts (LZ models) can have ratios as high
as 4-5.
HSBs, although quite resistant to tides,
have  in general $M_{dark}/M_{stars} \sim 2.5$ as they start off
with a ratio $M_{dark}/M_{stars}$ (within the optical radius) 
a factor of 2 lower than that in LSB models.

Contrary to common belief,
not all dSphs have extremely high dark matter content; most of them
have inferred central mass-to-light ratios (measured at the core radius)
between 5 and 20 (Mateo 1998), which, for the stellar 
mass-to-light ratios expected in objects dominated by old stars (($M/L)_{*B}
\geq 3$), implies dark matter halo masses within the range of our results.
Mass-to-light ratios in dEs turn out to be moderate, $\le 10$
(Carter \& Sadler 1990), in agreement with those of the remnants of HSBs.

Our evolutionary model is able to explain also the extreme dark matter
content of the faintest dSphs, Draco and Ursa Minor
(Lake 1990; Hargreaves et al. 1994a,b; Mateo 1998).
The GR8-like model almost preserves
the initial  $M_{dark}/M_{stars}$; with
$M_{dark}/M_{stars} \sim 12$ within $R_e$, the mass-to-light ratios would be
$\geq 36$ for $(M/L)_{*B}\geq 3$. 

\medskip
\epsfxsize=8truecm
\epsfbox{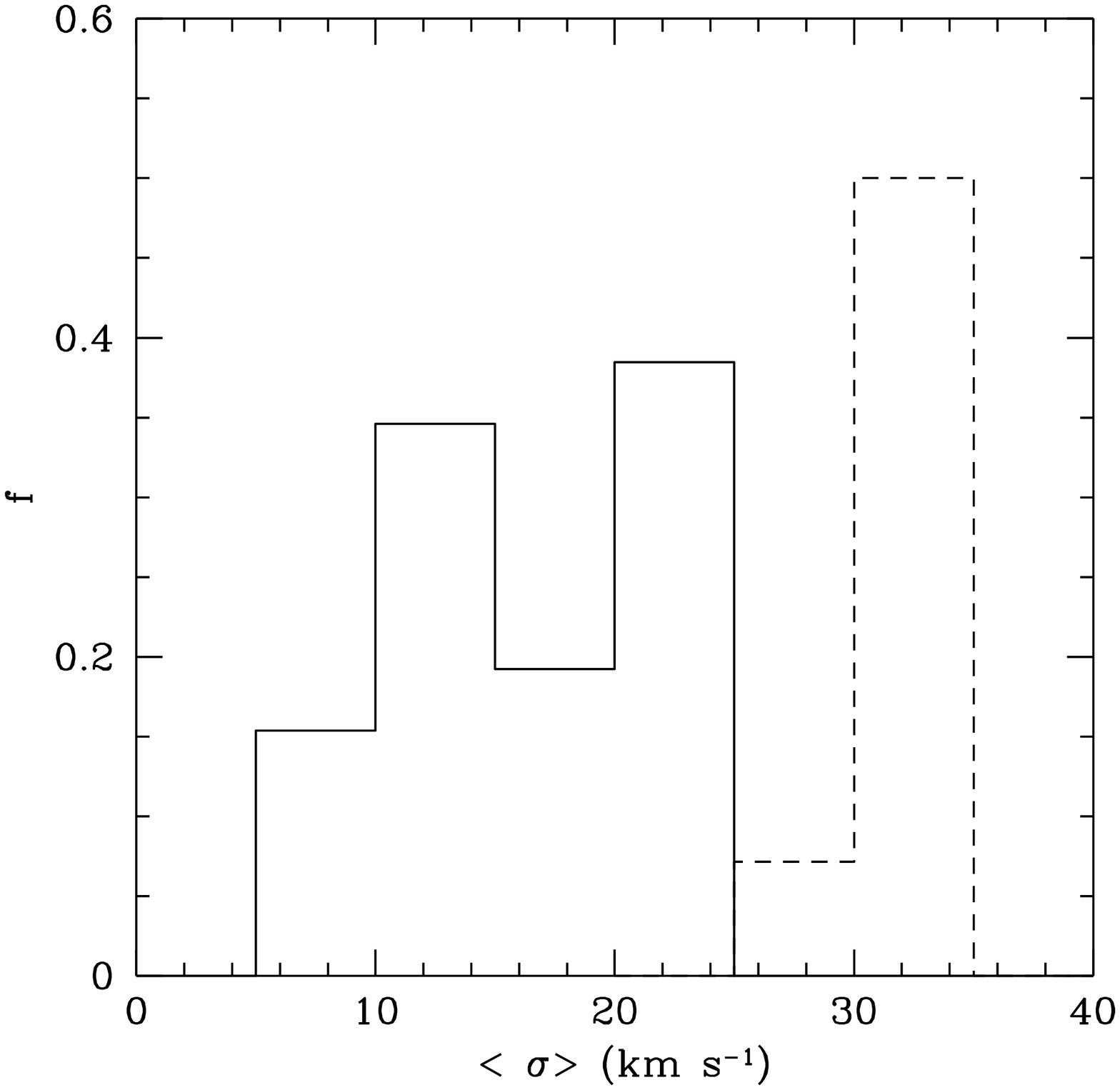}
\figcaption[f20.eps]{\label{fig:asymptotic}
\small{Histogram showing the fraction of the remnants, $f$, whose
average velocity dispersion falls within the values indicated 
on the x-axis: the solid line is
refers to the stellar remnants of LSBs, the dashed line is for the 
stellar remnants of HSBs.}}
\medskip

A dwarf as dense 
as GR8 needs to enter early into the Milky Way
halo to be transformed into a dSph, requiring a high number of tidal
shocks and thus a small orbital time. The hierarchical structure
formation scenario naturally meets these conditions because the
densest objects form, on average, earlier and are accreted by
the main halos on orbits with smaller radii (see section 3).
We have shown that the transformation occurs on a orbit with apocenter 
$\sim 100$ kpc, in agreement with the present distances of
Draco and Ursa Minor (Mateo 1998). 
As time goes on, lower-density dwarfs
will form and enter into an enlarged Milky Way halo: these will be on 
wider orbits and will be more akin to the standard LSB models, 
thus ending up in dSphs with lower mass-to-light ratios. An anti-correlation
between distance from the primary
and mass-to-light ratio should thus arise and is indeed
suggested by observations (Mateo 1998; Irwin \& Hatzidimitriou 1995).

Early-type dwarf galaxies in the Local Group show a certain spread in their
total central densities (including baryons and dark matter, see Mateo 1998).
Comparing these data  with the
densities of our remnants shows a remarkably good agreement.
The final central densities of the remnants of LBS are as low as
those of most dSphs (e.g. Sextans, Sagittarius and Fornax)
, namely $0.01 < {\rho}_0 < 0.1 M_{\odot}/pc^3$ (Mateo 1998)
The remnant of GR8 is comparable to the densest dSphs, Draco, Sculptor
and Ursa Minor, having $\rho_0 > 0.1 M_{\odot}/pc^3$ (Lake 1990).
Even higher densities are reached in the inner part of the remnants
of HSBs thanks to the highly concentrated stellar potential, as found
in dEs. However, central densities as high as those of NGC205 
($\rho_0 \sim 5 M_{\odot}/pc^3$, see Mateo (1998)) appear only in the
gas-dynamical runs (model HM2g), where the profile of the
stellar component steepens remarkably.

\medskip
\epsfxsize=8truecm
\epsfbox{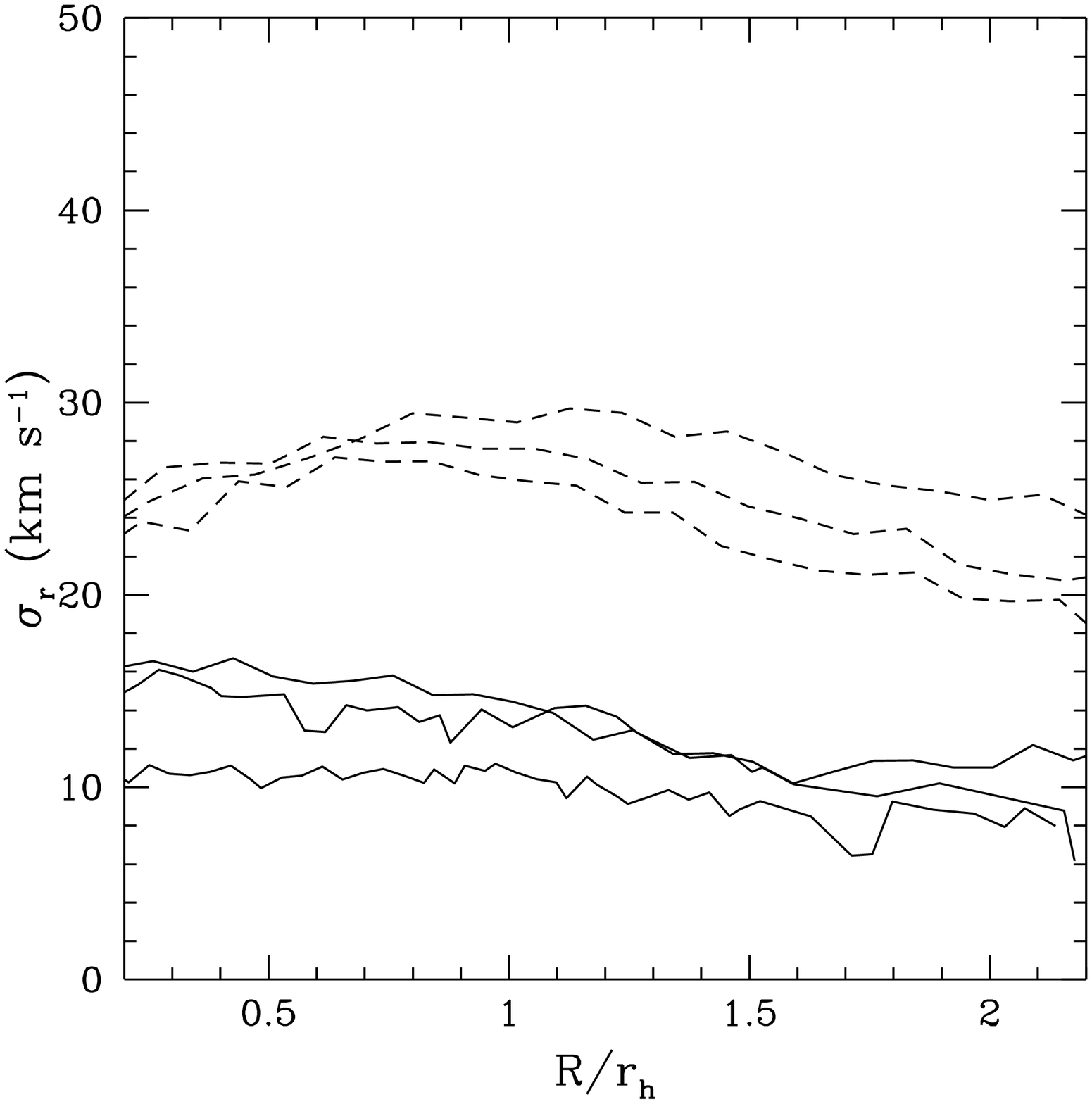}
\medskip
\epsfxsize=8truecm
\epsfbox{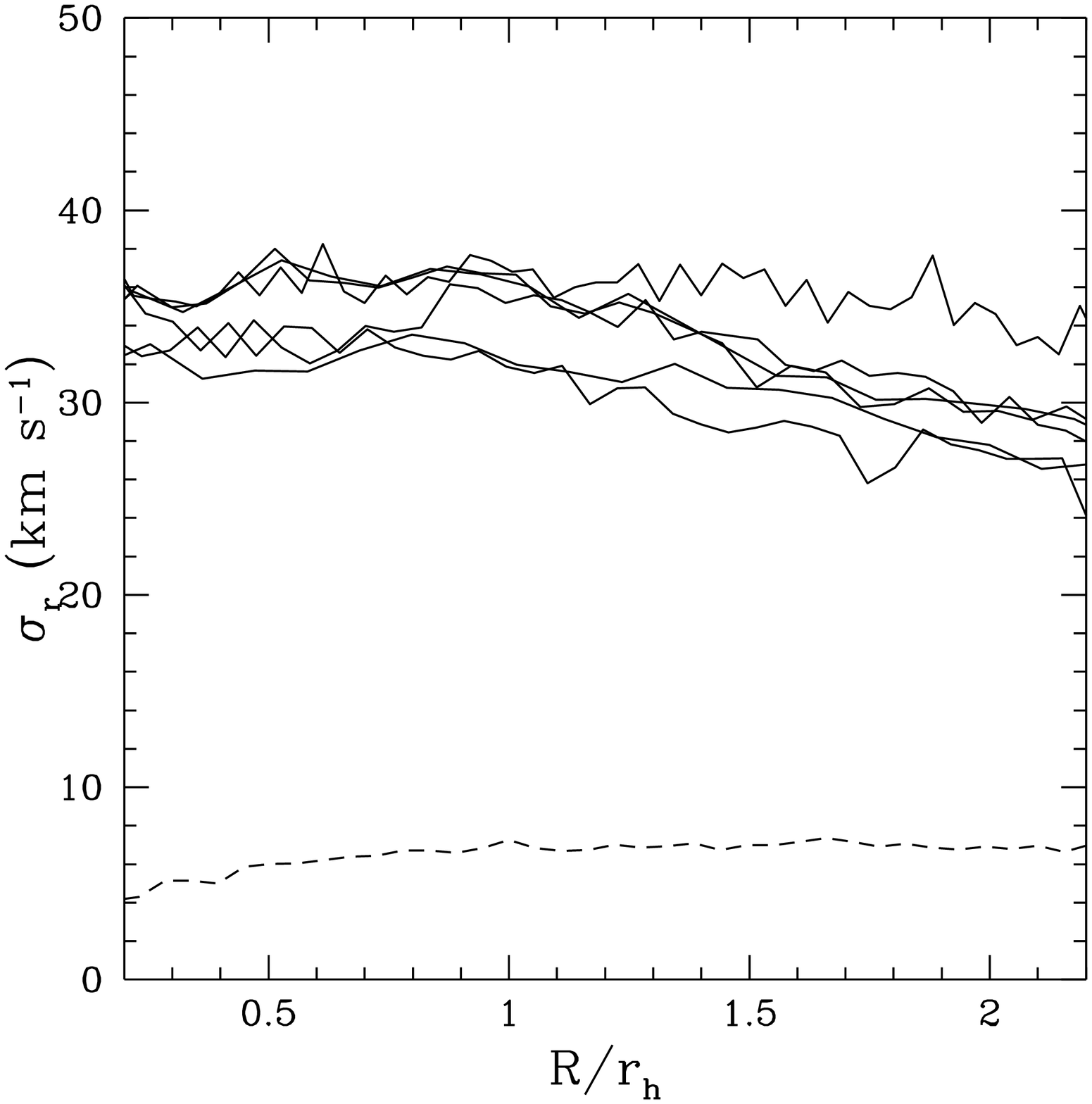}
\figcaption[f21b.eps]{\label{fig:asymptotic}
\small{Radial velocity dispersion profiles 
of LSB remnants (on top, the dashed lines are for LZ runs)
) and HSB remnants (bottom), along
with the profile for the GR8 model, represented by the dashed
line).}}
\medskip

It is important to note that the spread of the central densities of the
remnants is only slightly  larger than that of the initial models. 
Therefore, our ability to reproduce the vast range
of properties of early-type dwarfs results from the choice of initial
models spanning a wide range of central densities. 
The spread in the concentration $c$ in the initial models
is a factor of 4, which corresponds to a factor of
almost 20 in the initial central densities ($\rho \sim c^2$ for truncated
isothermal spheres). In cold dark matter models the typical concentration
of halos on a scale $M \simlt 10^{10} M_{\odot}$ is 
as high as that of our GR8 model. Cosmological N-Body simulations show that
the scatter of the concentration at a given mass scale is large 
(Bullock et al. 2000) but no as much as to encompass the extremely low
values chosen for our LSB models.
Interestingly, the latter values seem to be very common for small halos
formed in wark dark matter (WDM) cosmogonies 
(Bode, Ostriker, \& Turok 2000; Eke, Navarro, \& Steinmetz 2000).

\medskip
\epsfxsize=4truecm
\epsfbox{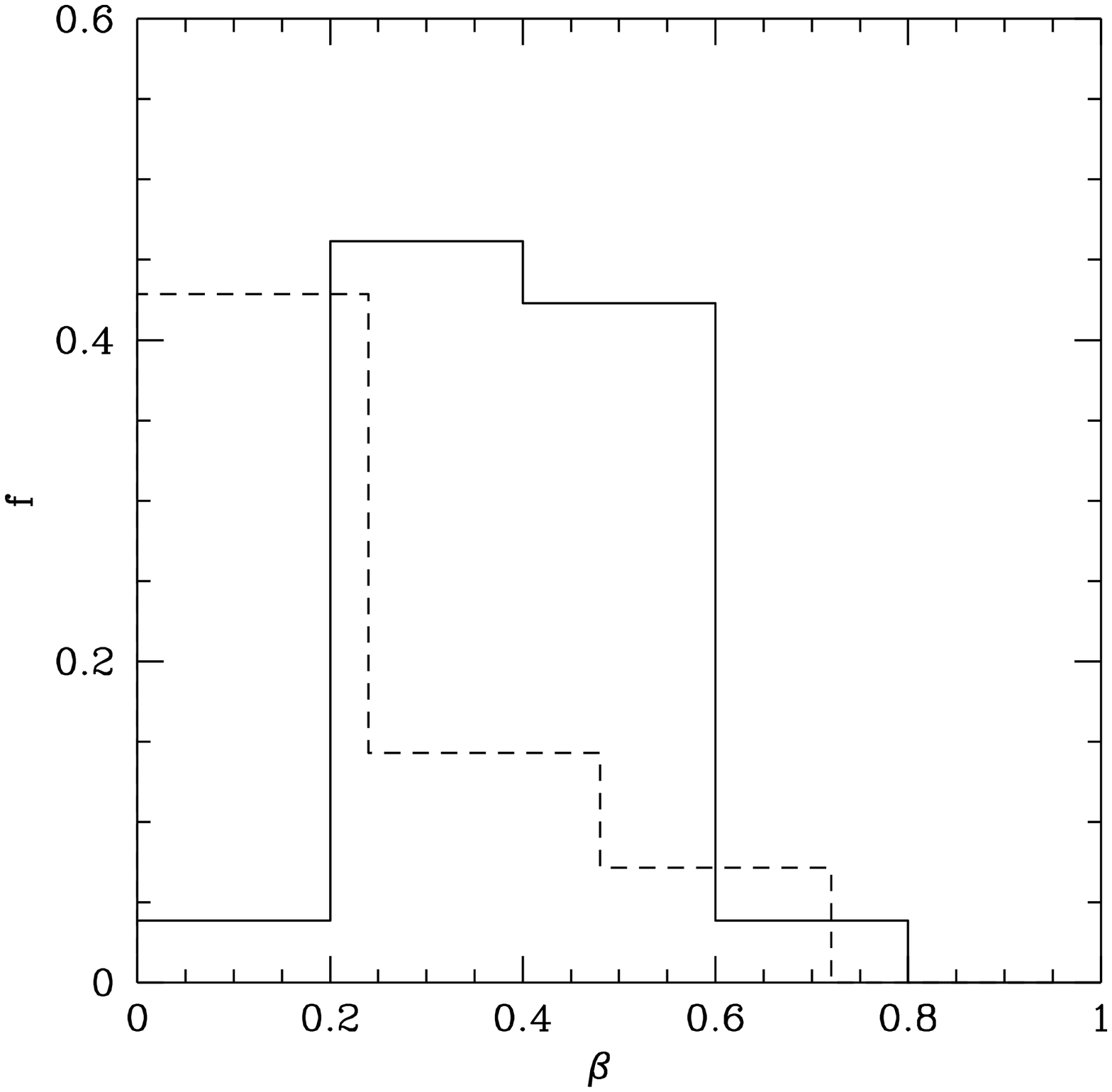}
\medskip
\epsfxsize=4truecm
\epsfbox{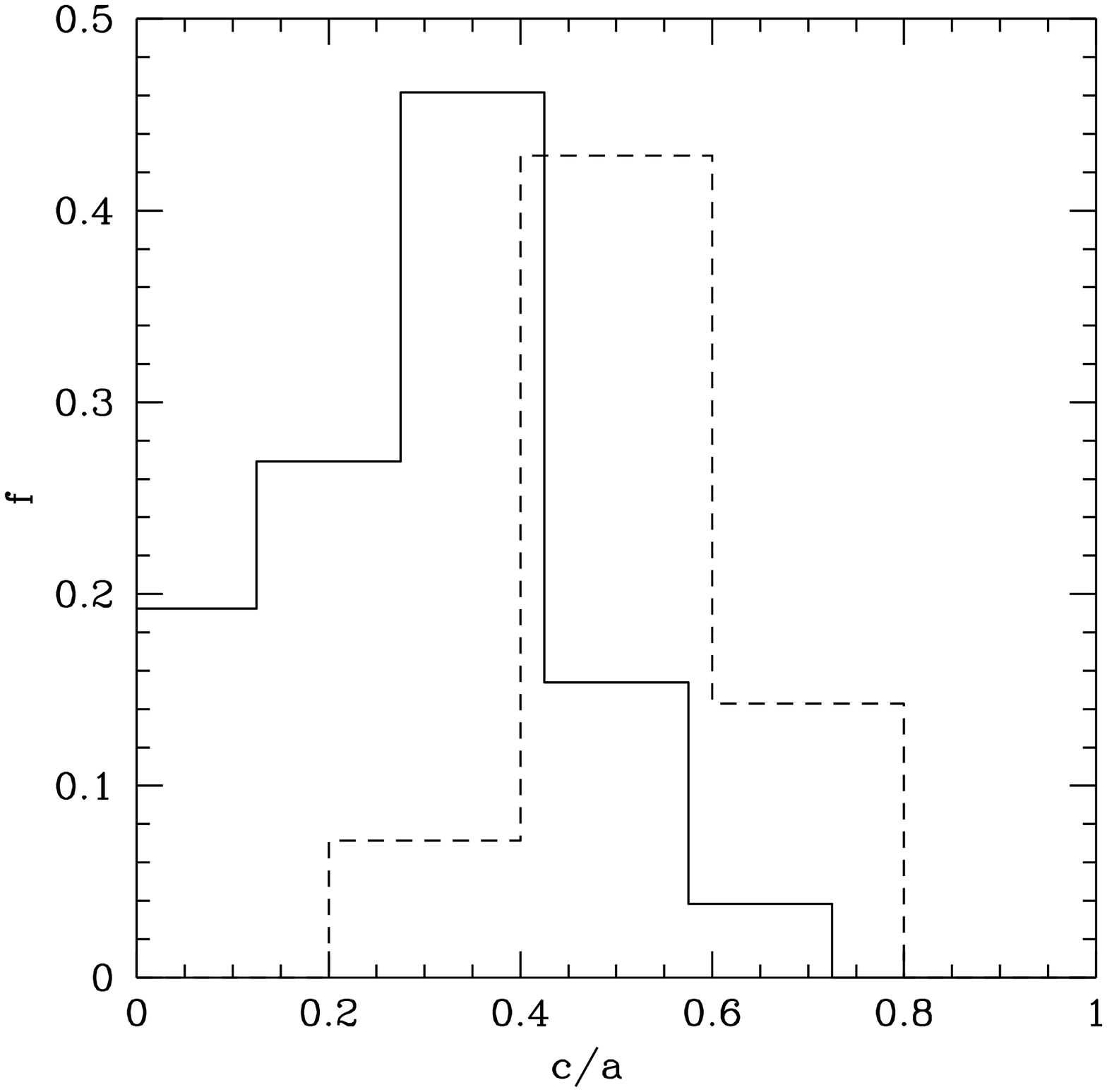}
\medskip
\epsfxsize=4truecm
\epsfbox{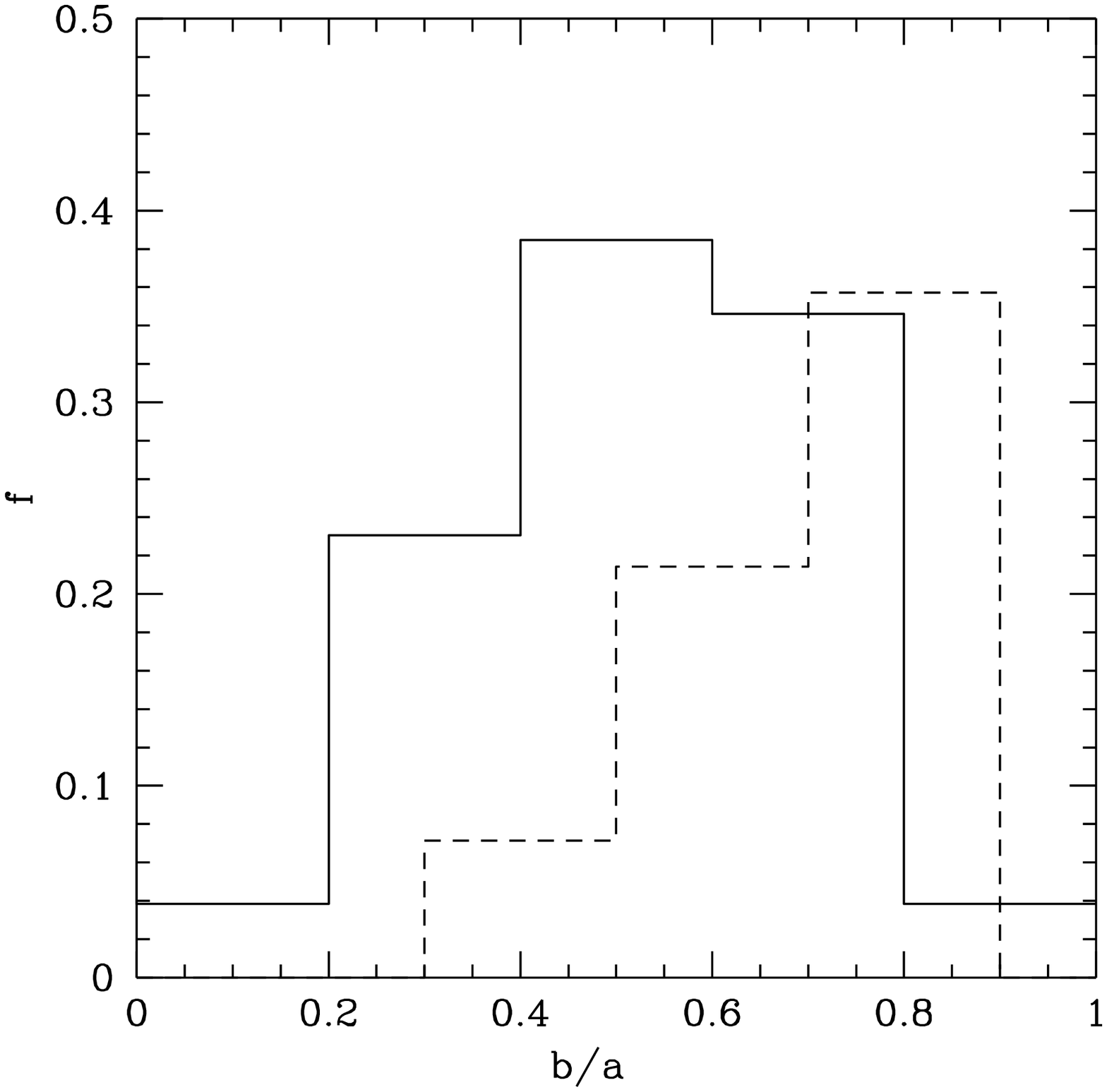}
\medskip
\epsfxsize=4truecm
\epsfbox{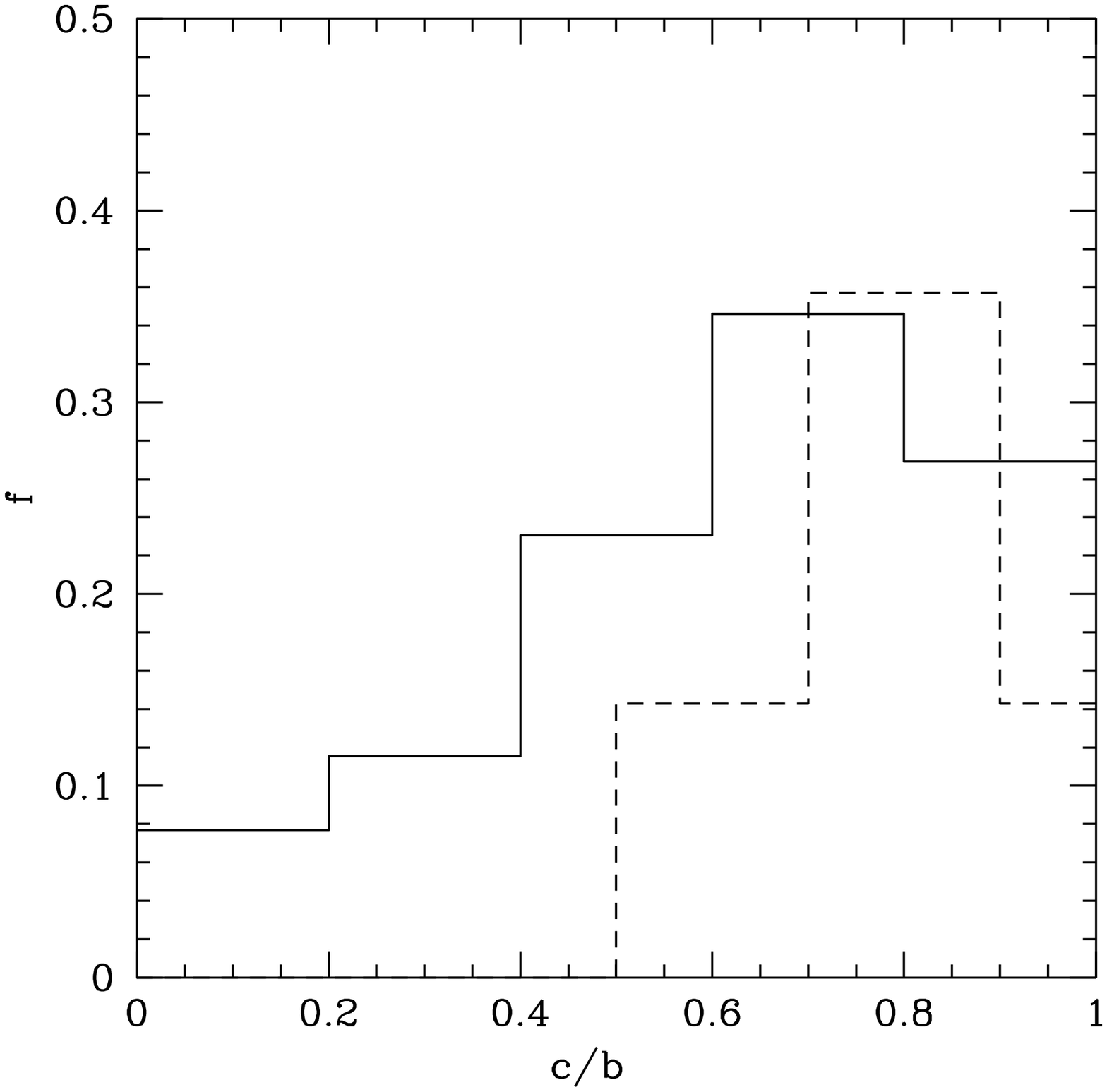}
\figcaption[f22d.eps]{\label{fig:asymptotic}
\small{Histograms showing the fraction of the remnants, $f$,
whose anisotropy parameter $\beta$ or
apparent ellipticities resulting from the ratios of two of their principal
axes fall within the values indicated
on the x-axis: the solid line
is for the remnants of LSBs, the dashed line is for the remnants of HSBs.}}
\medskip

\subsection{Apparent versus intrinsic $M/L$}

The issue of the high mass to light ratios found in some of the dSphs
is critical for understanding in general the nature of dark matter (Lake
1990). 
Many authors have pointed out that some steps
of the core fitting procedure used to derive $M/L$ for dSphs 
and/or tidal effects  could lead to overestimate them
(Piatek \& Pryor 1995 ; Irwin \& Hatzidimitriou 1995).

It is usually assumed that mass follows light in dSphs.
In Figure 24 we see the comparison
between final total mass and stellar mass profiles for three ``prototype''
models. For HSB and LSB remnants one would actually slightly
underestimate $M/L$ (by 10-30 \%) by assuming that mass follows light.
In contrast, the difference is remarkable for models as stiff as GR8, where
we would underestimate the dark matter mass by a factor $\sim 2$.

We can also ask whether the assumptions that (1) the measured velocity
dispersion can be interpreted as resulting from virial equilibrium
and that (2) such equilibrium is isotropic hold for our remnants.

Nearly all of our remnants have reached a nearly-stable equilibrium
by the end of the simulations.
However, in some of them a fair degree of rotation is present 
in addition to the velocity dispersion and this should be taken into account 
when estimating the virial mass of the object. 
In a few cases, in which $v/\sigma \ge 0.5$, one would underestimate
the true mass-to-light ratio by nearly a factor of 2 considering only
the contribution of the velocity dispersion to the kinetic energy
(because M $\sim <v^2>$). Therefore, if, for example, the rotation 
apparently detected in Ursa Minor by Hargreaves et al. (1994a) is
intrinsic, the  true $M/L$ could be even higher than that usually assumed.

\medskip
\epsfxsize=8truecm
\epsfbox{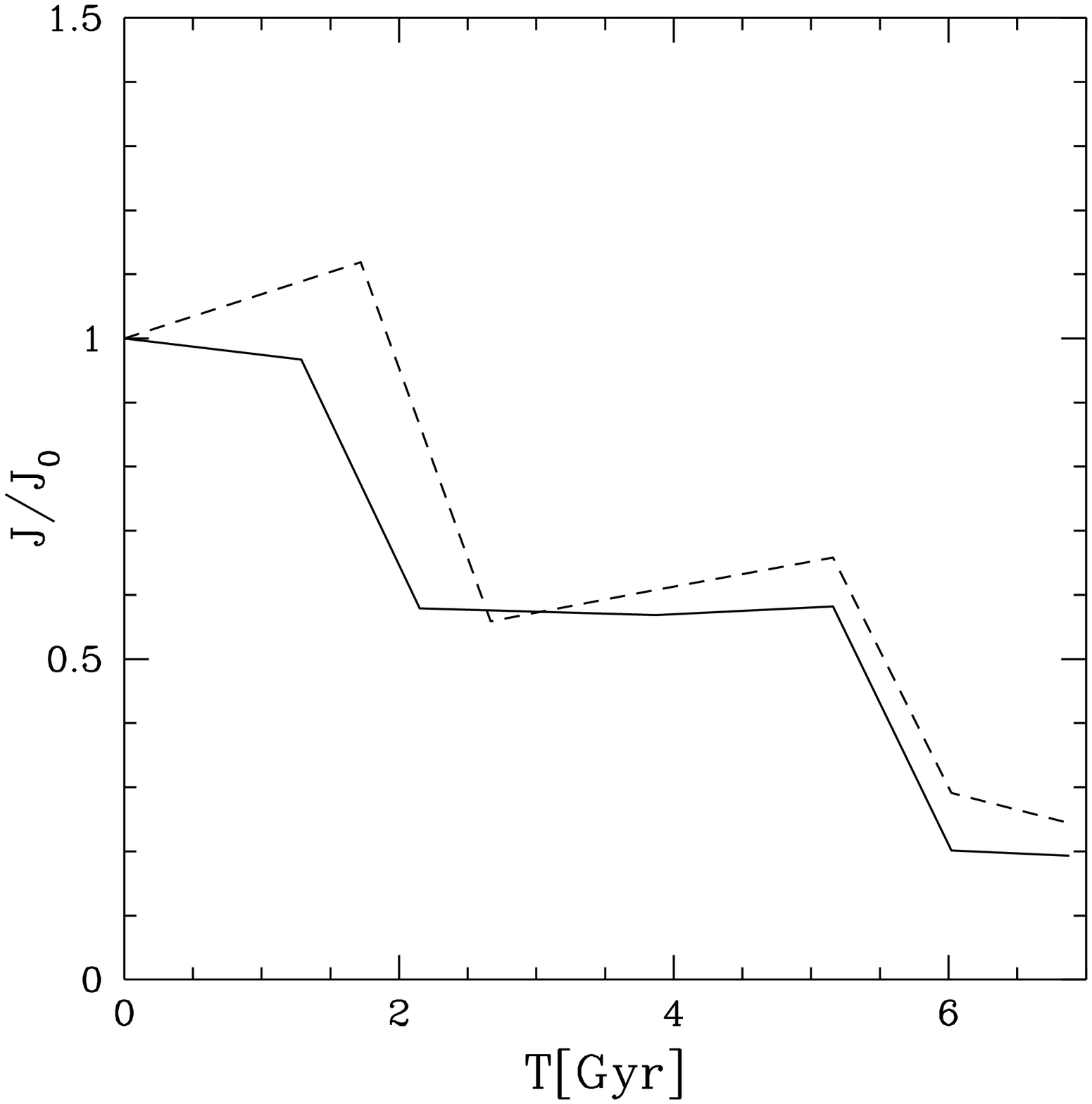}
\medskip
\epsfxsize=8truecm
\epsfbox{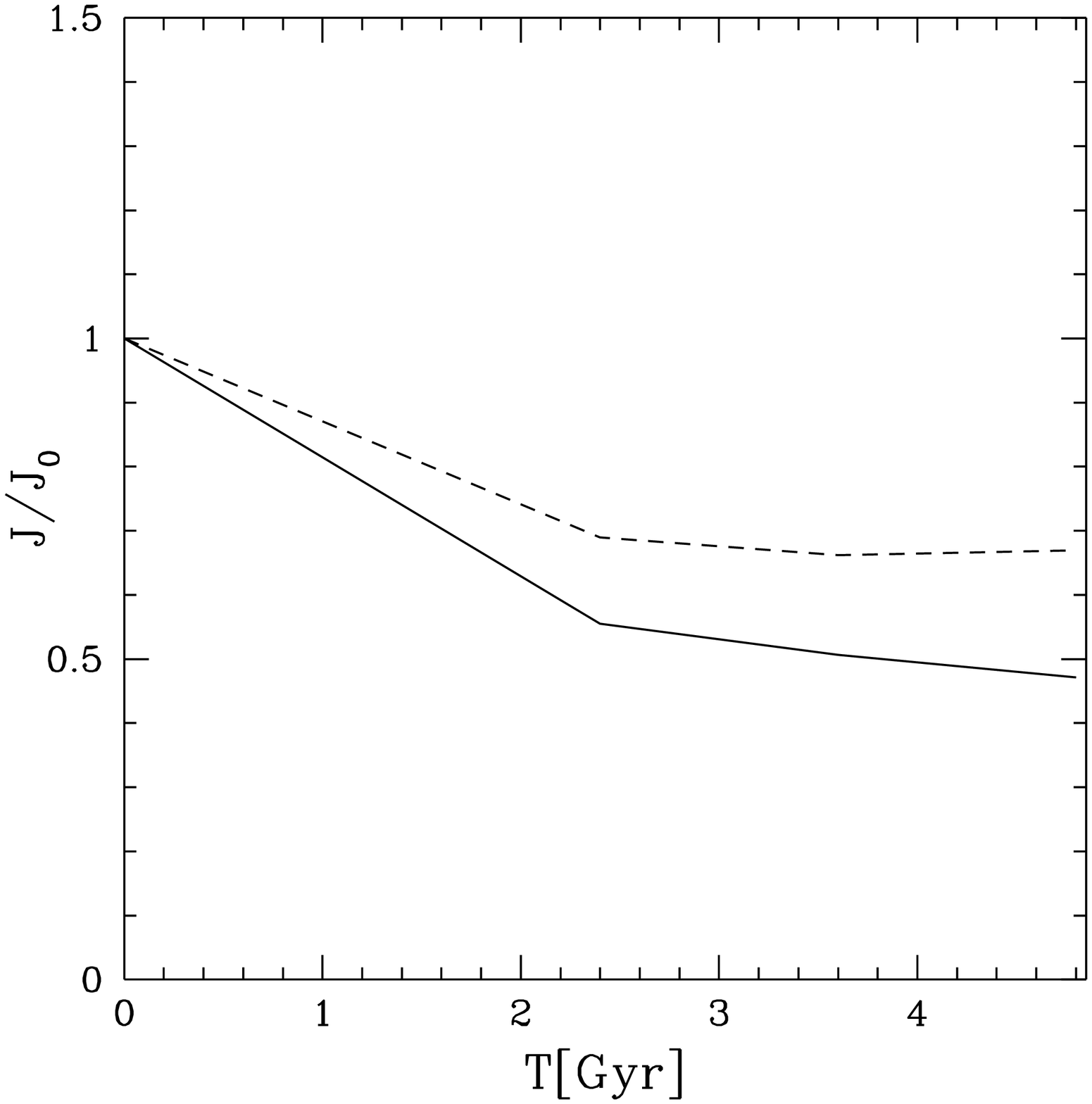}
\figcaption[f23b.eps]{\label{fig:asymptotic}
\small{Evolution of the specific angular
momentum of the bound stellar component in collisionless
(solid line) versus gasdynamical (dashed line) runs. On top
the results for an LSB run  are shown, while on bottom 
we plot those relative to an HSB run. The orbital parameters
are those those of runs L13 (LSBs) and run H08 (HSBs), listed in
Table 2 and 3.}}
\medskip

According to Figure 22 the maximum
value of the anisotropy parameter  
$\beta$ is $\sim 0.8$, which means that velocity dispersion
in the plane of the galaxy can be a factor $\sim 2$ higher than that normal
to the plane (here we are assuming that the velocity dispersions
along the two principal axes lying {\it within} the plane are comparable, as
found in most of the remnants); at most one would overestimate
the one-dimensional velocity dispersion by only $\sim 20 \%$ (from
$\sigma_{1d}=\sqrt{({\sigma_x}^2 + {\sigma_y}^2 + {\sigma_z}^2)/3})$,
which leads to an overestimate of a factor $\simlt 1.4$ for $M/L$
(from $\Delta {\sigma}^2 \sim \Delta M$) as an upper limit.
However, only $\simlt 5 \%$ of the remnants exhibit such high final radial
anisotropies.

A more interesting issue is provided by projection effects:
the velocity dispersion measured along the line
of sight (and thus  the
$M/L$) could be overestimated if the unbound, fast moving stars 
that make up the
tidal tails are included by chance in spectroscopic samples.
These ``extra tidal'' stars produce a radial
positive gradient in the velocity dispersion profile (Figure 25).
The larger variations are seen in the LSBs due to its more pronounced
mass loss.

\medskip
\epsfxsize=8truecm
\epsfbox{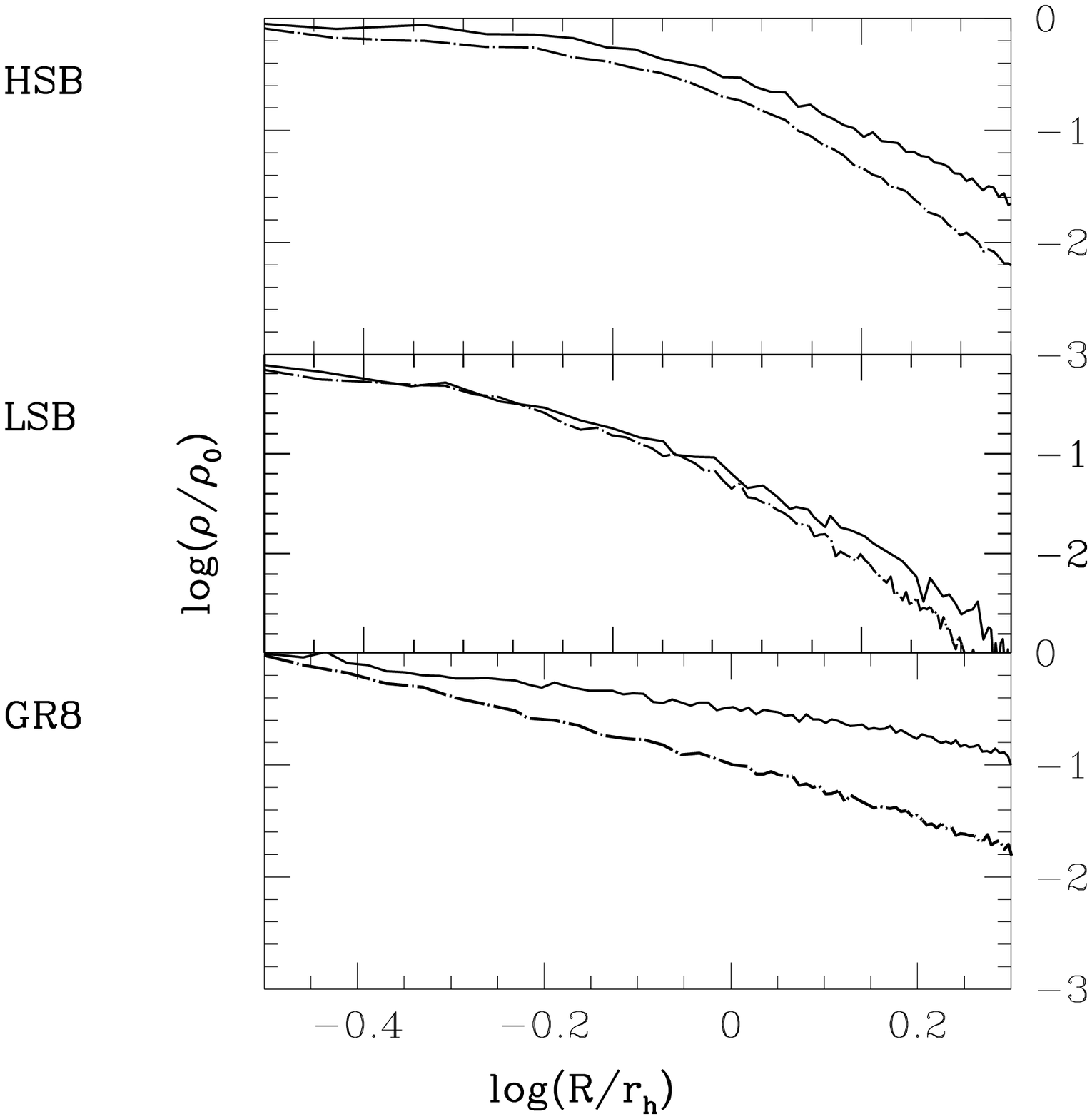}
\figcaption[f24.eps]{\label{fig:asymptotic}
\small{Comparison between final total mass (solid line) and stellar
mass profiles (dot-dashed line) for various remnants inside the final
tidal radius (the corresponding runs are H08, L10 and GR82). 
Densities are normalized to their central values
.}}
\medskip

As we can measure only projected distances, unbound stars, although 
lying far away from the center, could appear to be located
near the core of the dwarf.
From our simulations we can get an idea of how these projection 
effect can alter the observed kinematics.
We derived an upper
limit for the velocity dispersion overestimate for both LSBs and HSBs,
by selecting the ``worst case possible", namely two
runs with direct orbits (L01 and H01, see Table 2 and Table 3), 
in which the largest tails are produced. 
We viewed the remnant along the orbital plane close
to apocenter, where the stream density is at one of its density maxima
and the galaxy spends most of the time (at the turning points of 
the orbits the stream density is higher
because of conservation of phase space density, e.g. Helmi \& White 1999). 
We measure the average velocity dispersion of the bound and
unbound stars projected within a distance equivalent to $R_e$ 
and find the average of the dispersions of the two individual
samples by weighting according to the number of stars in each of them.

We find
that in this extreme case the velocity dispersion can be overestimated
by $\sim 40 \%$ for the LSBs, while for the HSBs the difference is of
a few percents. As a consequence, the mass-to-light ratio can be 
overestimated by a factor $\simlt 2$ for the LSBs, while essentially
no overestimate occurs for the HSBs (the latter statement would be 
even more true for dense dwarfs like the GR8 model).
The remarkable
difference in the apparent dispersions of LSBs vs. HSBs simply arises because 
in the former case a much larger fraction of the stars is located 
in the tidal streams that follow the original orbit.

\medskip
\epsfxsize=8truecm
\epsfbox{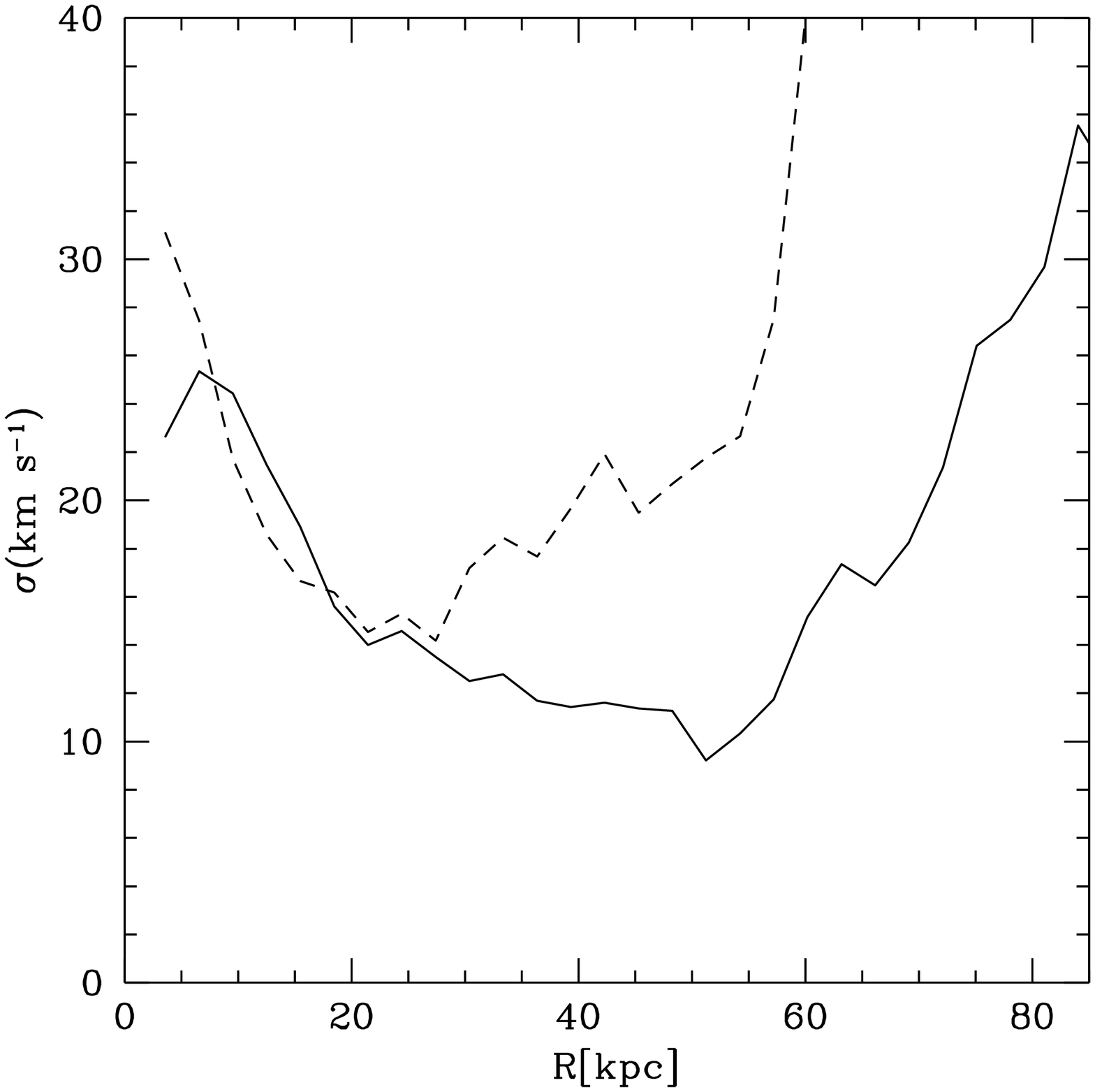}
\medskip
\epsfxsize=8truecm
\epsfbox{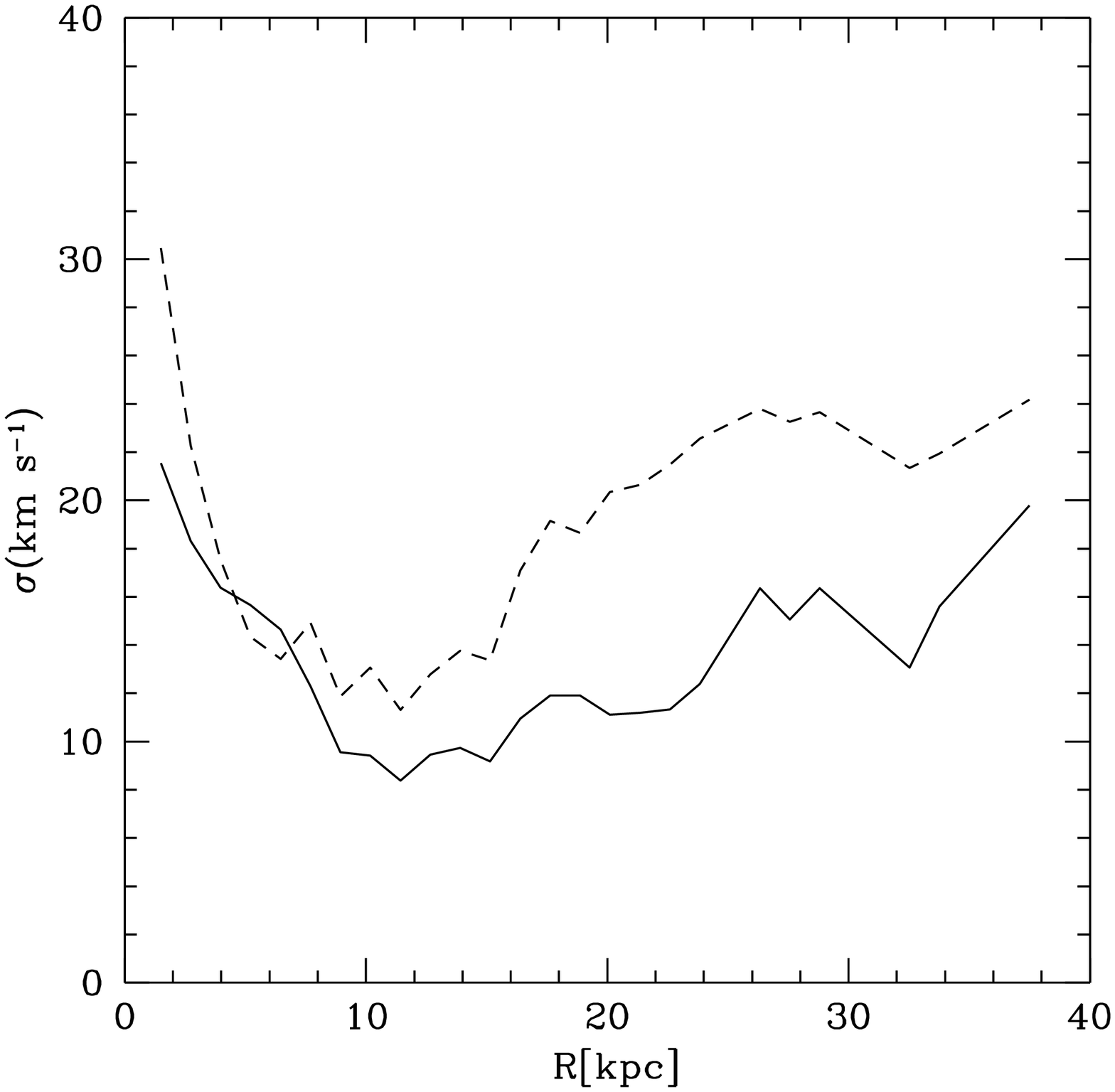}
\figcaption[f25b.eps]{\label{fig:asymptotic}
\small{Radial velocity dispersion (solid line) and tangential velocity
dispersion (dashed line) out to large radii for both an LSB (top) and 
an HSB remnant (bottom) with the same 
orbital parameters (run L01 and run H01).}}
\medskip

Although not negligible, such tidal effects 
cannot rule out the need of high dark matter contents in some of the 
dSphs: objects like Draco and Ursa Minor ($M/L > 25$) would still 
have $M/L > 12$ even with the maximum velocity dispersion
enhancement just described.
On the other end, the velocity gradient shown in Figure 25 
suggests that the rotation detected in Ursa Minor 
(Hargreaves et al. 1994a) might be only apparent.

\section{The nature of tidal stirring: the dependence on initial conditions}

So far we have based our description of the numerical results on the set 
of satellite models that we
consider more akin to observed dIrrs and we have placed them on
the typical orbits found in N-Body simulations of structure formation.
Furthermore, we have considered only one form of the primary potential, whose
global parameters were fixed again based on structure formation models.
In order to figure out how general the ``tidal stirring'' mechanism explored
in this paper is, in this section we will relax each of these hypotheses,
one at a time, investigating a much wider parameter space.

\subsection{Dependence on the initial structure of satellites}

As already mentioned in section (2.1), we have
built a special set of models which should be the least
prone to bar and/or bending
instabilities by varying $Q$, the initial disk thickness or the
core radius of the halo (see Table 1 for the models' parameters
and Figure 2 for the intrinsic stability properties).
We mostly placed such models on a direct orbit with moderate eccentricity
(apo/peri=4, $R_{peri} = 80$ kpc):
this configuration, in which the excited bar is barely
damaged by tidal stripping even in the case of the fragile LSBs,
is ideal to study the vertical heating caused by bar-buckling.

The results are illustrated in Figure 26. An LSB model with a
small core radius (run L21) undergoes a smaller increase of 
$\sigma_z$ (by about $\sim 50 \%$) compared to the standard 
reference case (run L02).
A weaker bar develops because the higher
halo central density damps the growth of $m=2$ modes and the 
buckling is thus reduced due to the lower radial anisotropy.
An analogous result is obtained with an HSB disk embedded in a more
compact halo (model HM1rc03), which has an initial rotation curve
very similar to that recently derived by Sofue (1999) for the LMC.
If placed on a nearly circular orbit (run H11) as implied by recent astrometric
measurements (Kroupa \& Bastian 1997), the kinematics remain disk-like
at $R > R_e$ (stripping is very modest), but a thick spheroidal component
develops after 5 Gyr. These results qualitatively agree with those
obtained by Weinberg (2000) in a detailed study of the 
response of the LMC to the tidal field of the Milky Way:
the thick spheroid could be responsible for some of the microlensing
events seen towards the LMC (Evans \& Kerins 2000).
In the models with $Q=4$ (run L23), namely considerably ``hotter'' than
the standard models,
the bar is also weak and $\sigma_z$ increases by
less than a factor of 2.

\medskip
\epsfxsize=8truecm
\epsfbox{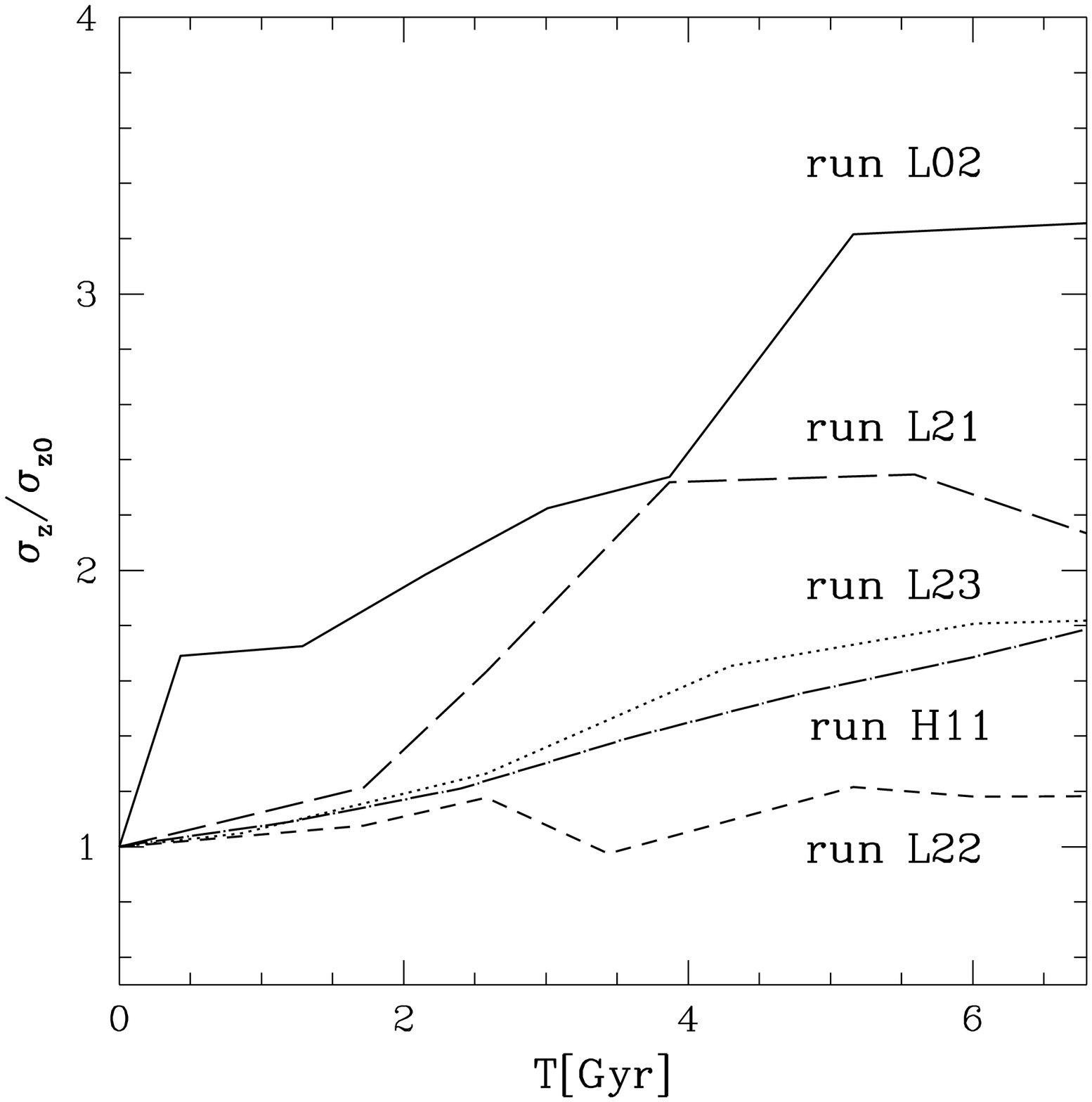}
\figcaption[f26.eps]{\label{fig:asymptotic}  
\small{Variation of the vertical velocity dispersion with respect to
the initial value in runs using models with different intrinsic stability
properties (see Tables 2,3 and 1 for the full description of, 
respectively, the runs and the satellite models employed).}}
\medskip

We then explore models with a larger scale height ($z_s = 0.3 r_h$)
and $Q=2$ (model LMH in run L22): the vertical dispersion increases 
by only $40 \%$, a factor $\sim 3$ less than in run L02. 
This is not surprising because in this model $\sigma_z$ is initially
2 times larger than usual, placing the dwarf well above the critical
radial anisotropy level necessary for the onset of bending (see Raha
et al. 1991).

Despite the different degree of evolution reported in these ``peculiar
models'' the final $v/\sigma$ inside the half-mass radius 
is always less than 1, with the exception of run L22.
However the latter case is rather unrealistic as 
late-type spiral and dwarf irregular galaxies show a tendency
towards very thin disks (see Van der Kruit \& de Grijs 1999). 

In general, our evolutionary model seems quite robust:
tidal shocks are too strong for the initial stability properties of 
the models to change substantially the final outcome of tidal stirring.

\subsection{Dependence on the orbital parameters}

In this section we will investigate to what extent the
effectiveness of tidal stirring depends on
the orbital parameters and, in particular, on the eccentricity
of the orbits. Although eccentric orbits
are expected for satellites in the framework of hierarchical
structure formation (Ghigna et al. 1998), so far
observational data on dSphs have been in favour of nearly circular
orbits (e.g. Schweitzer et al. 1995). We thus perform a number of runs
in which satellite models are placed on low-eccentricity orbits. 
LSB satellites on orbits with apo/peri $\sim 1.5$
maintain $v/\sigma > 1$ and a disk-like appearance even
after 7-8 Gyr if their pericenters are larger
than 150 kpc (run L08 and L17), irrespective of the disk/orbit orientation.
On the other end, if the orbits have equally low eccentricity but
considerably smaller pericenter (run L16 and H09)
, both LSBs and HSBs do transmute into
a spheroidal galaxy after a few orbits.
In the latter case the apocenter
($R_{apo} \sim 250$ kpc) is comparable to the present distance of 
the farthest dSph, Leo I, and more and stronger tidal shocks 
are suffered by the satellite
because the orbital time is shorter than in runs L08 and L17 
($\sim 3$ Gyr instead of $5$ Gyr)
and the pericenter distance are nearly a factor of 2 smaller.

However, compared to the standard runs employing
high-eccentricity orbits (e.g. run L1-L15 and H01-H08)
the velocity at pericenter is sensibly lower ($\sim 280$
km/s instead of $400$ km/s) and therefore the encounter takes place
in a ``less impulsive'' regime. Therefore, the satellite should
respond more adiabatically to the perturbation.
Evidently, the lower velocity is not  able to counterbalance the
effect of the large perturbing mass of the Milky Way halo
when the pericenter distance is sufficiently small. 
Therefore, orbital energy (or pericenter distance)
is at least as important as orbital eccentricity in setting
the effect of tides, as found also by Gnedin et al.(1999) using 
a semi-analytical approach.

\medskip
\epsfxsize=8truecm
\epsfbox{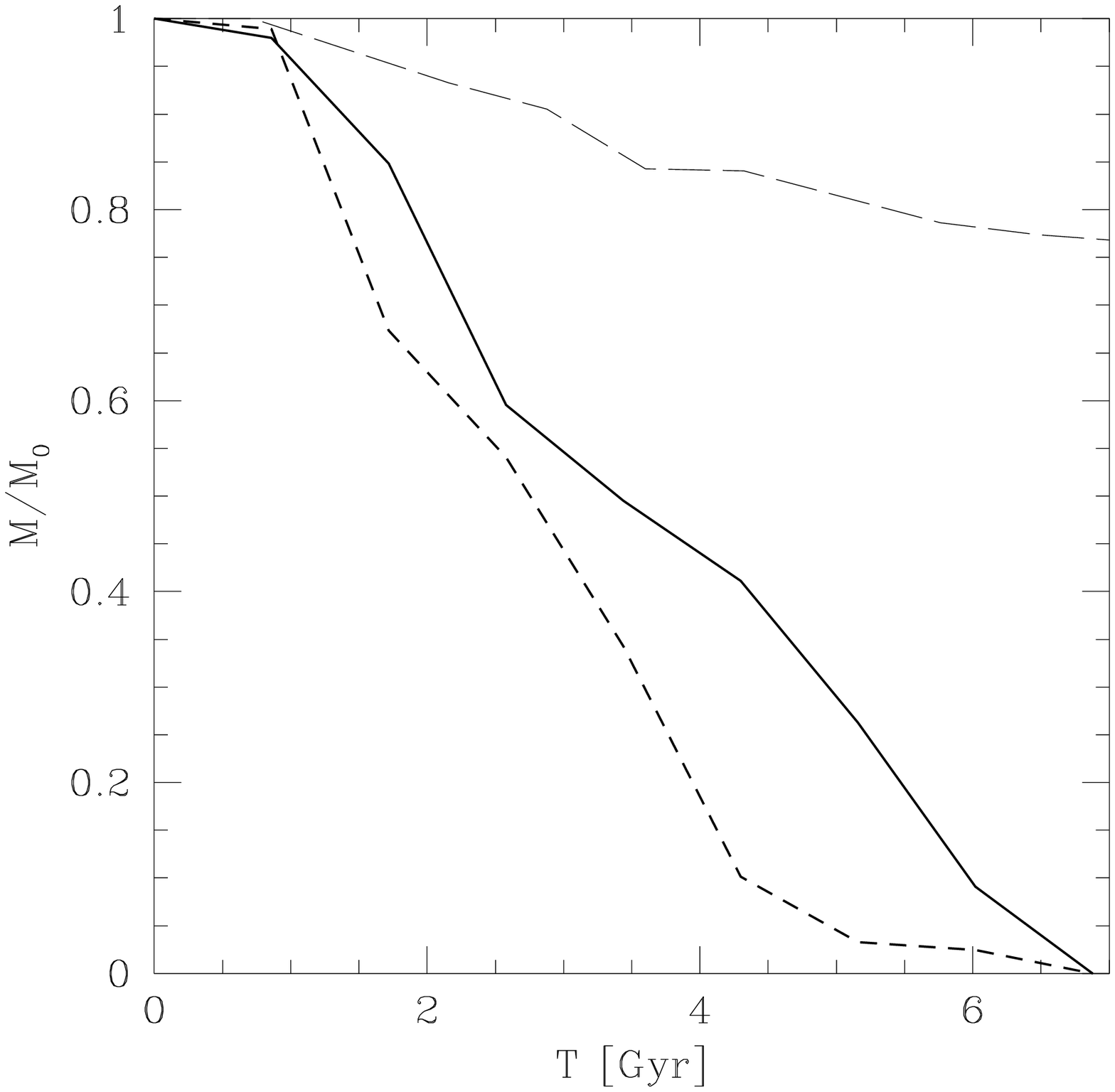}
\figcaption[f27.eps]{\label{fig:asymptotic}
\small{Tidal mass loss for satellites on low-eccentricity, tightly
bound orbits. The long-dashed line is for an HSB (run H10),
the solid line is for an LSB on the same
orbit (run L18) and the short dashed line is for an LSB on a circular orbit
(run L19).}}
\medskip

From our wide exploration of parameter space we can conclude that
the orbital time $T_{orb}$ is the most significant parameter and
that tidal stirring has been effective within
the past 10 Gyr only for satellites having $T_{orb}\le$ 4 Gyr,
which corresponds to a circular orbit with a radius $\simlt 0.5 R_{vp}$ 
($R_{vp}$ being the virial radius of the Milky Way).

Very dense satellites require smaller orbital times, allowing for
them {\it more and stronger shocks} (a smaller orbital time also implies
a smaller $t_{coll}$, see section 4.1), as shown
by the simulations of GR8; in the framework of hierarchical
structure formation this comes
out automatically because, on average, they formed earlier and thus fell into
the Milky Way halo when it had a smaller virial radius.
Thanks to the rescaling properties of halos (see section 2), 
the condition on the orbital time is automatically generalized
for every redshift $z$ if we assume that $R_{vp}$ is the virial
radius of the main halo {\it at the redshift of infall}.

The present distances of most
of the dSphs in the Milky Way or M31 systems are within $0.5 R_{vp}$ 
(assuming $R_{vp} \sim$ 400 kpc) and
should thus satisfy the constraint on the orbital time.
However, the farthest dSph,
Leo I, having a distance of about 270 kpc, probably lies
at the boundary of the ``allowed'' region.
Thus, if Leo I was tidally stirred by the Milky Way, it must be on
a fairly eccentric orbit in order to have a sufficiently small
orbital time despite its large apocenter distance. 
At more than
800 kpc from the Milky Way the Tucana dwarf (Gallart et al. 1999) could be
the greatest challenge for our model if future measurements of its internal
kinematics will confirm that it is a dSph in all respects.
A few satellites on nearly radial orbits with apocenters just inside 
the turnaround radius, namely far exceeding the 
virial radius of the primary halo, are indeed found in N-Body 
simulations (Ghigna et al. 1998).

\medskip
\epsfxsize=8.3truecm
\epsfbox{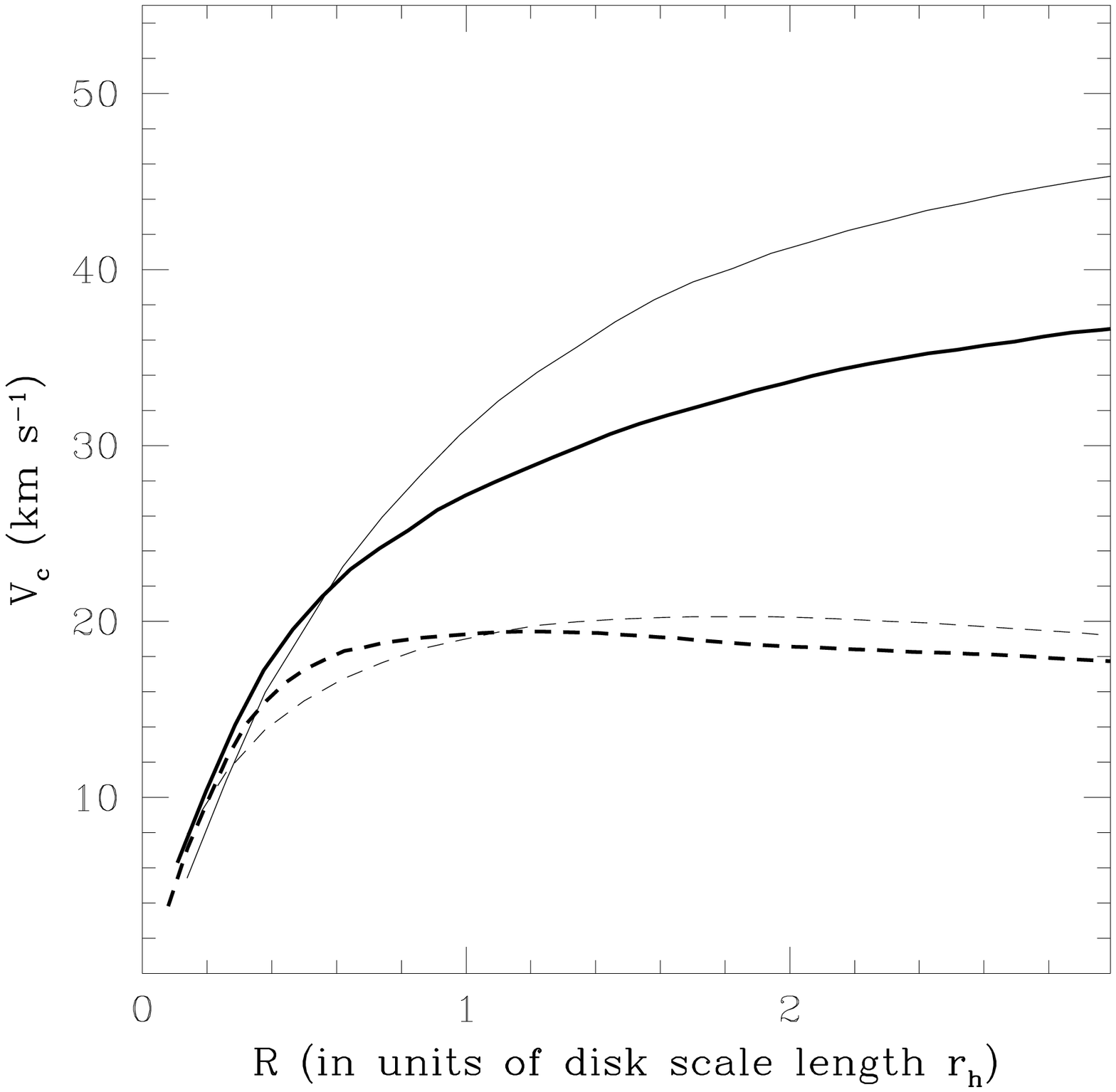}
\medskip
\epsfxsize=8truecm
\epsfbox{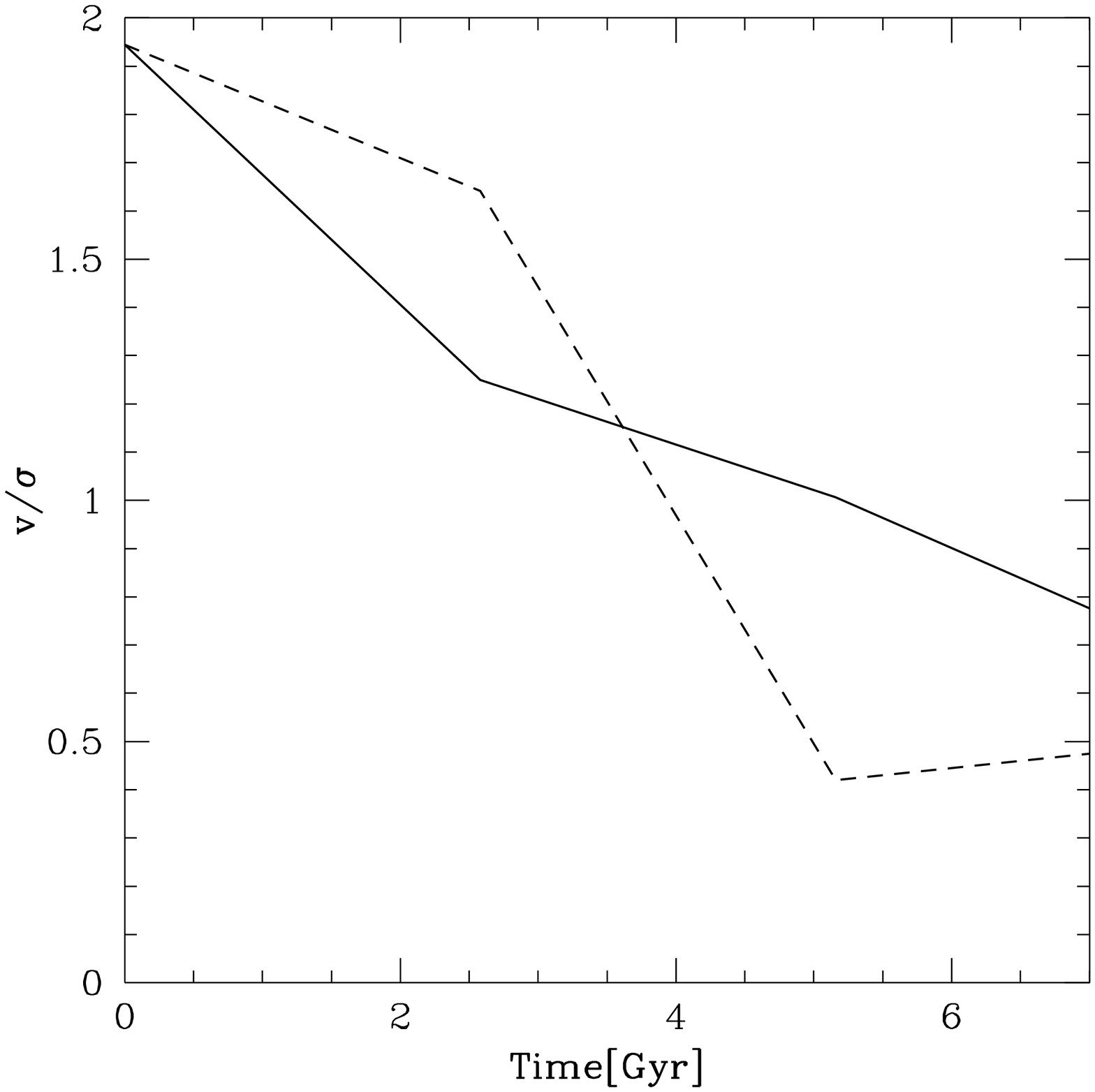}
\figcaption[f28b.eps]{\label{fig:asymptotic}
\small{In top panel: final
circular velocity profile of the LSB dwarf evolved in the ``minimal'' Milky
Way halo, in run L24 (thick lines), plotted
against the initial profile (thin lines). The contribution of the
stellar component is represented by the dashed lines. In Figure 6 an analogous
plot for a run in which the satellite has a similar orbit in the standard 
Milky Way halo (run L17) is shown. In bottom panel:
evolution of the $v/\sigma$ of the bound stellar component
for run L24 (solid line) and for run L17 (dashed
line), the latter using the standard model 
for the primary halo.}}
\medskip

We also performed a few runs to explore the regime of 
{\it very tightly bound orbits}, namely those with a corresponding circular
radius $\le 0.3 R_{vp}$.
In particular, we consider the case
of nearly circular and circular orbits with apocenters $< 130$ kpc
(runs L19, L20, LZ7, H10).
Orbits of this type would be consistent with the the present distances
of some of the Milky way satellites, the LMC, Sculptor, Ursa Minor, Draco and Sagittarius

What would be the fate of
a fragile LSB satellite on such  orbits?
Our simulations indicate that LSBs
are completely disrupted on nearly circular or even exactly circular
orbits at a distance of $\sim 75$ kpc (run L19 and L20).
The mechanism is simply one of continuous mass loss (Figure 27):
the disks of LSBs are so large that the tidal radius falls within them
at these short orbital distances and thus they are dramatically destabilized.
The system responds to strong
mass loss by expanding and thus reducing even further its binding energy
(Binney \& Tremaine 1987; Gnedin \& Ostriker 1999).
The destruction is faster on a circular orbit (Figure 27)
because the orbit-averaged tidal force is stronger.
This is in agreement with N-Body results on the survival of
cosmological halos on tightly bound orbits (Kravtsov \& Klypin  1999).
Instead, HSBs or LZ models (run H10 and LZ7) have disks lying 
inside their tidal radius and
thus lose mass only after tidal heating has sufficiently loosen
their potentials: they lose $20 \%$ of their stellar
mass and transform into a spheroidal but survive for several Gyr.

We conclude that low surface brightness satellites cannot
survive too close to the Milky Way or M31. Sagittarius might be the 
nearly-disrupted remnant of such a satellite: what remains
of the central part  of the LSB satellite, in run L19 and L20,
is recognizable as a density enhancement along the streams for 
a few Gyr and has a maximum elongation  $b/a=0.2$, similar to 
that inferred  for Sagittarius (Mateo 1998).
Stable satellites at short distances from the Milky Way must have
high central dark matter densities, as indeed seems to be the case for 
the nearby Draco and Ursa Minor dSphs (Mateo 1998; Van den Bergh 1999).

\subsection{Dependence on the primary halo potential}

A primary halo as massive and extended as adopted so far is expected
in structure formation models and is consistent with the observed
radial velocities of satellites once even the farthest
of the group , Leo I and Leo II, are taken into account
(Zaritsky et al. 1993; Wilkinson \& Evans 2000). A more conservative
approach is based on the use of the Magellanic Stream as halo mass 
tracer: the stream is likely bound to the Milky Way  and its origin
is explained as the result of gas being tidally stripped from the LMC
(Putman et al. 1998). Under the latter hypothesis the
Milky Way halo could extend out to only 50 kpc and could have a
total mass of only $5 \times 10^{11}$ solar masses (Little \& Tremaine 1983).
Assuming then an isothermal profile with a core radius of 4 kpc still yields
a circular velocity of 220 km/s at the solar radius, as observed.

We placed an LSB model on a bound eccentric orbit with apocenter 
of 220 kpc and pericenter of 25 kpc (run L24, see Table 2) in this
``minimal halo''. The apocenter of this orbit is still sufficiently large
to be consistent with the distance of the second most distant dSph satellite,
Leo II ($\sim 200$ kpc).
Due to the truncation of the potential the orbital time is larger than
for orbits with similar parameters in the standard potential 
($\sim 5$ Gyr instead of $\sim 3$ Gyr) and thus the satellite performs 
only one pericenter passage in 7 Gyr. The lower number of shocks as well 
as the longer time spent by the dwarf far from any significant 
influence of the tidal field produce a remnant with a still 
recognizable disk structure and $v/\sigma \sim 0.8$, larger than 
that of any dSph. In Figure 28 we see the evolution
of $v/\sigma$ for this run (L24) and also a comparison with run L17
(see Table 2), the latter corresponding to an orbit with 
similar apocenter and lower eccentricity placed in the standard halo model. 
In Figure 28 we can also see that the circular velocity profile
evolves much less than in the case of LSB satellites in the standard
primary halo (Figure 6), indicating that stripping is comparatively modest.
These results strengthen the point made in the last section,
namely that the orbital time is the key parameter
determining the transformation. Moreover, they highlight that
primary halos as massive and extended as those formed
in CDM cosmogonies are required to explain the
origin of {\it all} LG dSphs, out to the distances of Leo I and Leo II,
with tidal stirring.

\medskip
\epsfxsize=8truecm
\epsfbox{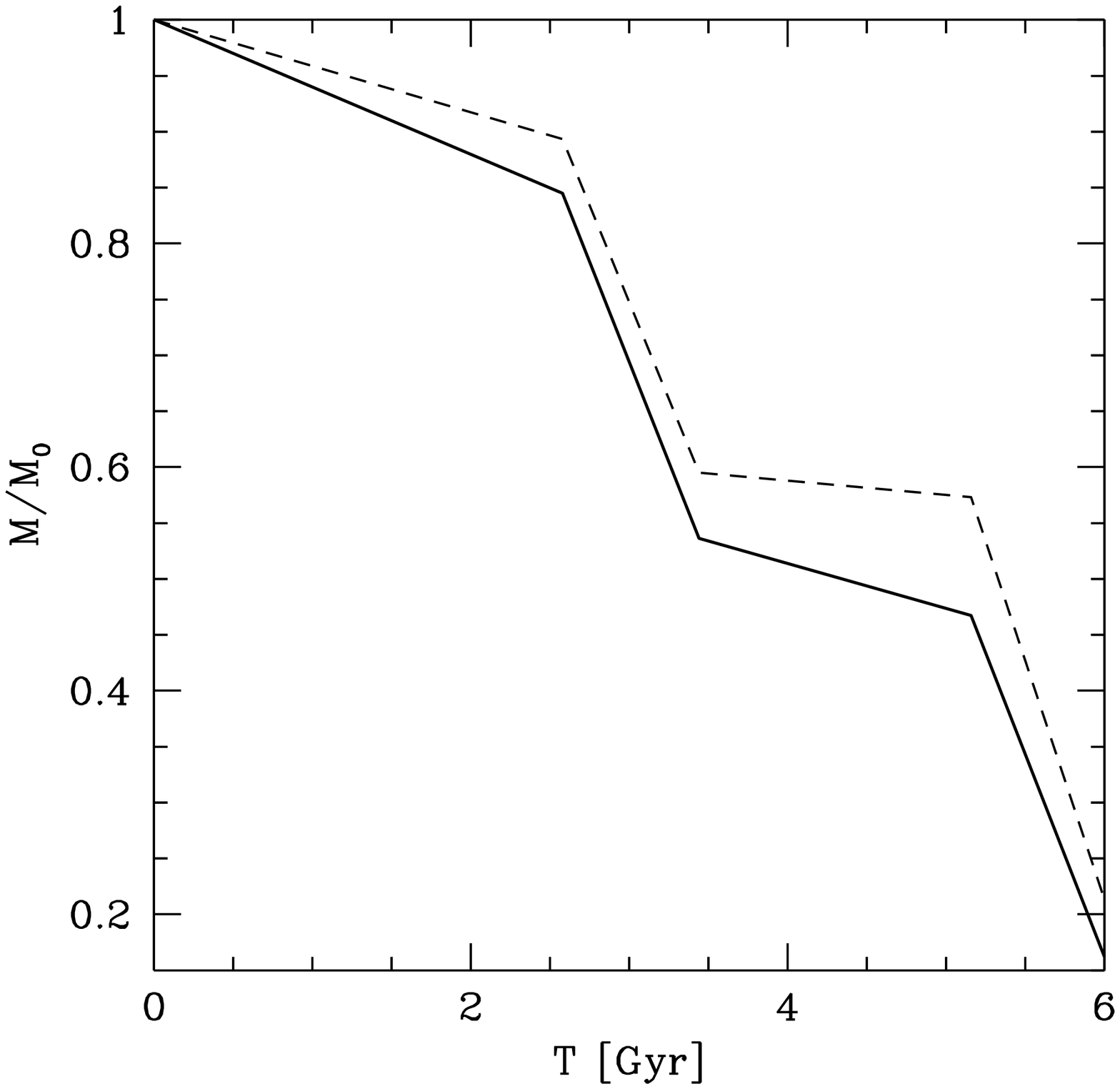}
\figcaption[f29.eps]{\label{fig:asymptotic}
\small{Evolution of the bound stellar mass for the LSB satellite 
when the disk potential is included in the primary
(solid line, run L25) or when only the dark matter halo of the
primary is responsible for the external perturbation
(dashed line, run L13). The same orbit was used in the two runs.
}}
\medskip

The plunging orbits chosen in many of our simulations penetrate quite close
to the center of the primary halo, at 30-40 kpc. The Milky Way and M31 have
disks that extend at least out to 15-20 kpc. However, their mass is
expected to be small compared to that of their halos, even if they
do not have $M_{dark}/M_{stars}$ ratios as high as those of 
dwarf galaxies (Persic \& Salucci 1997). In order to test the validity
of our results when a central disk component is added to the primaty,
we placed an LSB satellite (the most sensible to
variations in the tidal forces) on a orbit with pericenter of 40 kpc
in the usual isothermal halo and we added to the latter
a disk with a structure similar to that
of the Milky Way (i.e. with a mass $6 \times 10^{10} M_{\odot}$
and scale length of 3 kpc).
The disk was modeled using a Miyamoto-Nagai
potential

\begin{equation}
{\Phi_{M} (R,z)= - {GM \over { \sqrt{R^2 + {(a + \sqrt{z^2 + b^2})}^2}}}}
\end{equation}

where $R=\sqrt{r^2 - z^2}$ is the disk radius in cylindrical coordinates,
$a$ is the scale length ($= 3.5$ kpc)
and $b$ is the scale height. We set $b/a = 2$
(Binney \& Tremaine 1987).
Comparing the results obtained in this run (L25) with an analogous
run in the standard primary potential (L13) we see
only marginal differences: as illustrated in Figure 29, the mass
of the remnant after 6 Gyr is very similar, although the addition of the
a central disk component to the primary halo slightly enhances stripping.
We thus conclude that the tidal shock is always dominated by the
halo, even for quite small pericenter distances.

\section{Star formation history}

From the evolution of the gas surface density observed in the
gasdynamical runs (section 4.2) we can calculate the evolution
of the star formation rate (SFR) of the dwarfs using the Kennicutt law
(Kennicutt 1998). 
This is a phenomenological law, calibrated using large samples of 
disk galaxies,
from dwarf irregulars to large spiral galaxies, and
relates the {\it specific} star formation rate $dM_* /dtdA$ 
(measured in $M_{\odot}$ yr$^{-1}$ kpc$^{-2}$) to the gas
surface density $\Sigma_g$. At a given location in a galaxy we have:

\begin{equation}
{{dM_* \over {dt dA}} = \alpha {\Sigma_g \over t_{dyn}}}
\end{equation}

where $t_{dyn}$ is the dynamical time of the gas. 
The efficiency 
$\alpha$ is $\sim 0.1$ for the best fit indicated by Kennicutt (1998)
using a large dataset, although such data show a large
scatter, especially for faint galaxies:
for non-starburst dwarf galaxies, with SFRs 
$\le 0.1 M_{\odot}$ yr$^{-1}$, $\alpha$  ranges between 0.01 and 1.

We calibrate $\alpha$ for our model galaxies by requiring 
that the satellites, from their formation until the time they
enter into the Milky Way halo, have a constant star formation
rate high enough to produce the stellar mass of our initial models
(models LMg2 and HMg2, see Table 1).
Therefore $\alpha$ may vary depending on their formation epoch.
We consider two formation epochs for the dwarfs, either 12 or 15
Gyr ago, as the age of the oldest stars in dSphs falls within that range
(Grebel 1999, Hernand\'ez et al. 2000), while we always
assume that they entered into the Milky Way halo 10 Gyr ago. The 
SFRs and related parameters are reported in the caption of Figure 30.
With this procedure,  
the specific SFRs for LSB satellites turn out to be of the same order of those 
recently derived for local dIrrs like Sextans A and GR8 using HST photometry 
(Dohm-Palmer et al. 1997, 1998) and are pretty
close to those  derived for isolated LSB galaxies 
using N-Body/SPH simulations (Gerritsen \& de Blok 1999).
LSBs require high values of $\alpha$
because of the low surface density of their disks: instead, for HSBs the
efficiencies are closer to the ``best fit''
given by Kennicutt (1998), which is not surprising since
the galaxies in that sample are mostly normal, high-surface brightness
disks.

\medskip
\epsfxsize=8truecm
\epsfbox{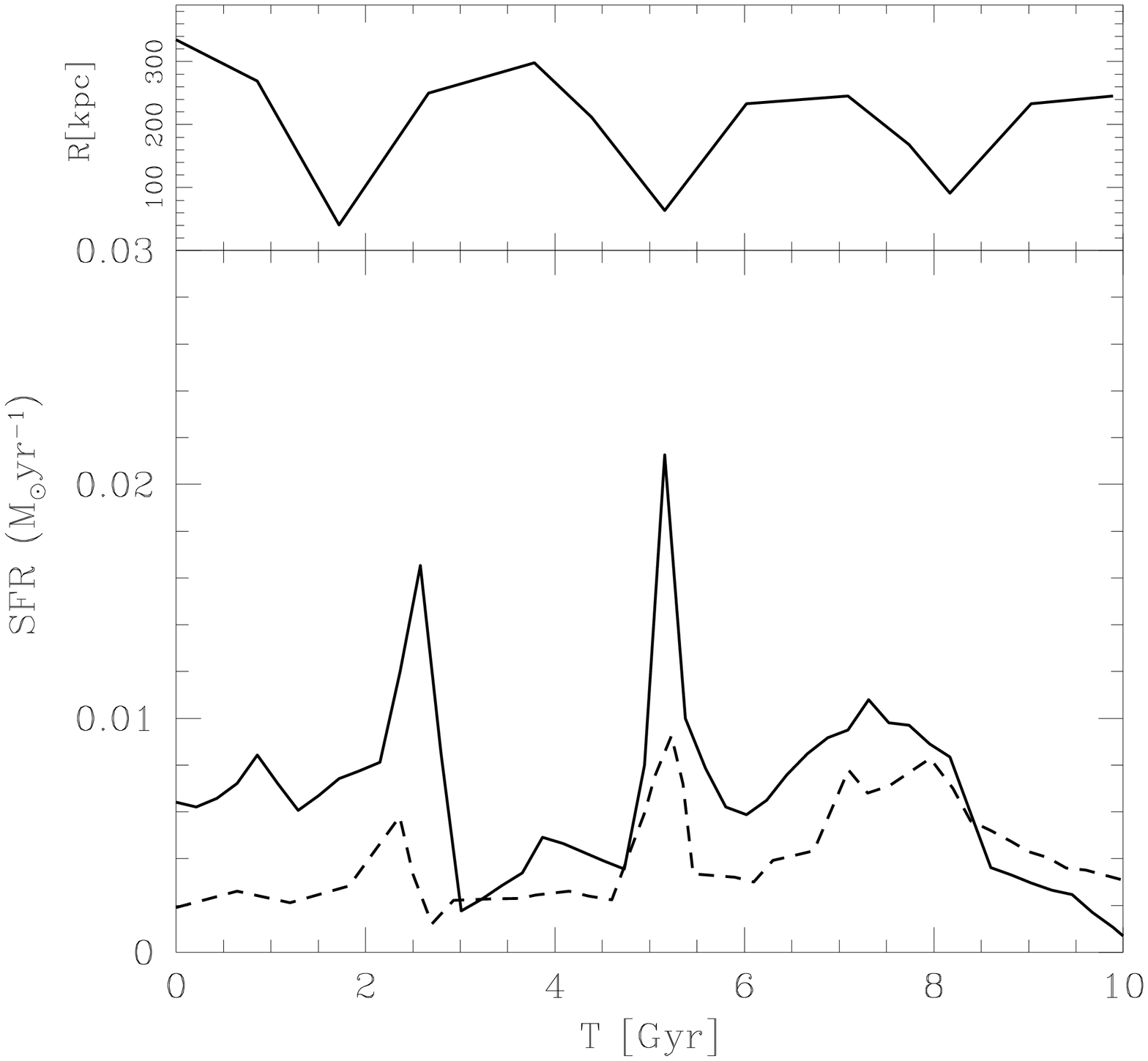}
\medskip
\epsfxsize=8truecm
\epsfbox{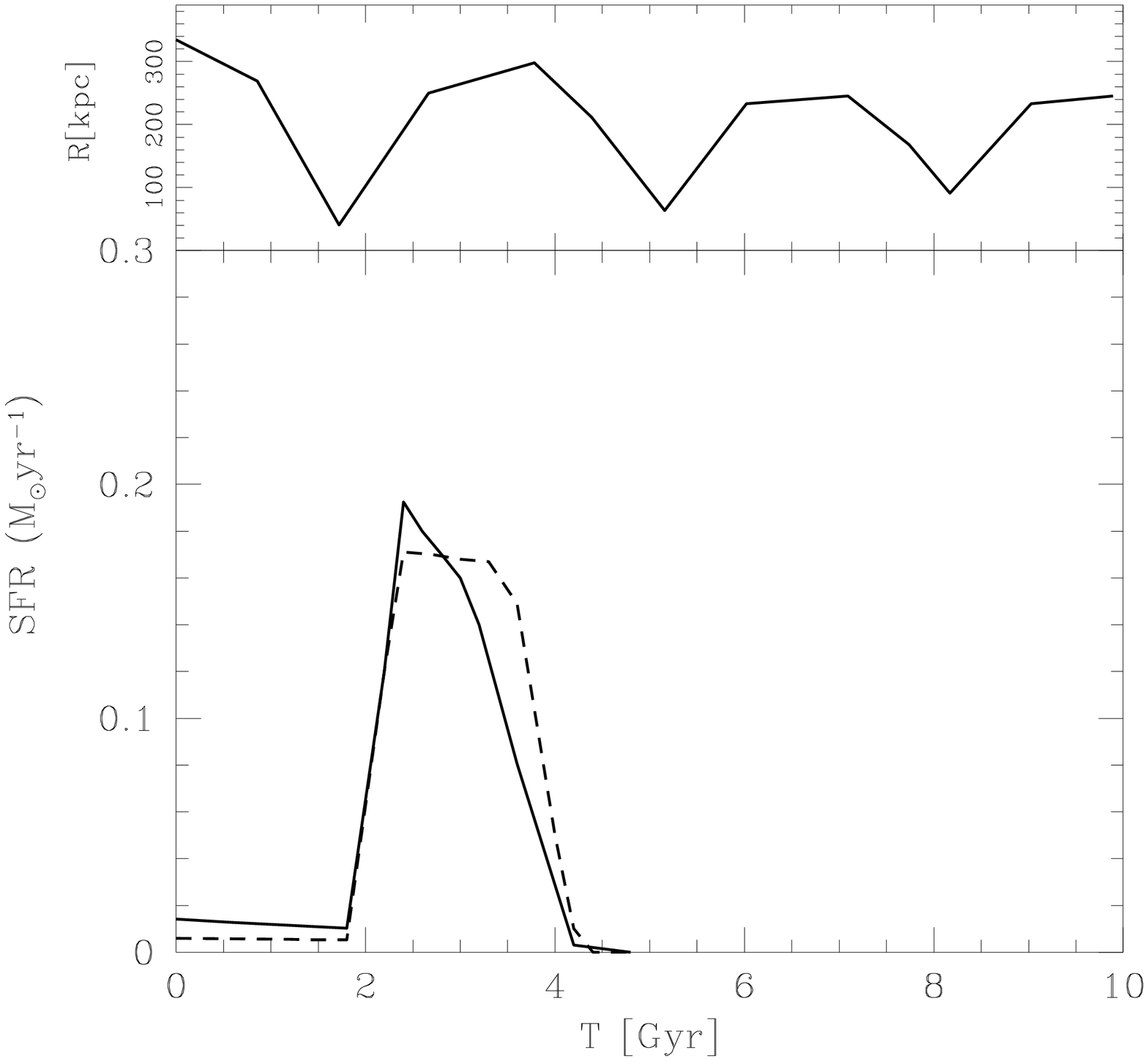}
\figcaption[f30b.eps]{\label{fig:asymptotic}
\small{In top panel: 
star formation history of the LSB satellite for two values of the
efficiency parameter $\alpha = 0.6$ (solid line) and $\alpha = 0.2 $ 
(dashed line). 
The specific SFR before first pericentric passage (i.e. before T=2 Gyr)
are $\sim 0.00025 M_{\odot}$  yr$^{-1}$kpc$^{-2}$ and $\sim 0.00075 M_{\odot}$ yr$^{-1}$ kpc$^{-2}$ for, respectively, $\alpha=0.2$ and $\alpha=0.6$, and the 
total SFR is always calculated integrating over the area of the remnant, 
which is $\sim 4$ kpc in radius.
In bottom panel: star formation history of the HSB satellite for 
$\alpha = 0.42$ (solid line), $\alpha = 0.18$ (dashed line). 
The specific SFR before
first pericentric passage is higher than that of LSBs because of the higher
surface density of the gas, notwithstanding the lower values of $\alpha$
(these are calibrated in order to obtain the same total SFR of the LSBs).
The satellites enter into the Milky Way halo at $T=0$ and 
are then followed for 10 Gyr.}}
\medskip

From the time at which we start the simulations up to the final time, we 
compute, for each output, the specific SFR as a function of radius calculating 
the values of $\Sigma_g$ and $t_{dyn}$ in equally spaced radial bins
(in calculating ${\Sigma}_g$ we always take into account the gas mass that
has already being consumed at the previous output in a given bin).
We then calculate the
average SFR in the dwarf integrating over its area (only out to $2r_h$
because the gas is rapidly stripped at larger radii) and weighting by the
gas mass in each bin.

The resulting star formation histories are shown in Figure 30.
In the case of the LSB  the star formation
rate has a periodic behavior, with peaks of activity quite well
correlated to each pericenter passage. The exact number and height 
of the peaks depends on the efficiency parameter $\alpha$: when this is high,
the gas is mainly consumed in the first two bursts and there is
only a small amount left when the satellite approaches the primary
for the third time. 
The small starbursts are separated
by a timescale of $\sim 3.5$ Gyr, roughly the orbital time 
of the dwarf galaxy.
Qualitatively, these star formation histories bear a striking resemblance 
to those recently derived for Leo I and Carina by Hernandez et al. (2000)
using HST color-magnitude diagrams. Do they also
correspond quantitatively? If we integrate the specific SFRs at 
the peaks obtained
for the LSB over the areas of Carina and Leo I (using directly the
size of the LSB remnant would be misleading as this corresponds more
to an extended dSph like Sextans), for a tidal radius of $\sim$ 750 pc 
(see Mateo 1998), we obtain, for  $\alpha = 0.2$, a SFR
between 150 and 350 $M_{\odot}$/Myr, pretty close to the values
of 150-200 $M_{\odot}$/Myr estimated by Hernand\'ez et al.
(2000). Instead, for a higher star formation efficiency the maximum SFRs are
sensibly higher than the observational estimate.

In the case of the HSB satellite, only one,
very stronger burst occurs after first pericentric passage, depleting
all the gas in less than $\sim 2$ Gyr, before the satellite completes
the second orbit.
Because gas consumption is always extremely fast compared to the orbital
time, the results do not depend sensitively on the efficiency
parameter $\alpha$.
The maximum star formation rate is almost a factor of 10 higher
than in the case of the LSB.
This second type of star formation history
is present also among observed early type dwarfs: Draco and Ursa Minor are well
known examples and even M31's dEs appear to have
had one major episode of star formation (Grebel 1998; Mateo 1998).

Supernovae feedback, which is not included in the derived star
formation histories, could have a significant effect, especially in 
the case of  HSBs, given their strong burst. 
Their star formation rate during the burst is of the
order of that measured in the dwarf galaxies analyzed by Martin (1999), 
which all show strong outflows of gas (even an LG dIrr, Sextans A, is 
included in such a sample sample). Mass loss rates of about $0.3 M_{\odot}$/yr
are thought to be associated with the prominent H$\alpha$ shells observed.
The velocity of the inflow of gas associated with the
bar is of the order of $\sim 10$ km/s, while the expanding shell
produced by supernovae explosions (which would occur almost instantly
compared to the dynamical time of the galaxy) would be moving at up to 100 km/s
(Meurer et al. 1992) and could in principle
reverse the motion of the gas. However, MacLaw \& Ferrara (1999), using
a high-resolution eulerian code to model the ISM in a star-bursting
dwarf within a massive dark matter halo, have shown that, unless the 
dwarf has a total mass as low as  $10^7 M_{\odot}$, the gas mass lost during
these outflows is only a few percent of the total mass even for 
starbursts as intense as $L_{\odot} \sim
10^{40}$ erg/s (comparable to the intensity calculated 
for the HSB satellite).
The holes produced in the gaseous
disk recollapse under the influence of the gravity of the galaxy and will
settle down again $3-4 \times 10^8$ yr after the explosion, switching on
again star formation. We can imagine
these process repeating several times as new supernovae explosions occur.
As a result, the same gas mass will be turned into stars on a longer
time scale than in our model: we should still see a first peak of activity,
but this will be  lower, being soon depressed by the first consistent
supernovae explosion, and will be then followed  by a less intense, extended
activity regulated more by feedback than by the bar inflow.
This might explain why in the dEs, after a first burst, some star 
formation took place even recently (Grebel 1999).

\medskip
\epsfxsize=8truecm
\epsfbox{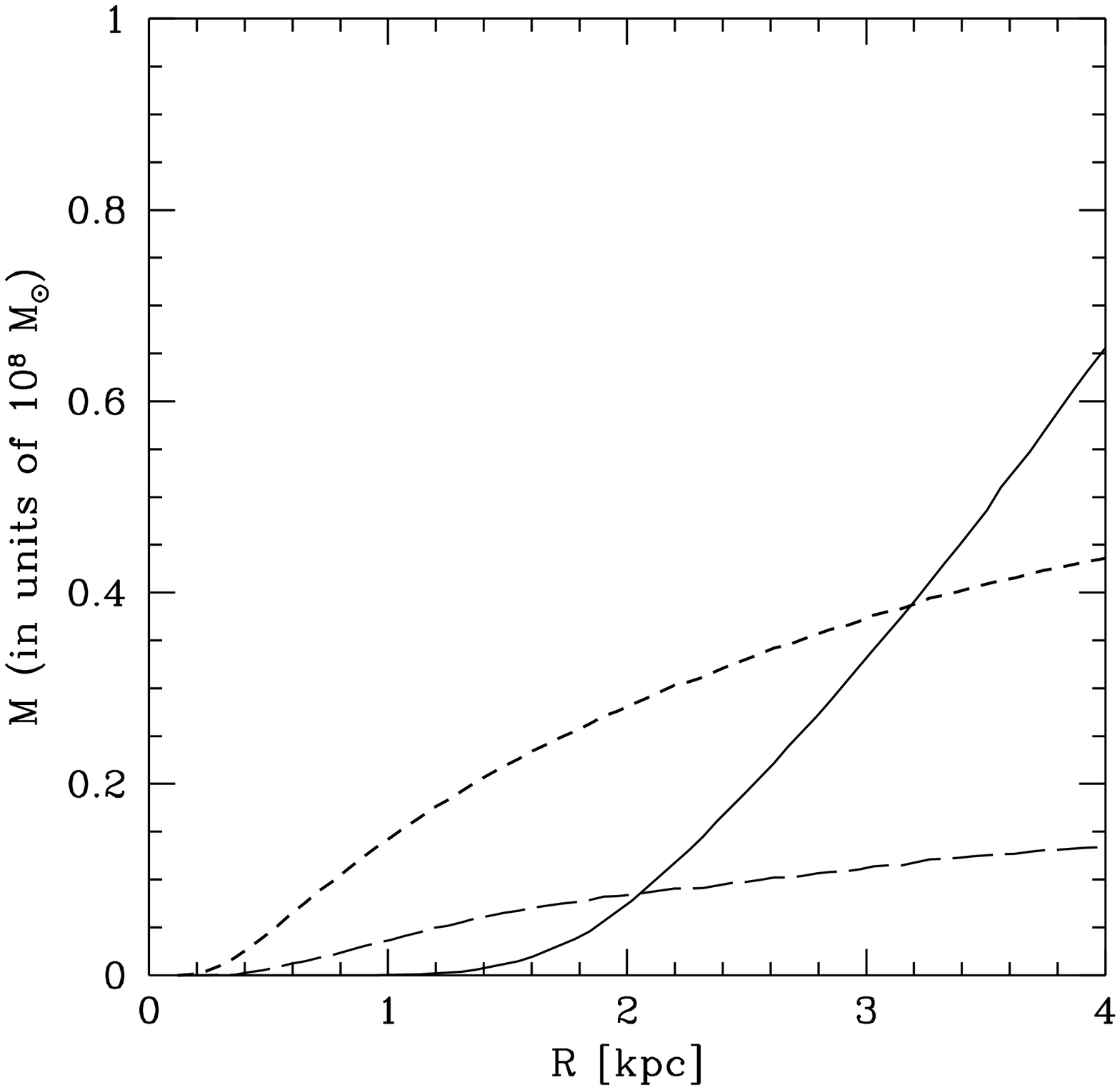}
\figcaption[f31.eps]{\label{fig:asymptotic}
\small{The evolution of the cumulative mass profile for the gas
in the LSB according to the derived star formation history.
We show the results in a region corresponding to the final size of 
the remnant in run L26. The solid line is used for the
initial distribution while the long-dashed and short-dashed lines  
represent the distribution after 10 Gyr for, respectively, $\alpha =0.6$ 
and $\alpha=0.2$.}}
\medskip

We can ask how the SFR history of the GR8-like dwarf would like if we
added gas. The disk stellar surface density of GR8 is pretty
close to that of the HSBs and this would be likely true even 
for the gaseous disk.
Therefore, the specific SFR would be necessarily close to that of HSBs.
A clear bar pattern appears after the second pericenter passage
and would likely produce a strong burst.
The star formation in Draco and Ursa Minor seems to have lasted $\sim 3$
Gyr according to Hernand\'ez et al. (2000) and this constraint would
be approximately satisfied if we adopt for GR8 a SF history similar to the HSB.
Hernand\'ez et al. (2000)
invoke feedback to explain the abrupt truncation of the star formation
for Ursa Minor: in our model this would result automatically from 
gas consumption. However, the GR8 model has a mass so low to fall
in the regime of substantial blow-away of the gas by supernovae 
according to Mac Law \& Ferrara (1999). Adding the effect of supernovae
would actually match the observations better, lowering the
intensity of the burst (those derived by Hernand\'ez et al. (2000) for
Draco and Ursa Minor are substantially weaker)
while still drastically reducing the star formation immediately
after it.

In section 4 we have seen that,
while remnants of HSBs have little scatter 
in their structural properties, remnants of LSBs have a
large scatter, being very sensible to the orbital configuration. On orbit
with large pericenters, for example, the bar instability is stronger
because stripping is less efficient: in this case
a star formation history closer to that of HSBs may result.
Therefore, a
strong  prediction of our model is that dSphs must have a large variety
of star formation histories, while dEs should have very similar histories.
The data seems to support this conclusion (Grebel 1999; Mateo 1998).

\medskip
\epsfxsize=5truecm
\figcaption[LSBgasdistrt160.ps]{\label{fig:asymptotic}
\small{Projected distribution of the stripped gas close to apocenter 
as seen from within the orbital plane in the case of the LSB dwarf
(run L26). 
The colour-code logarithmic density is shown (the darker the colour
the lower the density). The box is  30 kpc on a side. }}
\medskip

From the final density profiles of the gas (Figure 13) we can
argue that a compact stellar nucleus would form in the remnants of
HSBs. A crude conversion of the gas surface density directly into
stellar surface density implies that the central surface brightness 
should be boosted by $\sim 2$ magnitudes inside a scale of 500 pc
assuming the same $(M/L)_{*B}$ of the pre-existing stellar component
(this is a conservative choice shortly after the burst has started
because the blue luminosity would be higher at that time).
Such a central luminosity spike would probably resemble that associated 
to the blue nucleus in NGC205 (Ferguson \& Binggeli 1994).
The fraction of the newly formed stellar population with respect to the 
total stellar mass would be $\leq 50 \%$ in
both the LSBs and HSBs, similar to the initial gas fraction.
This is not surprising because gas and pre-existing
stars are stripped to the same extent and most of the gas that remains
bound is transformed into stars.
However, this means also that if the initial gas/stars ratio in the models 
were $> 1$, as in some dIrrs (e.g. SagDiG in the LG), 
the satellite could in principle form most of its stars during the interaction.

Due to the
bar driven inflow, young stars in the remnant would be principally
located in its central part.
This is what is observed in many LG early type
dwarfs, with Fornax, Carina, Sextans and NGC147 being well known examples
(Grebel 1999).
The much stronger bursts in HSB satellites should give rise to a higher metal
enrichment compared to LSB satellites because the supernovae rate should be
higher. As tidal stirring necessarily implies that the remnants
of the robust HSBs are more luminous than the remnants of 
the fragile LSBs, one would expect 
a correlation between metallicity and luminosity, similar to that observed
in the LG, where bright dEs are more metal-rich than dSphs (Mateo 1998).
On the other end, a higher metallicity spread should be present in dSphs
as a consequence of their star formation being more extended in time.
Radial metallicity gradients, with the more metal rich population 
sitting in the center of the galaxy, as observed in Fornax (Grebel 1999), 
are also a natural consequence of the present model.

\section{The final distribution of cold gas}

Figure 31 shows the final cumulative profile of the cold gas component
in the tidally stirred dwarfs once consumption by star formation is 
taken into account.
Only the results for the LSB are shown
as all the gas is consumed in the HSB, notwithstanding the chosen efficiency.
The final $M_{gas}/M_{stars}$ in the LSB remnant is $< 0.1$ for $\alpha = 0.2$
and $< 0.025$ for  $\alpha = 0.6$.
The limits derived for the LSB remnant for the lower value of $\alpha$,
which yields star formation rates closer to those observed for dSphs,
are slightly higher than the limits inferred for  
$M_{HI}/L$ in dSphs (Mateo 1998). However, our simulations lack 
heating sources, like supernovae, that could 
have ionized or even blown out some of the remaining gas.

\medskip
\epsfxsize=8truecm
\epsfbox{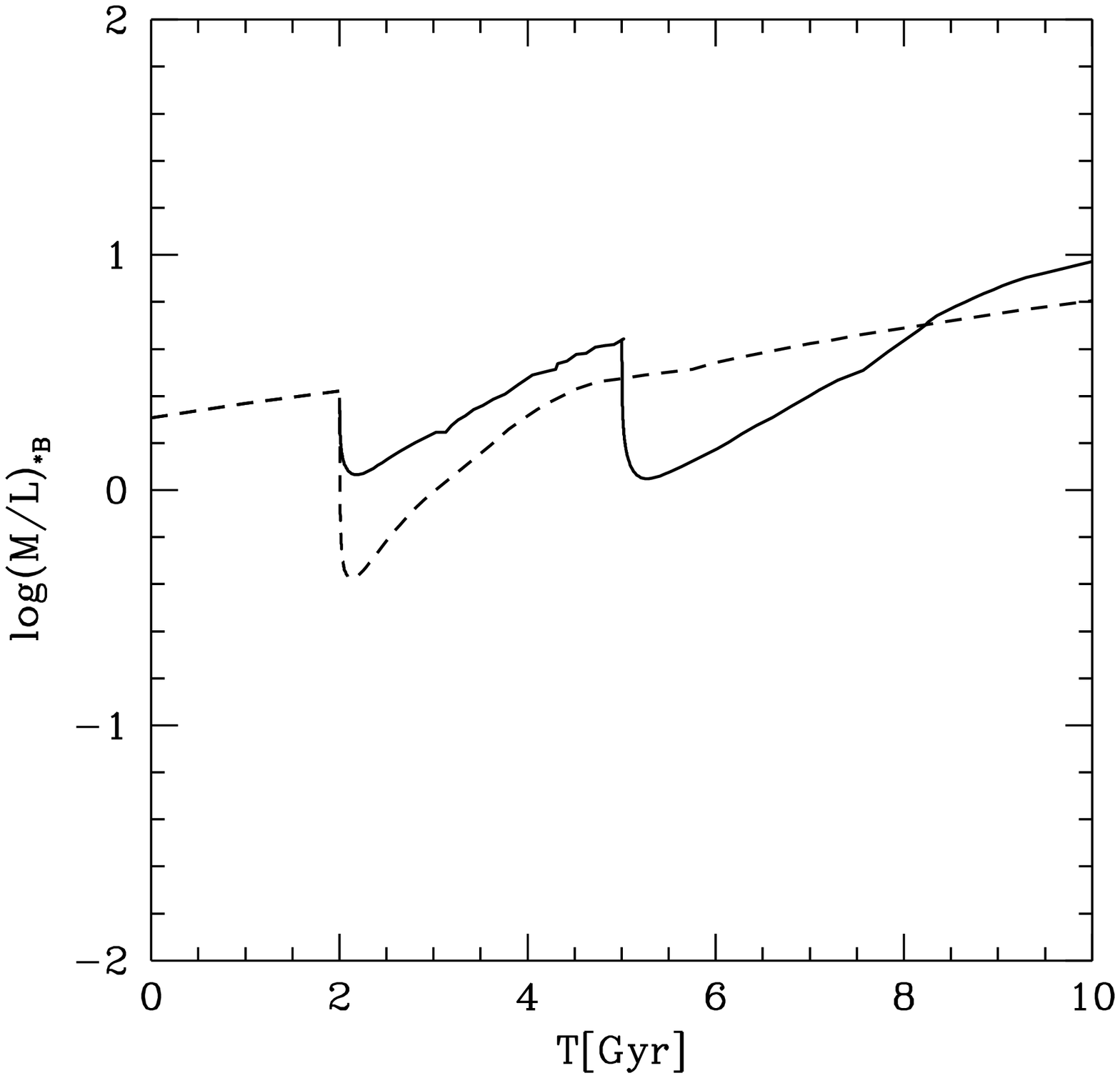}
\figcaption[f33.eps]{\label{fig:asymptotic}
\small{Evolution of the B band stellar mass-to-light ratio as obtained 
with stellar population synthesis models couled with the 
derived star formation histories for the LSB (solid line) and  for the
HSB (dashed line) satellite. The intensity of the
bursts was chosen according to the height of the peaks in the
star formation histories, while we used  $\tau_{02} = 4 \times 10^8$ yr and $
\tau_{03}= 1 \times 10^9$ yr (for the LSB) and
$\tau_{02} = 3 \times 10^8$ yr (for the HSB) as time constants
(see text, in particular equation (7).}}
\medskip

The final distribution of the unbound gas yields some interesting predictions.
Leading and trailing streams of gas form 
that parallel the stellar streams. These keep following
the satellite's trajectory in the primary potential for all the
duration of the simulations (more than 7 Gyr).
In the case of the LSB satellite the gaseous streams comprise more
than 60 \% of the total initial gas mass.
Such extra-tidal gas is never reaccreted by the dwarf galaxy, contrary
to recent claims (Lin \& Murray 2000): particles that end up in the
remnant have always been inside the final tidal radius, as was verified
by tracing back into time the position of the particles.
However, the streams tend to clump at apocenter, where the orbital 
velocity is low, and chance alignments along the line of sight are
possible, like in the case of the stellar streams (section 5.4).
When viewed from within the orbital plane, the streams 
{\it projected onto the plane of the sky} could look like
two gas clouds symmetrically distributed around the bound stellar component
(Figure 32).
The resulting configuration is very reminiscent of that of 
the outer HI distribution in Sculptor (Carignan et al. 1998).
In addition,  similar sistemic velocities would
be measured for the galaxy and the clouds, as these are moving on 
the same orbit: these results could explain the origin of the 
outer gas component recently discovered
around several LG dSphs (Blitz \& Robishaw 2000).

\section{Observable properties of the remnants}

Using stellar population synthesis models (Bruzual \& Charlot 1993)
we calculated the evolution of the luminosity and of
the stellar mass-to-light ratio of the remnants
according to the derived star formation histories.
We adopt synthesis models for stellar populations with a metallicity
of $1/4$ the solar value. 
Gerritsen \& de Blok (1999) find such low values for the metallicity
of LSB galaxies and dSphs in the Local Group have 
even lower metallicities in their older stellar populations (Mateo 1998).
Given a star formation rate $\Psi(t)$, the evolution of the total 
luminosity $L_B$ of the galaxy in the B band is given by:

\begin{equation}
L_B (t) = {\int_0 ^{t} {\Psi(t - \tau){{L}_B}_p (\tau) d \tau}}
\end{equation}

where ${{L}_B}_{p}(\tau)$ is the luminosity contributed by stars of age $\tau$.
We simply assume $\Psi(t) = \Psi_{1} + \Psi_{2}(t) + \Psi_{3}(t)$, where
$\Psi_{1}$ is the constant SFR prior to the first
pericenter passage, equal to $0.3 M_{\odot}$ yr$^{-1}$, and $\Psi_{2(3)}(t)=$ A exp$({-t/{\tau_0}_{2(3)}})$ 
represent the bursts ($\Psi_{3}(t)$=0 for the HSB). The
time and normalization constants of the bursts are derived by naively fitting
the star formation histories obtained for the highest value of
$\alpha$ (see previous section) in order to evaluate an upper
limit for the final luminosity.

In Figure 33 we can see the resulting evolution of the 
B band stellar mass-to-light ratio, $(M/L)_{*B}$,
for both the LSB and the HSB.
The final $(M/L)_{*B}$ are $\sim 5$ in both cases after 7-8 Gyr
(the earlier truncation of the star
formation in the HSB is compensated by the much stronger burst)
decreasing by more than a factor of 2 with respect to their initial
value.

We can now use such information to compute the luminosities of
the remnants from 
$(M/L)_{*B}$ and their final stellar masses. 
Being conservative, we consider the value of  $(M/L)_{*B}$ reached after 7 
Gyr, which is equal to 5, except in the case
of GR8, where we consider the value reached after 10 Gyr,
$(M/L)_{*B}=6.5$, since in our scenario we assumed a much earlier infall time
for the latter, corresponding to $z \geq 2$.
The surface brightness can be obtained as well: the resulting profiles 
were actually shown in section 5.
In section 5.3 we gave lower limits on the final value of $M/L$.
We can now compute more precisely the final $M/L$ by multiplying the obtained
$(M/L)_{*B}$ times the ratio $M_{dark}/M_{stars}$ found in the remnants.
The resulting $M/L$ are in the range  6-30, except
for the remnant of GR8, that would have  $M/L \geq 70$ ($M_{dark}/M_{stars}$ is
$\sim 12$), matching well the values inferred for Draco and Ursa Minor.

\medskip
\epsfxsize=8truecm
\epsfbox{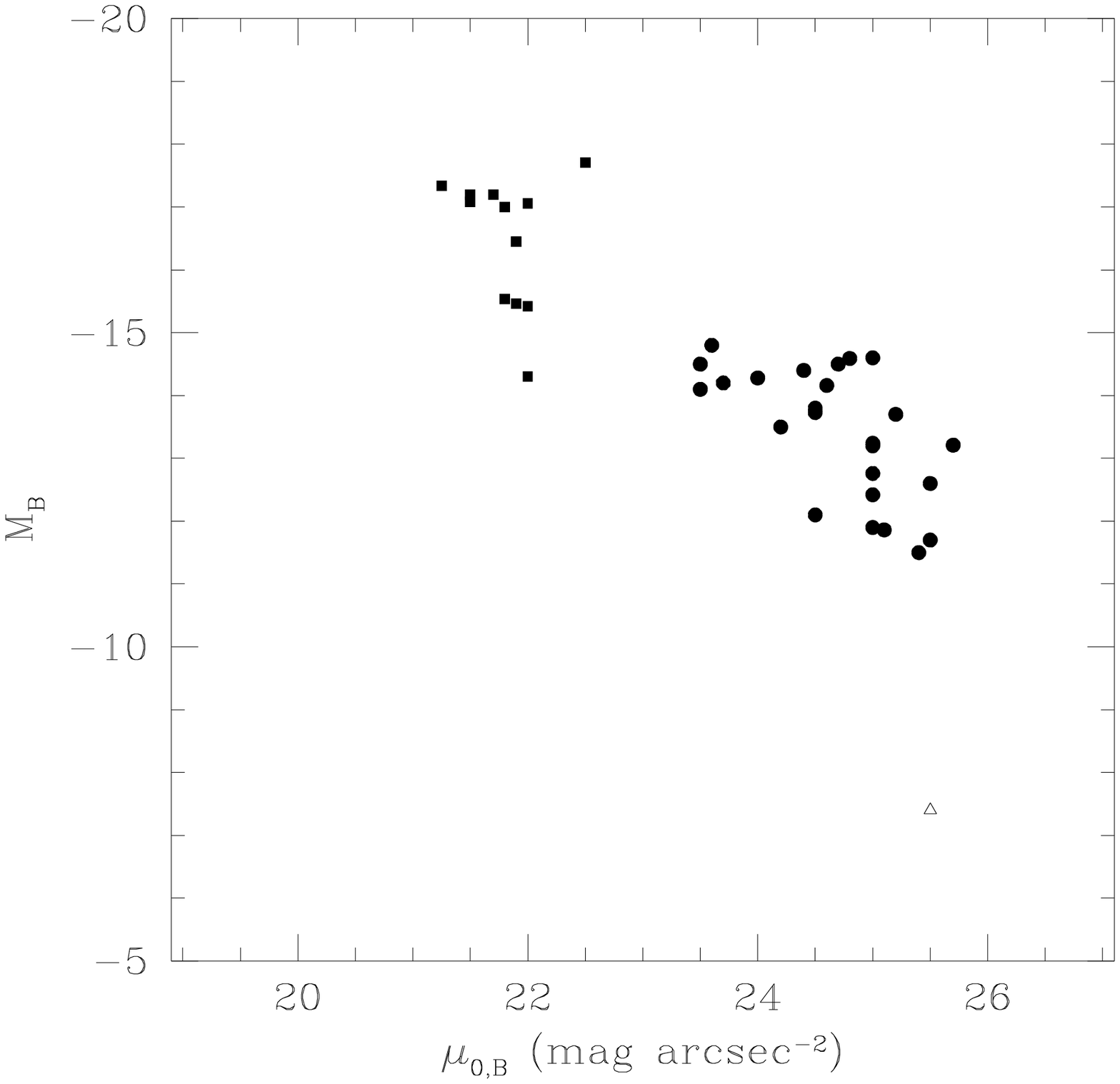}
\figcaption[f34.eps]{\label{fig:asymptotic}
\small{The $\mu - M$ relation for the remnants in our simulations. Filled
squares are used for remnants of HSBs, filled dots for those of LSBs and a triangle  is used for the remnant of GR8. B band magnitudes are measured at the 
Holmberg radius while $\mu_{0,B}$ is the central surface brightness.}}
\medskip

As shown in Mayer et al. (2001), it is possible to construct  
the Fundamental Plane for the remnants coupling the information
directly provided by the simulations with the model for the photometric
evolution described in this section. Such Fundamental Plane 
reproduces well that of observed early-type dwarfs (Ferguson \& Binggeli
1994). 
Here we show the correlation between luminosity and surface brightness, the
$\mu_B - M_B$ relation (Figure 34), which also matches well the observations
(Ferguson \& Binggeli 1994).
The range of surface brightness and luminosity spanned 
by the remnants of LSBs and HSBs corresponds to that spanned by, respectively, 
dSphs and dEs (see Mateo 1998).

\section{Discussion and Conclusion}

We have shown that disk-like dwarf galaxies resembling
nearby dIrrs can be transformed into dSphs and dEs
as a result of 2-3 strong tidal shocks at the pericenters of 
their orbits within the primary halo. Our model naturally
leads to the morphology-density relation observed in the Local Group;
present-day LG dIrrs, being close to the boundary of the LG zero velocity
surface (Van den Bergh 1999), have just decoupled from the Hubble Flow and
have not yet suffered any tidal forcing by the primary galaxies,
while dSphs are the descendants of dwarf disk galaxies that crossed
the boundary of the primary halos several Gyr ago.
The transformation results from a combination of mass loss by
tidal stripping plus removal of angular momentum and vertical heating
due tidally induced dynamical instabilities. The different impact
of mass stripping and of triggered instabilities in HSB dIrrs versus 
LSB dIrrs naturally explains the different structural properties of 
dEs versus dSphs.

The fact that, at least in the Local Group, 
the number of dEs is lower compared to that of dSphs
(only 3 around M31 against nearly 20 dSphs in the whole LG) 
can be understood in terms of the skewness of the number density 
of disk galaxies towards lower surface brightness, especially 
at low luminosities 
(Sprayberry et al. 1997; Bothun 1997; O'Neil \& Bothun 2000).
Photometric surveys of other nearby groups will be useful to
ascertain if the relative proportion of dSphs versus dEs is a general
trend among faint dwarf satellites. According to our model this
can be explained with a prevalence of objects with shallow halo profiles
(namely LSB dIrrs) among the progenitors of early-type dwarfs.
This might have an impact on our understanding of structure formation.
In fact, in CDM cosmogonies the average central density of halos on
the mass scale of dwarfs is higher than that assumed in LSB models 
(the GR8 model being a notable exception), the latter being closer to the
typical central density of small halos in WDM models 
(Bode et al. 2000; Eke et al. 2000). 

On the other end, the origin of the extremely dark matter dominated dSphs,
Draco and Ursa Minor, is well explained assuming a CDM scenario.
In fact, in the latter model one expects that
the dwarf satellites forming at high redshift have a typical central
dark matter density as high as that of the faint dIrr
GR8. These small galaxies are very robust to tides, but if they fall
at $z \ge 2$ into the primary halo
their orbital times are so small that they can double
the number of pericenter passages relative to the satellites
infalling at $z \le 1$. By $z=0$ they transmute into tiny spheroidal
galaxies but maintain very high dark matter contents, just like Draco and
Ursa Minor; one can actually use such high density dwarfs
to place a lower limit  on the mass of the dark matter particle 
in WDM models (Dalcanton \& Hogan 2000).

In this paper we have shown that {\it tidal stirring} 
is effective if satellites have sufficiently low orbital
times, notwithstanding the moderate eccentricities of their orbits.
LSB satellites on circular orbits very close to the Milky Way  can even disrupt
due to their large, low density disks and will produce long-lasting 
density enhancements that resemble the dwarf spheroidal Sagittarius.
However, the farthest dSphs, Leo I and Leo II, must be moving on fairly
eccentric orbits (apo/peri $> 3$) 
in order to be transformed. These distant dwarfs
place also a constraint on the structure of the primary halo, which
must be as large and massive as expected in hierarchical models
for them to have sufficiently small orbital times.

Our evolutionary mechanism is able to explain
the difference in gas content between dwarf spheroidals and dwarf irregulars
as well as the recent observations of extra-tidal gas around some LG dSphs.
In addition, we obtained both periodic star formation histories
and single-burst star formation histories for, respectively, LSB and HSB
satellites, reproducing the variety observed among LG early-type dwarfs.
With tidal stirring operating in a hierarchical structure formation scenario,
we expect dSphs closer to the primary galaxies to have stars typically
older then those in more distant ones because they should have, on average,
an earlier infall time and thus an earlier
bursting phase:Van den Bergh (1994) has pointed out that there 
are indications of such a correlation among LG dSphs.
Van den Bergh (2000) and Grebel (1999)
have recently noted that some peripheral 
globular clusters in LG dSphs (e.g. Fornax) are younger (less than 10 Gyr)
than globular clusters in the outer halo of the Galaxy.
In our scenario they could have formed in the strongest phase of the tidal
shock, immediately following pericenter passage, but in this case we
would expect to find them quite  close to the center: subsequently, dynamical friction can only take them even closer to the center, apparently 
leading to a conflict with observations. 
However, Oh, Lin \& Richer (2000) have found 
that tidal mass loss provoked by the primary galaxy would cause 
a diffusion of the clusters' orbits towards the outer regions 
as they preserve their  radial and azimuthal actions
during the gradual decline of the potential of the dwarf. 
Therefore, we can argue that tidal stirring might provide a
mechanism that explains, at the same time, 
the formation of globular clusters in dSphs and their 
present spatial distribution.

The stars stripped in both the HSB and LSB should be prevalently old
(with an age $> 7$ Gyr), as we know that the new star formation takes place in the inner, bound regions of the satellite. As a result,
we expect the outer stellar halo of the Milky Way to be abundant in old,
low metallicity stars. If, as suggested by the results of the LSB runs,
we assume that Carina, Fornax, Leo I, Leo II,
Sculptor, Sextans and Sagittarius have typically lost $\sim 60\%$ of 
their original stellar mass, 
we find out that dwarf spheroidals should have contributed at least $10^9$
solar masses to the old stellar halo. These should be still recognizable
as part of distinct streams tracing the orbits of the satellites, with
a maximum surface brightness (at the turning points of
the orbits) of just 30 mag
arcsec$^{-2}$ (in the B band).  Spectroscopic evidence for stellar streams
from the dSph Carina has been claimed (Majewski et al. 2000)
and the northern stream of Sagittarius or a new stream associated 
to another disrupted nearby dwarf has recently been observed 
(Martin\'ez-Delgado et al. 2000).
Future astrometric missions, like SIM and GAIA (Gilmore et al. 1998; 
Helmi et al. 2000) should reveal such faint features and will also carry 
out high-quality measurements of proper motions for many satellites 
of the Milky Way, thus providing a test for the orbital
configurations used in this model.

Stellar streams projected on the plane of the sky close to the
line of sight can produce
subtle observational effects in the case of fragile LSB satellites, thereby
inflating the measured velocity dispersions.
Their signature should be an apparent rotation of
the dwarf and could have been observed in Ursa Minor (Hargreaves et al. 1994a).
Instead, velocity dispersion 
anisotropy cannot inflate significantly the measures
of the mass-to-light ratios.

From the evolution of the stellar mass-to-light ratio, it turns out
that HSB satellites have $(M/L)_{*B}$ as low as $0.4$ during the bursting
phase: the high blue luminosity would originate mainly from the central part
of the galaxies, which could thus resemble the
blue compact dwarfs(BCDs) observed by Guzm\'an et al. (1997) at
intermediate redshift ($z \sim 0.5- 0.7$). 
The velocity widths of these galaxies
correspond to those of the largest among our HSB satellites (between
50 and 75 km/s) and the total blue luminosities would also
be comparable (model HM1 reaches a luminosity of about $M_B=-20$
in this phase). Interestingly, Guzm\'an et al. (1997) conclude
that fading would transform BCDs into objects similar to NGC205
and the other dEs after several Gyr and that none of them
could fade so much as to resemble dSphs. The latter conclusion 
is very reminiscent of our claim that HSB dIrrs are the progenitors of dEs
and not of the faint dSphs. 
Future investigations aimed at searching for massive
galaxies in the vicinity of blue compact dwarfs may shed light on 
the connection with tidally stirred dwarfs.

\acknowledgments

The authors thank G.Gavazzi, C.Gallart, M.Vietri, R.Bower, A.Helmi 
and S.Ghigna for stimulating discussions.
Simulations were carried out at the CINECA (Bologna) and ARSC 
(Fairbanks) supercomputing centers.

\newpage

\hoffset -2cm
\begin{table}
\begin{center}
\begin{tabular}{c|c|c|c|c|c|c|c|c|c|c} \hline \hline
Model & $M_{200}/M_P$ & $M_d/M_{200}$ &
$(M_{200}/M_d)_{opt}$ & $r_c/r_t$ & $r_t$ & $r_h$ & $V_c$ & ${{\mu}_0}_B$ &
$Q_{opt}$ & $X_2$ \\ \hline
&  &  & $(M_{\odot}/L_{\odot})$ &
 & (kpc)  & (kpc) & (km s$^{-1})$& & & \\ \hline

HM1 (HSB) & 0.032 & 0.016 & 
6 & 0.02  & 140 & 2  & 75 & 22 & 2 & 2 \cr
HM2 (HSB) & 0.01 & 0.016 & 
6 & 0.02 & 95 &  1.35 & 50 & 22.5 & 2 &  2 \cr
HMg2 (HSB)  & 0.01 & 0.016 & 
6 & 0.02 & 95 &  1.35 & 50 & 22.5 & 2 &  2 \cr
HZ (HSB) & 0.013  & 0.016  &  
6 & 0.02  & 49.5 & 0.7  & 75 & 21.5 & 2 & 2 \cr
HM1rc03 (HSB) & 0.032 & 0.016 & 
10 & 0.006 & 140 &  2 & 75 & 22 & 2 &  2.7 \cr
LM1 (LSB)  & 0.032 & 0.016 & 
12  & 0.036 & 140 & 4.8 & 75 & 24 & 2 & 3  \cr
LMH (LSB,$z_s=0.3r_h$)  & 0.032 & 0.016 & 
12  & 0.036 & 140 & 4.8 & 75 & 24 & 2 & 3 \cr
LM2  (LSB) & 0.01 & 0.016   & 12  & 0.036 & 95 &
3.2 & 50 & 24.5 & 2 & 3 \cr
LMg2 (LSB) & 0.01 & 0.016   & 12  & 0.036 & 95 &
3.2 & 50 & 24.5 & 2 & 3 \cr
LZ (LSB) & 0.013 & 0.016    & 12  & 0.036 & 140 &
1.69 & 75 & 23.5 & 2 & 3 \cr
LM1rc03 (LSB)  & 0.032 & 0.016  &
15  & 0.0108 & 140 & 4.8 & 75 & 24 & 2 &  5 \cr
LM1Q4 (LSB)  & 0.032 & 0.016  &
12  & 0.036 & 140 & 4.8 & 75 & 24 & 4 & 3 \cr
GR8  & $1.38 \cdot 10^{-4}$ & $2.174\times 10^{-3}$ & 32 & 0.014 & 7.8 &
0.076 &  17 & 22.5  & 2  &  6 \cr
\hline
\end{tabular}
\end{center}
\medskip
\caption{Models of satellites}
The structural parameters of the dwarf galaxy models are shown.
$M_p$ is the mass of the Milky Way halo, 
$M_d$ is the mass of the satellite's disk, 
while $M_{200}$ is the virial mass of the halo of the satellite;
$r_t$ is the truncation radius 
of its halo (= $R_{200}$), $r_h$ is its disk scale length and
$r_c$ is its halo core radius. The circular velocity $V_c$ is 
measured at the virial radius of the satellite (i.e. $V_c = V_{200}$).
The mass-to-light ratio at the optical
radius (= $3 r_h$) is ${(M_{200}/M_d)}_{opt}$ and ${{\mu}_0}_B$ is the B band
central surface brightness in mag arcsec$^{-2}$.
Model LMg2 and HMg2 have a gaseous component in the disk, with mass
$M_g = 0.35 M_d$.
The stability parameter $X_2$ is estimated at $r_h$,
while $Q$ is estimated at the optical radius.
\end{table}

\newpage

\begin{table}
\begin{center}
\centering
\begin{tabular}{c|c|c|c|c} \hline \hline
run & model & $R_{peri}$ & $R_{apo}$ & $\theta$ \\   \hline
&  & (kpc) & (kpc) & degrees  \\ \hline

L01 & LM1 & 40 & 360 & 0 \cr
L02 & LM1 & 80 & 360 & 0 \cr
L03 & LM1 & 40 & 360 & 180 \cr
L04 & LM1 & 40 & 360 & 40 \cr
L05 & LM1 & 40 & 360 & 63 \cr
L06 & LM1 & 80 & 360 & 63 \cr
L07 & LM1 & 40 & 360 & 125 \cr
L08 & LM1 & 40 & 360 & 150 \cr
L09 & LM1 & 240 & 360 & 63 \cr
L10 & LM2 & 40 & 360 & 0 \cr
L11 & LM2 & 80 & 360 & 0 \cr
L12 & LM2 & 80 & 360 & 180 \cr
L13 & LM2 & 40 & 360 & 40 \cr
L14 & LM2 & 40 & 360 & 63 \cr
L15 & LM2 & 80 & 360 & 90 \cr
L16 & LM2 & 40 & 360 & 130 \cr
L17 & LM2 & 125 & 250 & 63 \cr
L18 & LM2 & 200  & 360 & 90 \cr 
L19 & LM1 & 75   & 120 & 63 \cr
L20 & LM1 & 75  & 75 & 63 \cr
L21 & LM1rc03 & 80 & 360 & 0 \cr
L22 & LMH & 80 & 360 & 0 \cr
L23 & LM1Q4 & 80 & 360 & 0 \cr
L24 & LM1 (minimal Milky Way potential) & 25 & 220 & 40 \cr
L25 & LM1 (Miyamoto disk included) & 40 & 360 & 40 \cr
L26 & LMg2 & 40 & 360 & 40 \cr
\hline
\end{tabular}
\end{center}
\medskip
\caption{Simulations}
Initial conditions of the runs.
We indicate the galaxy model used,
the pericenter distance $R_{peri}$, the apocenter distance $R_{apo}$, 
and the angle between the spin of the disk and the
orbital angular momentum ($\theta$).
\end{table}

\begin{table}
\begin{center}
\centering
\begin{tabular}{c|c|c|c|c} \hline \hline
run & model & $R_{peri}$ & $R_{apo}$ & $\theta$ \\   \hline
&  & (kpc) & (kpc) & degrees \\ \hline

LZ1 & LZ & 40 & 360 & 0 \cr
LZ2 & LZ & 80 & 360 & 0 \cr
LZ3 & LZ & 40 & 360 & 180 \cr
LZ4 & LZ & 40 & 360 & 40 \cr
LZ5 & LZ & 40 & 360 & 63 \cr
LZ6 & LZ  & 40 & 360 & 125 \cr
LZ7 & LZ  & 75 & 120 & 63 \cr
H01 & HM1 & 40 & 360 & 0 \cr
H02 & HM1 & 80 & 360 & 0 \cr
H03 & HM1 & 40 & 360 & 180 \cr
H04 & HM1 & 40 & 360 & 40 \cr
H05 & HM1 & 40 & 360 & 63 \cr
H06 & HM1 & 40 & 360 & 125 \cr
H07 & HM1 & 40 & 360 & 150 \cr
H08 & HM2 & 40 & 360 & 40 \cr
H09 & HM2 & 125 & 250 & 40 \cr
H10 & HM2 & 75 & 120 & 63 \cr
H11 & HM1rc03 & 50 & 120 & 0 \cr
H12 & HMg2 & 40 & 360 & 40 \cr
HZ1 & HMz & 40 & 360 & 0 \cr
HZ2 & HMz & 40 & 360 & 180 \cr
HZ3 & HMz & 40 & 360 & 40 \cr
HZ4 & HMz & 40 & 360 & 63 \cr
HZ5 & HMz & 40 & 360 & 125 \cr
GR81 & GR8 & 40 & 360 & 63 \cr
GR82 & GR8 & 12 & 110 & 63 \cr
\hline
\end{tabular}
\end{center}
\medskip
\caption{Simulations}
Initial conditions of the runs.
We indicate the galaxy model used,
the pericenter distance $R_{peri}$, the apocenter distance $R_{apo}$, 
and the angle between the spin of the disk and the
orbital angular momentum ($\theta$).
\end{table}

\end{document}